\documentclass[10pt,journal,compsoc]{IEEEtran}
\ifCLASSOPTIONcompsoc
  \usepackage[nocompress]{cite}
\else
  \usepackage{cite}
\fi

\ifCLASSINFOpdf
   \usepackage[pdftex]{graphicx}
   \usepackage{subfigure}
   \usepackage{booktabs}
   \usepackage{bbding}
   \usepackage[flushleft]{threeparttable}
   \usepackage{mathrsfs}
    \usepackage{float}
    \usepackage{ragged2e}
    \usepackage{color}
    \usepackage{multirow}
    \usepackage{diagbox}
    \usepackage{subfiles}
\else
\fi
\usepackage{amsmath}
\usepackage{algorithmic}

\usepackage{array}

\begin{document}
%
\title{VNVC: A Versatile Neural Video Coding Framework for Efficient Human-Machine Vision}
%
%
%
%

\author{Xihua~Sheng,
        Li~Li,~\IEEEmembership{Member,~IEEE,} Dong~Liu,~\IEEEmembership{Senior~Member,~IEEE}, and Houqiang Li,~\IEEEmembership{Fellow,~IEEE}
\IEEEcompsocitemizethanks{\IEEEcompsocthanksitem Date of current version \today. (Corresponding author: Li Li). 
\IEEEcompsocthanksitem The authors are with the CAS Key Laboratory of Technology in Geo-Spatial Information Processing and Application System, University of Science and Technology of China, Hefei 230027, China (e-mail: xhsheng@mail.ustc.edu.cn; lil1@ustc.edu.cn; dongeliu@ustc.edu.cn; lihq@ustc.edu.cn). \protect\\}
}

%
%

\markboth{Journal of \LaTeX\ Class Files}%
{Sheng \MakeLowercase{\textit{et al.}}: VNVC: A Versatile Neural Video Coding Framework for Efficient Human-Machine Vision}
%



\IEEEtitleabstractindextext{%
\begin{abstract}
\justifying
Almost all digital videos are coded into compact representations before being transmitted. Such compact representations need to be decoded back to pixels before being displayed to humans and – as usual – before being enhanced/analyzed by machine vision algorithms. Intuitively, it is more efficient to enhance/analyze the coded representations directly without decoding them into pixels. Therefore, we propose a versatile neural video coding (VNVC) framework, which targets learning compact representations to support both reconstruction and direct enhancement/analysis, thereby being versatile for both human and machine vision. Our VNVC framework has a feature-based compression loop. In the loop, one frame is encoded into compact representations and decoded to an intermediate feature that is obtained before performing reconstruction. The intermediate feature can be used as reference in motion compensation and motion estimation through feature-based temporal context mining and cross-domain motion encoder-decoder to compress the following frames. The intermediate feature is directly fed into video reconstruction, video enhancement, and video analysis networks to evaluate its effectiveness. The evaluation shows that our framework with the intermediate feature achieves high compression efficiency for video reconstruction and satisfactory task performances with lower complexities.
\end{abstract}

}

\maketitle

\IEEEdisplaynontitleabstractindextext

%
\IEEEpeerreviewmaketitle

\IEEEraisesectionheading{\section{Introduction}\label{sec:introduction}}


%
%
%
%
\IEEEPARstart{V}{ideos} contributed to 75\% of all Internet traffic in 2017,
and the percentage reached 82\% in 2022~\cite{barnett2018cisco}. A digital video is composed of several frames with each frame as a static image. For a typical 1080p video in the standard dynamic range, 24 bits are
used to represent the color \emph{(r,g,b)} of each pixel, and the size of one frame can be approximated as 6\emph{Mbytes}.  For a video of 30 frames per second, the bit rate can
be as high as 180\emph{Mbytes} per second without compression. Therefore, encoding videos into compact representations with smaller sizes is an urgent requirement to reduce the transmission cost. After receiving the transmitted compact representations, they need to be decoded back to the pixel domain before being displayed to humans.  \par
In addition to being decoded for display, the coded video representations also need to be enhanced/analyzed by downstream human and machine vision algorithms in various video applications such as online meeting, autonomous driving, and smart city~\cite{gao2021recent}. Video enhancement aims to improve video quality for a better human visual quality,  such as video denoising~\cite{tassano2019dvd,tassano2020fastdvdnet}, video super-resolution~\cite{tian2020tdan,wang2019edvr, pan2021deep}, and video deblurring~\cite{pan2023cascaded,zhu2022deep}. Video analysis refers to analyzing the video scenes, such as action recognition~\cite{lin2020tsm,carreira2017quo,feichtenhofer2019slowfast}, video object detection~\cite{zhu2017flow,sun2020semantic}, video multiple object tracking~\cite{pang2021quasi,zhang2022bytetrack}, and video object segmentation~\cite{yang2021aot, caelles2017one}. To perform these downstream human and machine vision algorithms, the coded representations usually need to be fully decoded back to pixels first. To perform machine vision tasks more efficiently, JPEG-AI~\cite{ascenso2023jpeg, ascenso2021jpeg}, which is a neural image coding standard for human and machine vision, proposes to enhance/analyze the coded representations directly without fully decoding them into pixels. However, in the existing literature, there is no neural video coding framework with similar functionalities to JPEG-AI.\par

Therefore, in this paper, our goal is to propose a versatile neural video coding framework for efficient human and machine vision (VCHM). To make the proposed video coding framework versatile for both human and machine vision, we put forward three requirements similar to JPEG-AI~\cite{ascenso2023jpeg, ascenso2021jpeg} for this framework, i.e., it should: 1) use a single bitstream for both human and machine vision; 2) achieve high compression efficiency for video reconstruction and satisfactory performance for human and machine vision tasks; 3) enable directly performing human and machine vision algorithms without decoding the coded representations into pixels. The first requirement is to make the framework versatile to human and machine vision tasks.  The second requirement is to reduce transmission and storage costs while keeping high reconstruction quality and task performance. The third requirement is to save more computational complexity for human and machine vision tasks. There is no existing solution for VCHM that meets all these requirements.\par

The existing solutions of VCHM can roughly be divided into three categories. Solutions in the first category focus on the compression of deep features of neural networks. Videos are firstly converted to features by the task networks on the front end. The features are packed into a video sequence and compressed by video codecs or directly compressed by feature codecs~\cite{duan2015overview,duan2018compact} to compact representations. The compact representations are decoded into features and used by the task networks on the server end. Although this kind of solution can offload part of the computation from servers to front-end devices, they need to transmit an additional bit stream when human visualization is required. Although generative models can be used  to reconstruct videos from the deep features~\cite{duan2020video}, the generated videos are of low quality, especially at high bit rates, which cannot meet the requirement of high human visual quality.\par

Solutions in the second category explore to use a scalable bit stream~\cite{Xia2020AnEC,jin2022semantically,sun2020semantic, choi2022scalable} to support the hybrid human-machine vision. High-level features, such as the output of the task networks (class IDs and bounding boxes for objects)~\cite{jin2022semantically,sun2020semantic}, and low-level features, such as motion and content information of objects and background, are extracted and compressed separately. The bit streams of high-level features and low-level features are organized and combined into a scalable bit stream. Only part of the scalable bit stream needs to be transmitted and decoded depending on the machine vision tasks at the server end. This kind of solution enables using a single bit stream for both human and machine vision. However, the scalable coding paradigm may decrease the compression efficiency for video reconstruction.  \par

\begin{figure*}[htb]
  \centering
   \includegraphics[width=0.88\linewidth]{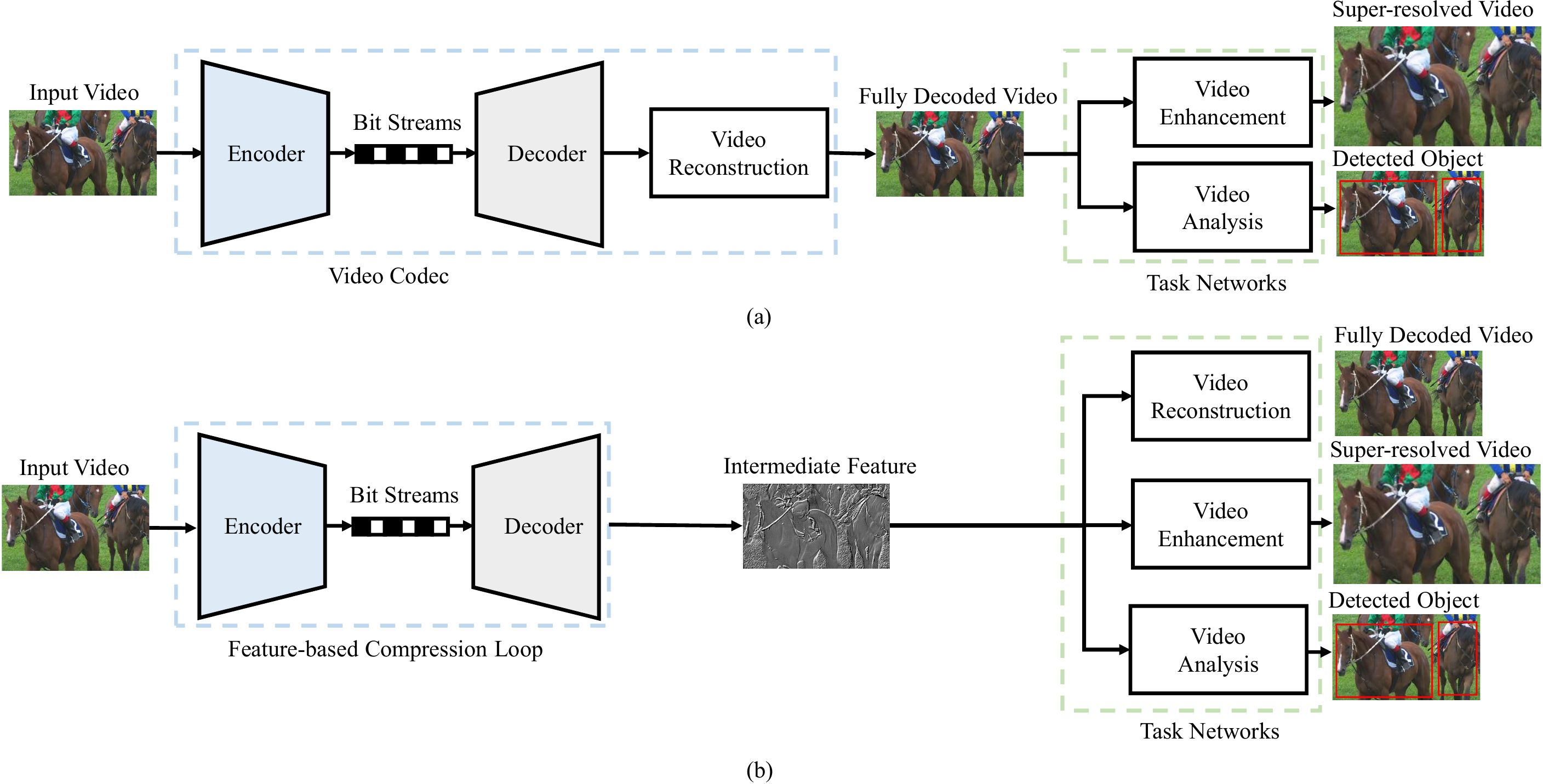}
   \caption{(a) Existing video codecs need to compress input video into bit streams and decompress the bit streams to a fully decoded video in the pixel domain first. Then, video enhancement and video analysis algorithms are performed on the pixel-domain decoded video. (b) Overview of our proposed versatile neural video coding (VNVC) framework. The framework is comprised of a feature-based compression loop and task networks. The feature-based compression loop encodes the input video into bit streams and decodes them to intermediate features. Obtaining the partially decoded intermediate features, the task networks allow users to choose to perform video reconstruction, video enhancement, such as video super-resolution, and video analysis, such as video object detection. }
   \label{fig:framework}
\end{figure*}

Solutions in the third category need to decode the bit stream back to pixels first and then perform machine vision algorithms. As illustrated in Fig.~\ref{fig:framework} (a), this kind of solution needs to compress the videos into bit streams first. Then the bit streams are decompressed back to the pixel domain and fed into the task networks. This kind of solution can achieve high compression efficiency for video reconstruction. However, it cannot meet the requirement of directly performing machine vision tasks without decoding the bit stream into pixels. Some earlier neural video codecs~\cite{lu2020end,lin2020m,hu2022fvc} may have the potential to meet the requirement but their compression efficiency is inferior to the traditional video codecs. Recent conditional coding-based neural video codecs~\cite{li2021deep,sheng2022temporal,shi2022alphavc,canfvc,li2022hybrid} rely on learning temporal contexts and have caught up with or even surpassed traditional video codecs towards compression efficiency under certain conditions. However, learning temporal contexts from previous frames makes their decoding highly coupled with video reconstruction. It is infeasible to apply them to vision tasks without fully decoding the bit stream to pixels. \par

To meet the above-mentioned three requirements at the same time, we propose a versatile neural video coding framework (VNVC).
As illustrated in Fig.~\ref{fig:framework} (b), our proposed framework contains a feature-based compression loop and downstream task networks. In the feature-based compression loop, neither the encoder nor the decoder needs to obtain pixel-domain reconstructed frames. The input frame is encoded into a single bitstream and partially decoded to an intermediate feature that is obtained before performing reconstruction. The intermediate feature can be referenced to estimate motion vectors by a cross-domain motion encoder-decoder and learn temporal contexts by a feature-based temporal context mining module for encoding (resp. decoding) the following frames. In the task networks, the partially decoded intermediate features can be reconstructed back to videos (pixels) or enhanced/analyzed directly. We evaluate our proposed framework on various well-established benchmark datasets.  The evaluation results show that our proposed framework can achieve high compression efficiency for video reconstruction and satisfactory performance for video enhancement and video analysis tasks with lower complexities.\par

Our contributions are summarized as follows:
\begin{itemize}
    \item We propose a versatile neural video coding framework that uses a single-bitstream, compact coded representation, targeting video reconstruction, video enhancement, and video analysis tasks simultaneously.

    \item We propose to partially decode the compact coded representation into an intermediate feature that can be directly fed into the downstream video enhancement and video analysis algorithms, without the need for pixel-domain reconstructed videos.

    \item Compared with pixel-domain reconstructed videos, with the partially decoded intermediate features, much decoding complexity can be saved and satisfactory task performance can be achieved for downstream video enhancement and video analysis tasks.
    
    \item Compared with pixel-domain reference frames, with the partially decoded intermediate features as reference, much encoding complexity can be saved.

\end{itemize}
The remainder of this paper is organized as follows. Section~\ref{sec:related_work} gives a brief review of related work. Section~\ref{sec:overview} and Section~\ref{sec:opti} introduce the architectures and training strategies of our framework. Section~\ref{sec:setup} and Section~\ref{sec:experiment} present the experimental setup and experimental results of the framework. Section~\ref{sec:ablation} analyzes the influence of proposed key technologies. Section~\ref{sec:conclusion} concludes this paper.

\section{Related Work}\label{sec:related_work}
\subsection{Video Coding for Hybrid Human-Machine Vision}
Existing solutions of video coding for hybrid human-machine vision can be divided into three categories. \par
In the first category, solutions focus on feature compression. Features play an important role in machine vision tasks. In view of the necessity of transmitting compact descriptors of features, MPEG has finalized the standardization of SIFT-like \cite{lowe2004distinctive} feature descriptors for visual search (CDVS) (ISO/IEC15938-13)~\cite{duan2015overview} in Sep. 2015 and combination of SIFT-like feature descriptors and deep learning-based feature descriptors for video analysis (CDVA) (ISO/IEC15938-15)~\cite{duan2018compact} in Jul. 2019 for efficient and effective image/video retrieval and analysis. With the deployment of learning-based applications, Chen et al.~\cite{PMID:31562087} proposed to compress the deep features of neural networks. However, even for machine vision tasks, further human analysis is required when the computer recognizes abnormal actions or detects the target object. These feature coding-based schemes only focus on feature compression while ignoring human visualization.  Duan et al.~\cite{duan2020video}  leveraged the generative models to reconstruct videos from features. However, the generated videos are of low quality, especially at a high bit rate, which cannot obtain a high human visual quality. \par

In the second category, solutions try to use a  scalable bit stream to support human and machine vision simultaneously. Xia et al.~\cite{Xia2020AnEC} proposed to connect signal-level and task-level compact descriptors in a scalable manner. Only part of the scalable bit stream needs to be transmitted and decoded depending on subsequent tasks.
Jin et al.~\cite{jin2022semantically} extended the scalable bit stream to a semantically structured bit stream. Apart from the intermediate layer features of task networks, they also organized the output of task networks, such as bounding boxes of objects, into the bit stream. Choi and Bajić~\cite{choi2022scalable} proposed a scalable video coding framework that supports object detection through its base layer and human vision via its enhancement layer. This kind of solution enables using a single bit stream for both human and machine vision. However, the scalable coding paradigm may decrease the compression efficiency for video reconstruction. \par
In the third category, solutions compress the videos using video codecs and then fully decode the bit stream into pixels for human and machine vision tasks. The anchor of a relative standard--video coding for machine (VCM)~\cite{zhang2021use}--developed by MPEG adopts this pattern. This kind of solution can achieve high compression efficiency for video reconstruction. However, it is inefficient for them to decode the bit stream into pixels before performing machine vision algorithms. Much computational burden is consumed for pixel-domain video reconstruction and the speed of vision tasks is decreased.\par

To make the video coding framework versatile for both human and machine vision, we follow the requirements of JPEG-AI~\cite{ascenso2023jpeg,ascenso2021jpeg}, which is the first image coding standard based on machine learning for both human and machine vision, and propose three strict requirements for video coding: 1) use a single bit stream for both human and machine vision; 2) achieve high compression efficiency for video reconstruction and satisfactory performance for human and machine vision tasks; 3) enable directly performing human and machine vision algorithms without decoding the bit stream into pixels. As far as we know, in the existing literature, there is no video codec that simultaneously meets these requirements.\par

\subsection{Neural Video Coding}
Recently, neural video coding has achieved great success. Existing neural video codecs can be roughly categorized into four classes. In the first class, schemes~\cite{lu2020end,hu2022fvc,wu2018video} follow a residual coding pattern. Motion-compensated prediction is first performed in the pixel domain or feature domain. Then the residue from the prediction is compressed by an autoencoder. In the second class, schemes~\cite{liu2020conditional,mentzer2022vct} apply existing image codecs to video compression. Each frame is compressed into a compact representation by an image codec. Temporal entropy models are then built to exploit the temporal correlations between each compact representation. In the third class, schemes~\cite{Habibian_2019_ICCV,sun2020high} use the 3D convolution-based autoencoder to compress multiple video frames directly. The schemes in these three classes may have the potential to apply to video enhancement and analysis without the need to obtain fully decoded videos in the pixel domain.
Nevertheless, their compression efficiency is inferior to that of traditional video codecs~\cite{sullivan2012overview,wiegand2003overview,bross2021overview}. \par
To further increase the compression ratio of neural video codecs, schemes in the fourth class used a conditional coding paradigm to replace the residue coding in the first class.  Li 
et al.~\cite{li2021deep} proposed the first conditional coding-based neural video codec. They learn temporal contexts from the previously decoded frame and then make the codec exploit the temporal correlation automatically from the temporal contexts. Hoet al.~\cite{canfvc} extended conditional coding to motion compression and Shi et al.~\cite{shi2022alphavc} further used it for I frame coding. The conditional coding paradigm has made the neural video codecs catch up with or even surpass traditional video
codecs under certain conditions~\cite{li2021deep,sheng2022temporal,shi2022alphavc,canfvc,li2022hybrid}. However, learning temporal contexts from the previously decoded frame makes it infeasible to directly perform machine vision algorithms without fully decoding the bit stream into pixels. Although Sheng et al. \cite{sheng2022temporal} (our baseline TCMVC) proposed to learn temporal contexts from a decoded reference feature, the feature is very close to the pixel-domain reconstructed frame (the input feature of the last convolutional layer of the decoder).  Even though the decoded feature may be used for human and machine vision tasks directly, only the computational cost of the last convolutional layer can be saved. 

\begin{figure}[t]
  \centering
   \includegraphics[width=\linewidth]{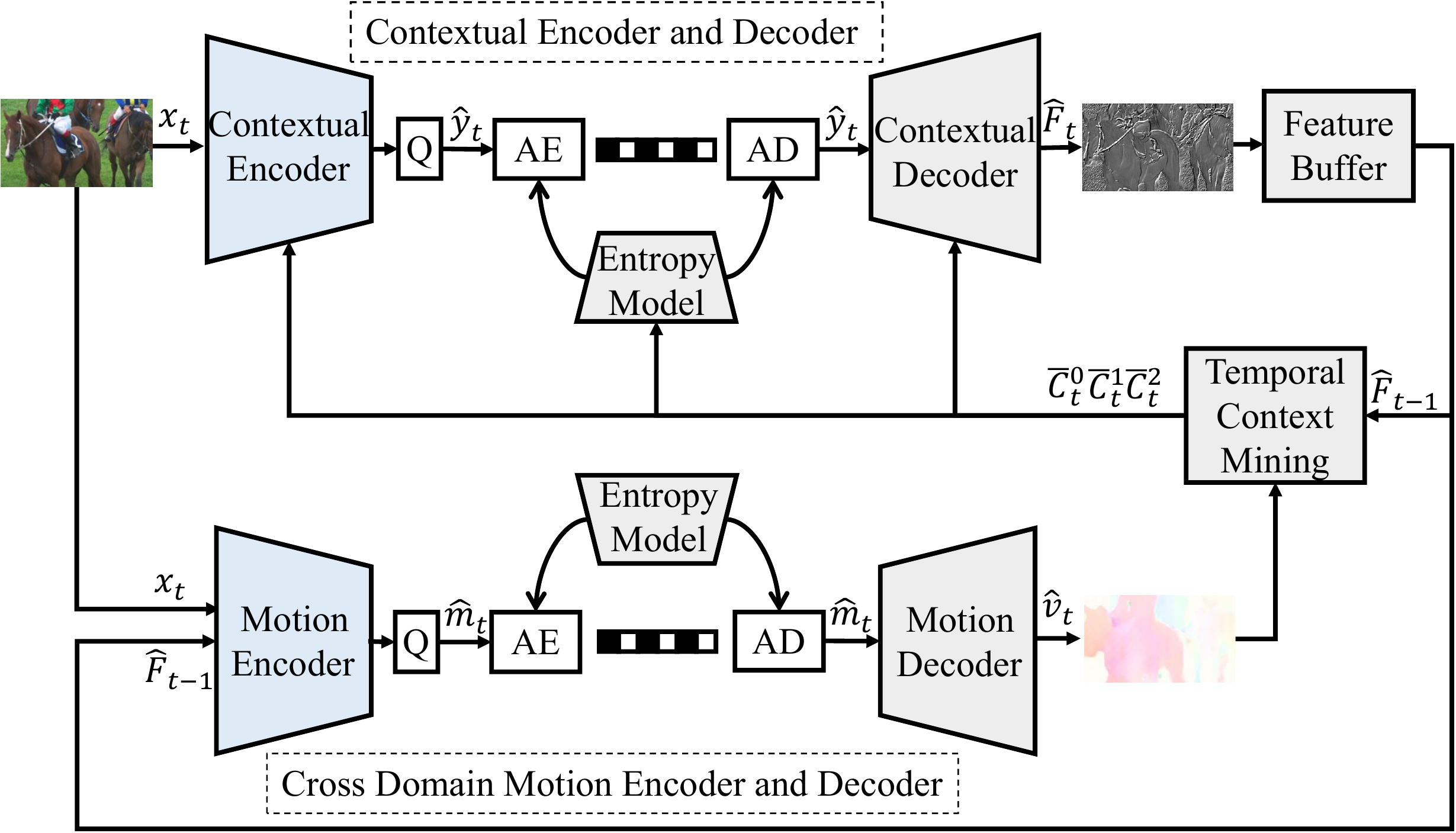}
   \caption{Architecture of our proposed feature-based compression loop. In the feature-based compression loop, neither the encoder nor the decoder needs to obtain pixel-domain reconstructed frames. The input frame $x_t$ is encoded into a bitstream and partially decoded to intermediate features $\hat{F}_t$ by a contextual encoder and decoder. The feature $\hat{F}_t$ is regarded as a reference to help compress the next frame $x_{t+1}$. The feature-based temporal context mining module is used to learn multi-scale temporal contexts from previous intermediate feature $\hat{F}_{t-1}$. The temporal contexts $\bar{C}_{t}^{0}$, $\bar{C}_{t}^{1}$, $\bar{C}_{t}^{2}$ are used to improve the compression efficiency of the contextual encoder and decoder. The cross-domain motion encoder and decoder are proposed to jointly estimate, compress, and reconstruct the motion vector (MV) $\hat{v}_t$ from the input frame $x_t$ and previous intermediate feature $\hat{F}_{t-1}$. ``Q" refers to the quantization operation. ``AE" and ``AD" refer to the arithmetic encoder and decoder, respectively.}
   \label{fig:feature_based_compression_loop}
\end{figure}
\section{Framework Architecture}\label{sec:overview}
We illustrate an overview of our proposed versatile neural video coding framework in Fig.~\ref{fig:framework}. The framework is comprised of a feature-based compression loop and downstream task networks. The feature-based compression loop is used to encode one frame into compact representations and partially decode to an intermediate feature. The task network allows users to perform multiple tasks directly from the intermediate feature, such as video reconstruction, video enhancement, and video analysis.

\subsection{Feature-based Compression Loop}
The architecture of the feature-based compression loop is shown in Fig.~\ref{fig:feature_based_compression_loop}. 
In the feature-based compression loop, neither the encoder nor the decoder needs to obtain pixel-domain reconstructed frames. It mainly includes a contextual encoder-decoder, a feature-based temporal context mining module, a cross-domain motion encoder-decoder, quantization, and entropy models.\par
\subsubsection{Contextual Encoder and Decoder} The contextual encoder and decoder are based on an autoencoder with a hyper-prior structure~\cite{balle2018variational}.
As illustrated in Fig. 3, the contextual encoder compresses an input frame $x_t$ into a compact coded representation $\hat{y}_t$ with the size of $H/16 \times W/16 \times M$. $H$ is the height of $x_t$ and $W$ is the width of  $x_t$. $M$ is the number of channels of the compact representation and is set to 96 in our framework.
In the encoding procedure, multi-scale temporal contexts $\bar{C}_{t}^{0}$, $\bar{C}_{t}^{1}$, $\bar{C}_{t}^{2}$ learned by a feature-based temporal context mining module are channel-wise concatenated into the contextual encoder to reduce temporal redundancy. Inversely, the contextual decoder decompresses the compact coded representation $\hat{y}_t$ back to an intermediate feature $\hat{F}_t$. $\hat{F}_t$ is an N-channel feature with the same resolution to $x_t$. N is set to 64 in our framework. In the decoding procedure, the multi-scale temporal contexts are also re-filled into the contextual decoder. We regard  $\hat{F}_t$ as a reference feature in the feature-based compression loop to help compress the next frame $x_{t+1}$. We can also directly feed it into various downstream human and machine vision tasks without the need for pixel-domain reconstructed videos.\par
\begin{figure}[t]
  \centering
   \includegraphics[width=\linewidth]{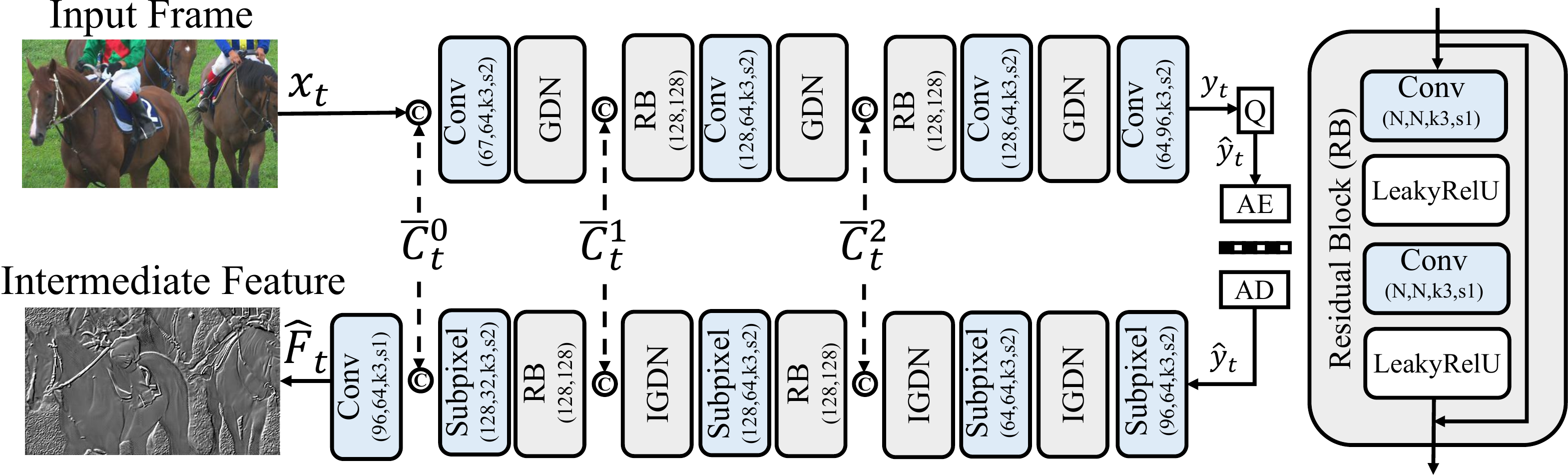}
   \caption{Architecture for the contextual encoder-decoder.}
   \label{fig:contextual encoder-decoder}
\end{figure}
\begin{figure}[t]
  \centering
   \includegraphics[width=0.95\linewidth]{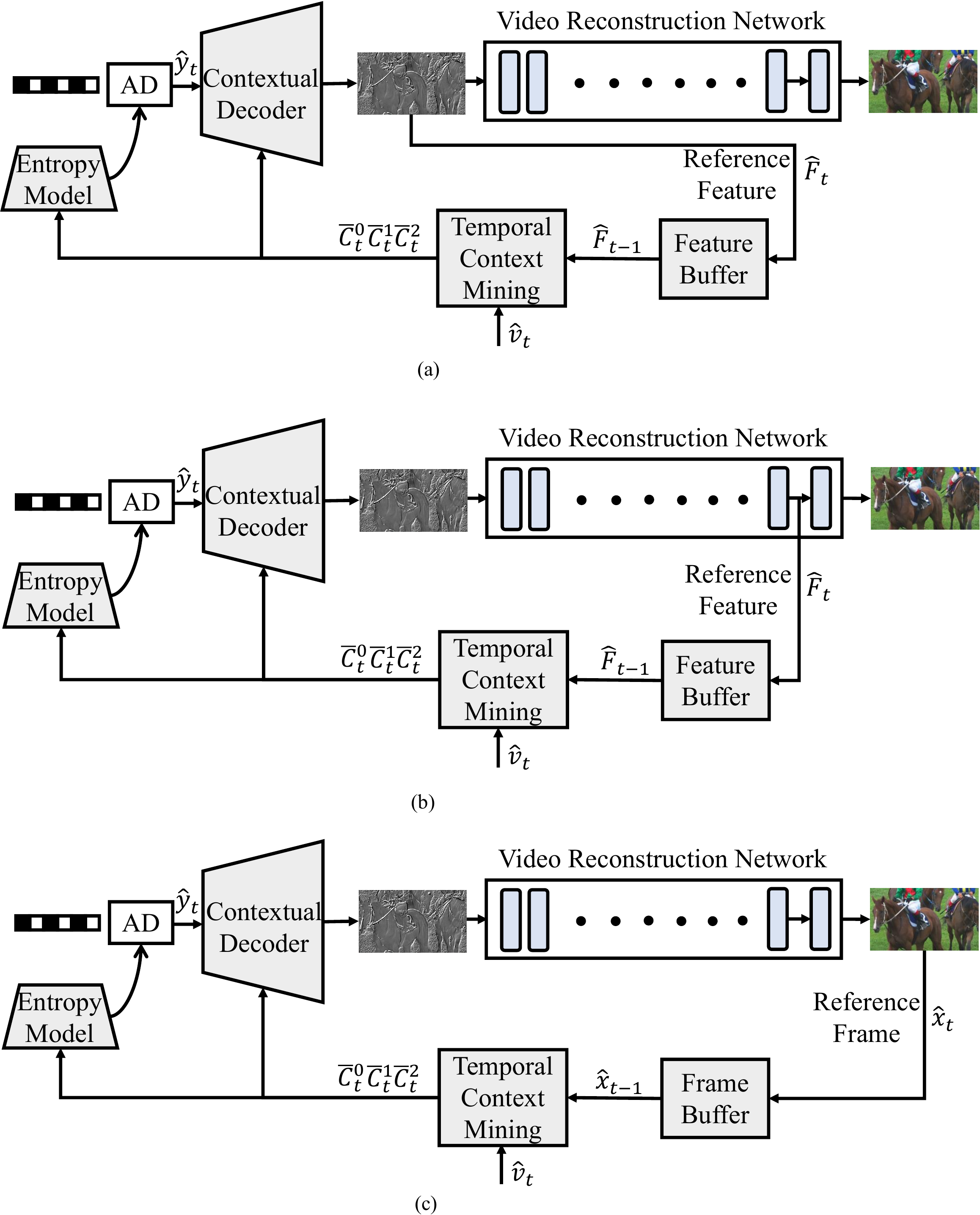}
   \caption{Difference between the ways of learning temporal contexts. (a) Our proposed scheme learns temporal contexts from a reference intermediate feature $\hat{F}_{t-1}$ before being fed into the video reconstruction network. (b) Our baseline, TCMVC~\cite{sheng2022temporal}, regards the input feature of the last convolutional layer of the video reconstruction network as a reference feature and learns temporal contexts from the reference feature $\hat{F}_{t-1}$. (c) Schemes like DCVC~\cite{li2021deep} learn temporal contexts from the previously decoded pixel-domain frame $\hat{x}_{t-1}$.
}
   \label{fig:TCM}
\end{figure}
\subsubsection{Feature-based Temporal Context Mining}One of the major reasons for the high compression performance of conditional coding-based neural video codecs is that they take advantage of temporal contexts. As illustrated in Fig.~\ref{fig:TCM} (c), schemes like DCVC~\cite{li2021deep} learn temporal contexts from the previously decoded frames $\hat{x}_{t-1}$. When oriented to human and machine vision tasks, the decoder must obtain the pixel-domain reconstructed frames first and then perform downstream vision tasks. Our baseline, TCMVC~\cite{sheng2022temporal}, regards the input feature of the last convolutional layer of the video reconstruction network as a reference feature and learns temporal contexts from the reference feature $\hat{F}_{t-1}$. However, as illustrated in Fig.~\ref{fig:TCM} (b), the feature is still very close to the pixel-domain reconstructed frame. Even though the decoded feature may be used for human and machine vision tasks directly, only the computational cost of the last convolutional layer can be saved. 
In this work, as illustrated in Fig.~\ref{fig:TCM} (a), we propose to learn temporal contexts from an intermediate feature $\hat{F}_{t-1}$ before being fed into the video reconstruction network. If we directly feed the intermediate feature into downstream human and machine vision tasks, all the computational complexity of the video reconstruction network can be saved. \par
To make the temporal contexts contain richer temporal information, we follow TCMVC~\cite{sheng2022temporal} to learn multi-scale temporal contexts $\bar{C}_{t}^{0}$, $\bar{C}_{t}^{1}$, $\bar{C}_{t}^{2}$ using a feature-based temporal context mining module. They have the same resolution, half of the resolution, and quarter of the resolution as the input frame $x_t$, respectively. Their numbers of feature channels are the same as $\hat{F}_t$, e.g., 64.
They are channel-wise concatenated into the contextual encoder-decoder to exploit temporal correlations. The detailed architecture of the feature-based temporal context mining module can be found in~\cite{sheng2022temporal}.\par

\subsubsection{Cross-Domain Motion Encoder and Decoder}
\begin{figure}[t]
  \centering
   \includegraphics[width=\linewidth]{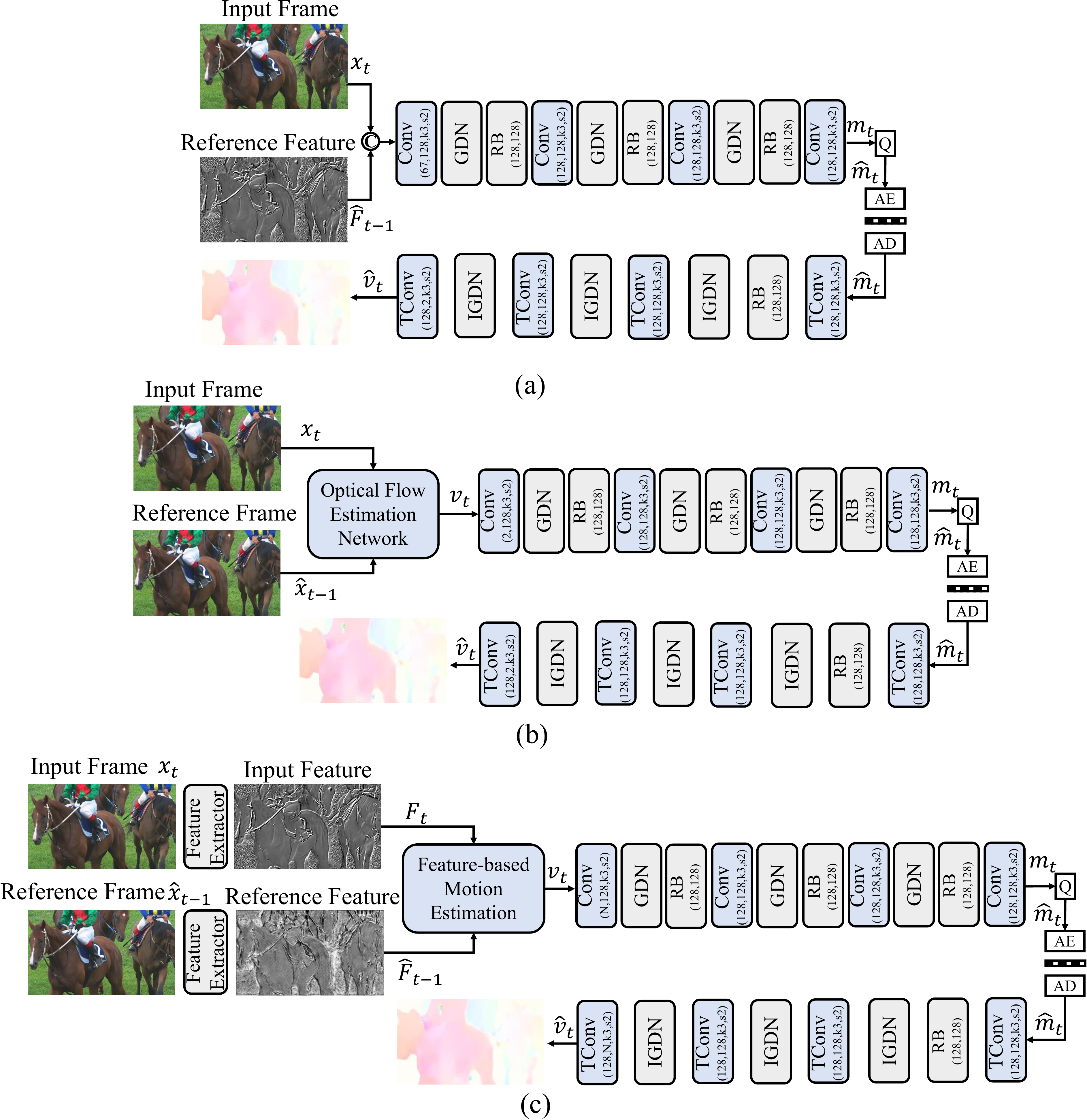}
   \caption{Difference between the ways of motion estimation and compression. (a) In our proposed cross-domain motion encoder-decoder, the input frame $x_t$  and the reference intermediate feature $\hat{F}_{t-1}$ can be directly fed into the motion encoder without the need for obtaining pixel-domain reconstructed frames and separate motion estimation modules. 
   (b) Schemes like DVC~\cite{lu2020end}, DCVC~\cite{li2021deep}, and TCMVC~\cite{sheng2022temporal} use a separate optical flow estimation network to estimate pixel-wise motion vectors between the input frame $x_t$ and the reference frame $\hat{x}_{t-1}$. (c) Schemes like FVC~\cite{hu2022fvc} transform $x_t$ and $\hat{x}_{t-1}$ into the feature domain first. Then, feature-based motion estimation modules are used to to estimate feature-based motion information. 
   }
   \label{fig:mv_encoder_decoder}
\end{figure}

As illustrated in Fig.~\ref{fig:mv_encoder_decoder} (b), a separate optical flow estimation network~\cite{ranjan2017optical,ilg2017flownet} is commonly used to estimate pixel-wise motion vectors in the existing motion estimation and compensation-based neural video compression schemes. These schemes require the encoder to obtain pixel-domain reconstructed videos. Although some feature-based neural video codecs~\cite{hu2022fvc} use feature-domain motion estimation, such as deformable convolutions~\cite{dai2017deformable}, as illustrated in Fig.~\ref{fig:mv_encoder_decoder} (c), they also need to obtain pixel-domain decoded frames first and then convert them into the feature domain. 
In this work, we propose a cross-domain motion encoder-decoder based on an autoencoder with a hyper-prior structure~\cite{balle2018variational}. As illustrated in Fig.~\ref{fig:mv_encoder_decoder} (a), we remove the separate motion estimation module and combine the motion estimation and compression together. The reference intermediate feature $\hat{F}_{t-1}$ and the input frame $x_t$ are directly fed into the motion encoder and compressed into a compact representation with the size of $\hat{m}_t$ with $H/16 \times W/16 \times K$. $K$ is the number of channels of the compact representation and is set to 128 in our framework. Then, the motion decoder decodes $\hat{m}_t$ back to the reconstructed motion vector $\hat{v}_t$. 
Our proposed cross-domain motion encoder-decoder efficiently reduces the encoding complexity for encoder-side pixel-domain video reconstruction and separate motion estimation modules. \par
\subsubsection{Quantization and Entropy Modeling} Towards quantization, we add the uniform noise~\cite{DBLP:conf/iclr/BalleLS17} to replace the quantization at the training stage and perform the exact rounding operation at the testing stage. Towards entropy modeling, we use the factorized entropy model~\cite{DBLP:conf/iclr/BalleLS17} for hyper-prior and the Laplace distribution~\cite{balle2018variational} to model the contextual and motion compact representations $\hat{y}_t$ and $\hat{m}_t$. When estimating the mean and scale of the Laplace distribution of $\hat{y}_t$, we generate a temporal prior~\cite{sheng2022temporal} from the multi-scale temporal contexts $\bar{C}_{t}^{0}$, $\bar{C}_{t}^{1}$, $\bar{C}_{t}^{2}$ to exploit the temporal information. We do not apply the auto-regressive entropy model~\cite{NEURIPS2018_53edebc5,sheng2022temporal} for parallel decoding.

\subsection{Task Networks}\label{task_network}
Obtaining the intermediate feature $\hat{F}_t$ decoded by the contextual decoder, users can convert it into the pixel domain for video reconstruction. When oriented to human and machine vision tasks, such as video enhancement and video analysis, users can directly enhance/analyze the intermediate feature $\hat{F}_t$ without the need for pixel-domain video reconstruction.\par
\subsubsection{Video Reconstruction Network} 
\begin{figure}[t]
  \centering
   \includegraphics[width=0.9\linewidth]{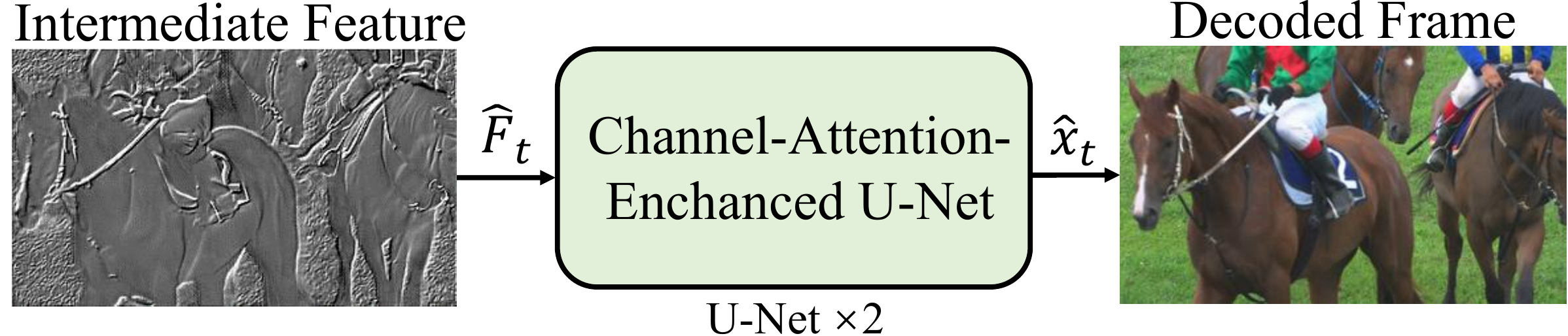}
   \caption{Architecture for the video reconstruction network. Two channel-attention-enhanced~\cite{8701503} U-Nets~\cite{xia2017w, li2022hybrid} make up the video reconstruction network. The detailed architecture can be found in the supplemental material.}
   \label{fig:frame_generator}
\end{figure}

When users ask to display the video, we can feed $\hat{F}_t$ into a video reconstruction network to generate the pixel-domain decoded frame $\hat{x}_t$.  As illustrated in Fig.~\ref{fig:frame_generator}, the video reconstruction network is comprised of two channel-attention-enhanced~\cite{8701503} U-Nets~\cite{xia2017w, li2022hybrid}. Its large receptive field can improve the reconstruction capability effectively~\cite{li2022hybrid}. Different from~\cite{sheng2022temporal,li2022hybrid}, we do not feed the temporal context $\bar{C}_{t}^{0}$ into the video reconstruction network to reduce the computational complexity. The detailed architecture for the video reconstruction network can be found in the supplemental material. \par 

\subsubsection{Video Enhancement Networks} When users ask to improve the quality of decoded videos, we directly enhance $\hat{F}_t$. We take video denoising and video super-resolution as examples.\par
\begin{figure}[t]
  \centering
   \includegraphics[width=\linewidth]{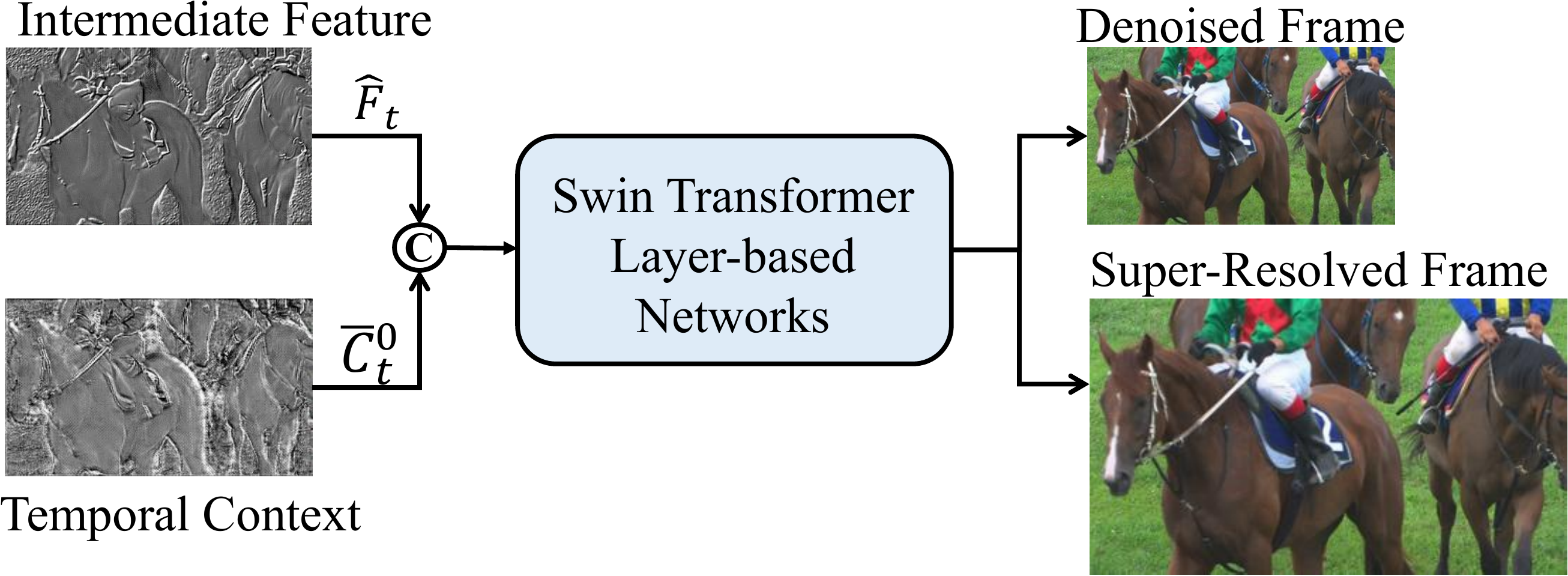}
   \caption{Architecture for video denoising and video super-resolution networks. Following SwinIR~\cite{Liang2021SwinIRIR}, Swin Transformer Layer~\cite{liu2021Swin} is used as the backbone. The detailed architecture can be found in the supplemental material.}
   \label{fig:denoise_sr}
\end{figure}

For video denoising, we focus on the case of non-blind and blind additive white Gaussian noise (AWGN). 
We feed the decoded noisy intermediate feature $\hat{F}_t$ into a video denoising network. As illustrated in Fig.~\ref{fig:denoise_sr}, we follow SwinIR~\cite{Liang2021SwinIRIR} to use Swin Transformer~\cite{liu2021Swin} as the backbone to construct the video denoising network. Existing video enhancement schemes commonly design additional temporal alignment modules to align the adjacent frames, so that more temporal information can be exploited. However, in our proposed framework, temporal alignment has been done in the feature-based temporal context mining module. We can directly feed the temporal context $\bar{C}_{t}^{0}$ into the network to exploit the temporal correlation without the need for additional temporal alignment modules.\par
For video super-resolution (SR), we focus on $4\times$ video SR with Bicubic (BI) degradation and real-world video SR.
We feed the decoded low-resolution intermediate feature $\hat{F}_t$ into a video SR network. As illustrated in Fig.~\ref{fig:denoise_sr}, Swin Transformer~\cite{liu2021Swin} is also used as the backbone. To exploit the temporal information of the previous frame, we also feed the temporal context $\bar{C}_{t}^{0}$ into the network.\par
\subsubsection{Video Analysis Networks.}
When users need to analyze the video scenes, we directly analyze $\hat{F}_t$. We take video action recognition, video object detection, video multiple object detection, and video object segmentation as examples. \par

For video action recognition,  we use the temporal shift model (TSM)~\cite{lin2020tsm} to perform action recognition.  As shown in Fig.~\ref{fig:action_recognition}, given the sampled decoded intermediate features $\hat{F}_1$,...,$\hat{F}_K$, we use TSM with ResNet-50~\cite{he2016deep} backbone to process each of the features individually, and the predicted class scores are averaged to give the final prediction. To make ResNet-50 in TSM enable extracting high-level features of the intermediate features, its input channel number is changed from 3 to 64, i.e., the channel number of  $\hat{F}_t$.  Apart from this change, we have not made any other changes to TSM. \par

For video object detection, as shown in Fig.~\ref{fig:VID}, we apply the sequence level semantics aggregation (SELSA) model~\cite{wu2019sequence} with Faster R-CNN~\cite{7485869} detector to detect objects. In Faster R-CNN, we use ResNet-101 as the backbone.  Similarly, to make ResNet-101 enable extracting the high-level features of each intermediate feature $\hat{F}_t$,  its input channel number is changed from 3 to 64, i.e., the channel number of  $\hat{F}_t$. In addition,  different from the original SELAS model, which aggregates the temporal information of 14 neighboring frames, we only aggregate the temporal information of a single previous frame to achieve an online detection scheme.\par

For video multiple object tracking, as presented in Fig.~\ref{fig:MOT}, we apply the Quasi-Dense Tracking (QDTrack)~\cite{qdtrack_conf} with Faster R-CNN~\cite{7485869} detector to track objects. In Faster R-CNN, we use ResNet-50 as the backbone. Similarly, we change the input channel number of  ResNet-50 from 3 to 64 to make ResNet-50 enable extracting the high-level features of each intermediate feature $\hat{F}_t$.\par

For video object segmentation, as shown in Fig.~\ref{fig:VOS},  we apply Associating Objects with Transformers Large version (AOTL)~\cite{yang2021aot} with ResNet-50 backbone for segmentation. Similarly, we change the input channel number of ResNet-50 from 3 to 64 to make AOTL segment the intermediate feature $\hat{F}_t$.

\section{Framework Training}\label{sec:opti}
\subsection{Training for Feature-based Compression Loop}\label{sec:opti_loop}
We adopt a step-by-step progressive training strategy~\cite{sheng2022temporal} to train the feature-based compression loop. 
We first train the cross-domain motion encoder-decoder. We use the decoded motion vector $\hat{v}_t$ to generate a predicted frame $\tilde{x}_{t}$. Then we use the following rate-distortion cost to train the motion encoder-decoder,
\begin{equation}
    L_m=\lambda d_m(x_t, \tilde{x}_t) + R_m,
\label{equ:mv_RD}
\end{equation}
where $d_m(x_t, \tilde{x}_t)$ denotes the distortion between the input frame $x_t$ and the predicted frame $\tilde{x}_t$. 
$R_m$ represents the bit rate used for encoding the quantized motion compact representation $\hat{m}_t$ and its hyper-prior. Note that, the predicted frame is only used in the training stage and there is no need to generate the predicted frame in the testing stage.\par
\begin{figure}[t]
  \centering
   \includegraphics[width=0.7\linewidth]{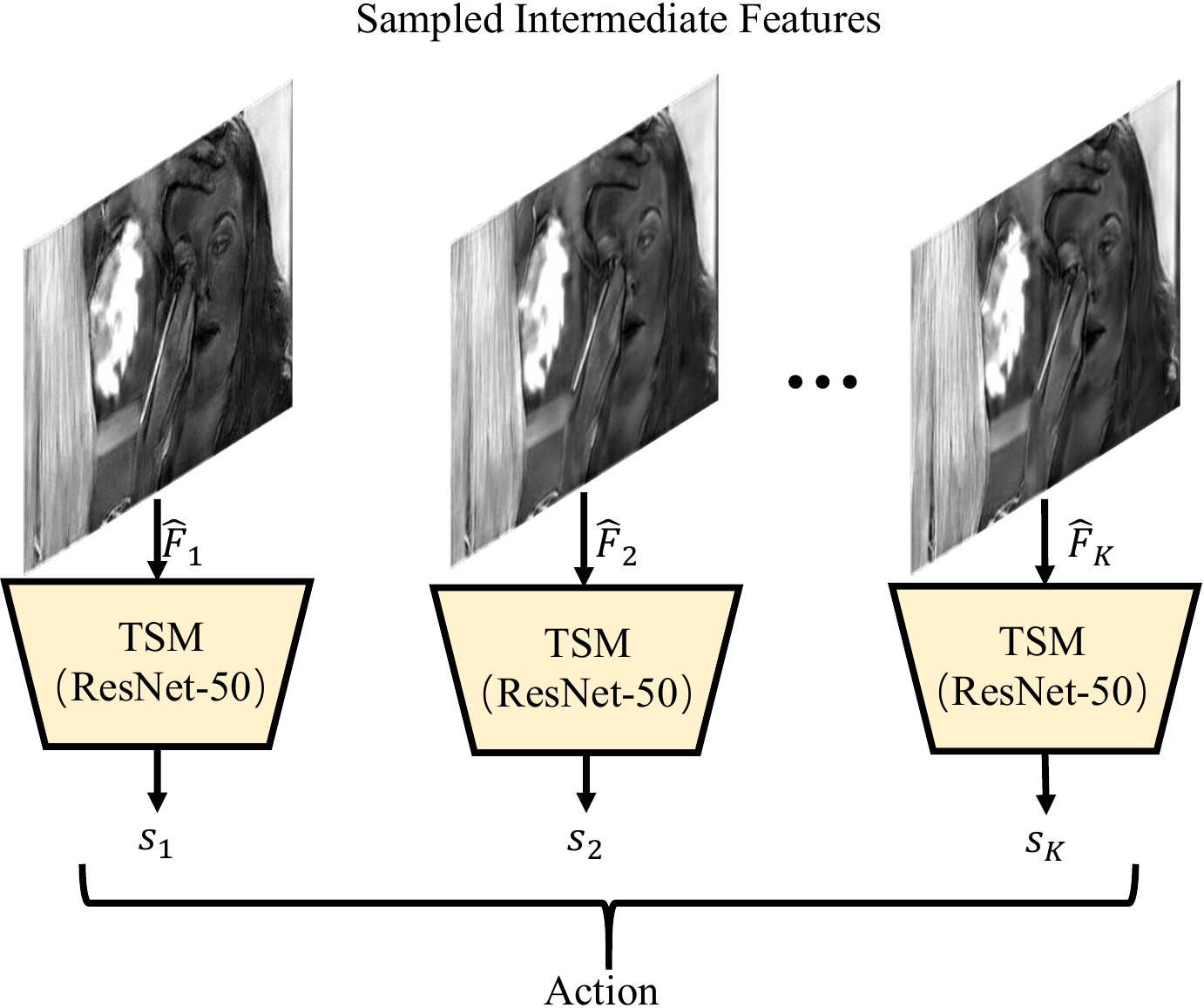}
   \caption{Architecture for video action recognition. Temporal shift model (TSM)~\cite{lin2020tsm} with ResNet-50 backbone is used as the recognizer.}
   \label{fig:action_recognition}
\end{figure}
\begin{figure}[t]
  \centering
   \includegraphics[width=0.7\linewidth]{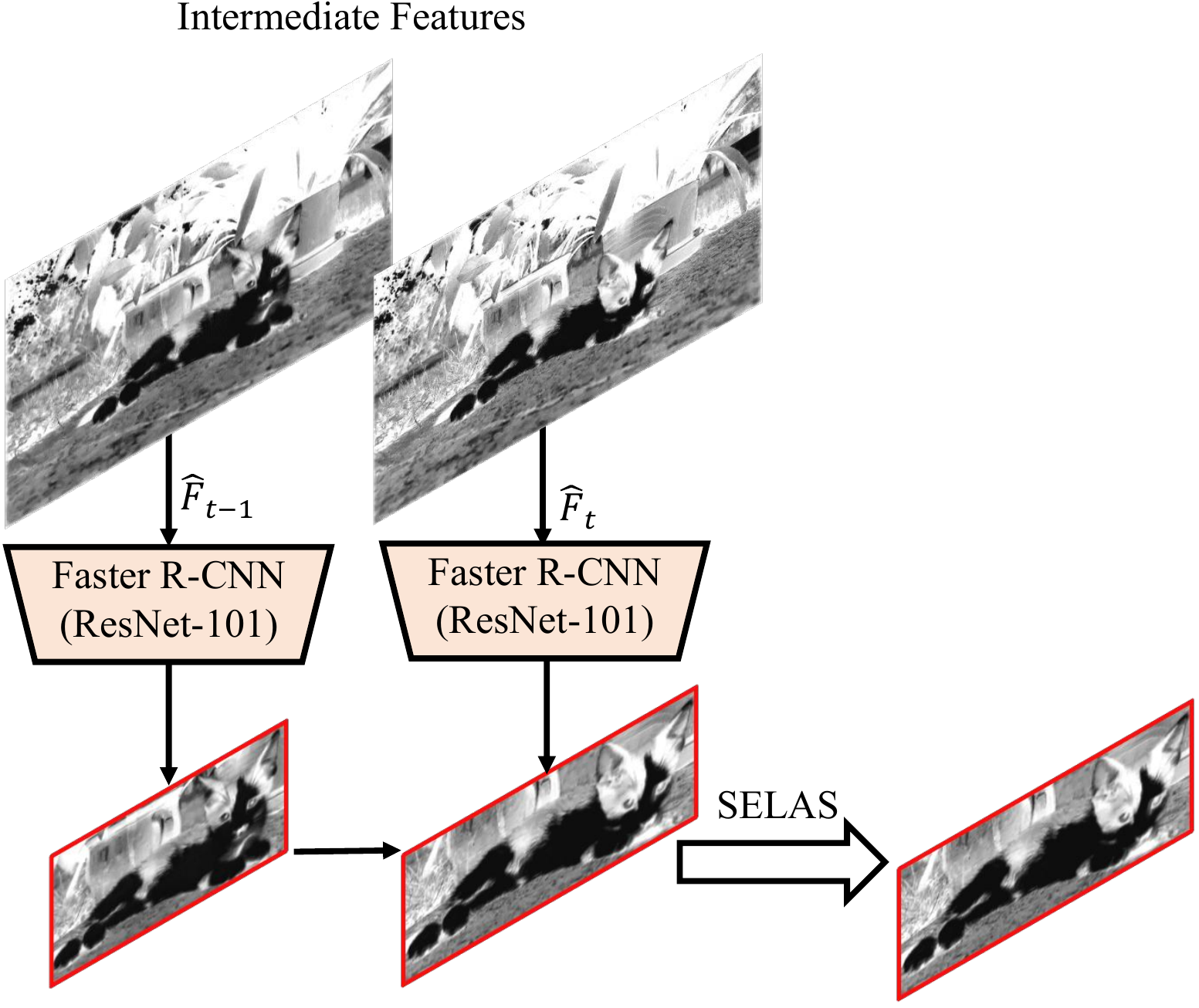}
   \caption{Architecture for video object detection. Sequence level semantics aggregation (SELSA)~\cite{wu2019sequence} with Faster R-CNN~\cite{7485869} detector is used for detection. In Faster R-CNN, ResNet-101 is used as the backbone.}
   \label{fig:VID}
\end{figure}
\begin{figure}[t]
  \centering
   \includegraphics[width=0.45\linewidth]{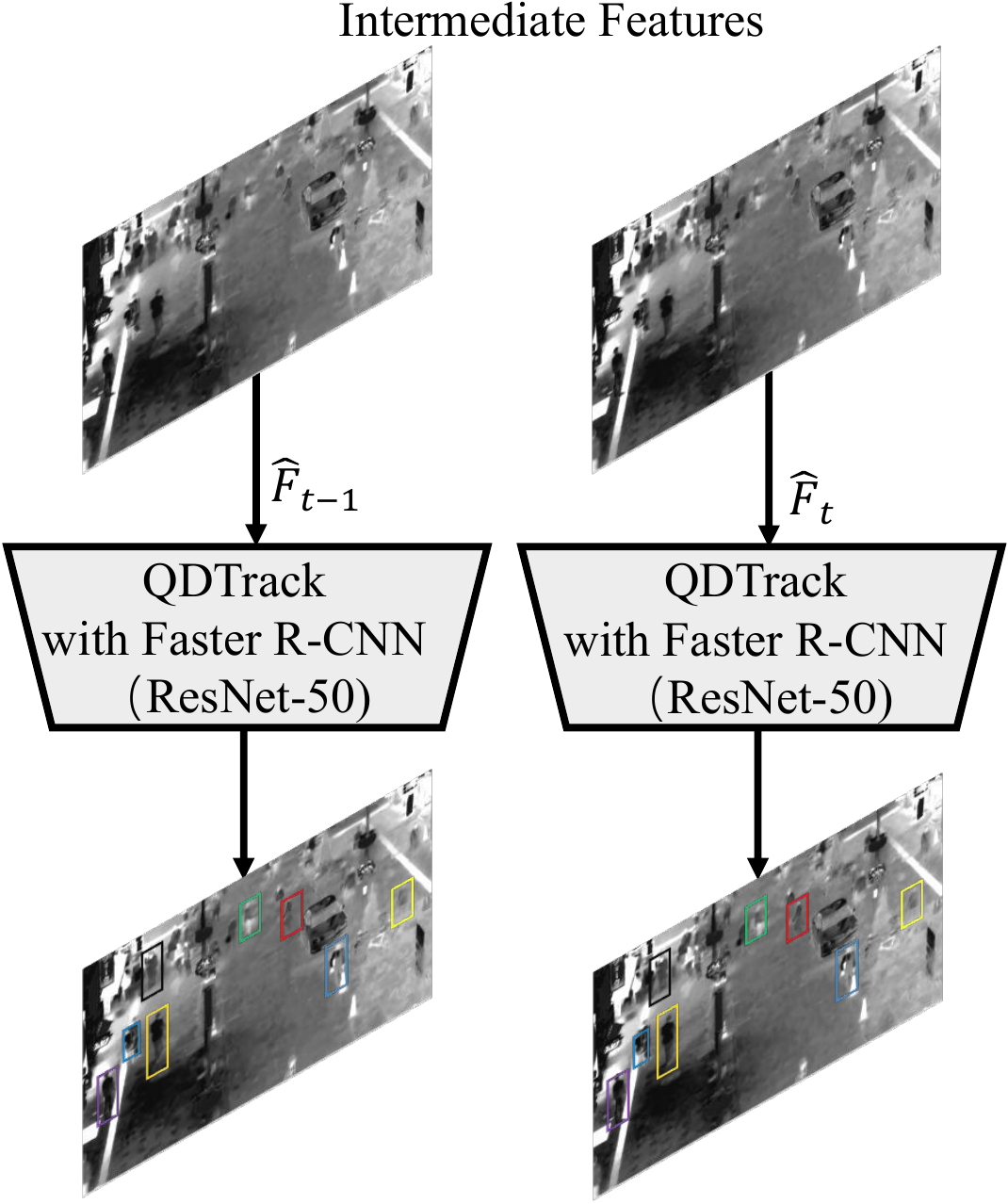}
   \caption{Architecture for video multiple object tracking. Quasi-Dense Tracking (QDTrack)~\cite{qdtrack_conf} with Faster R-CNN~\cite{7485869} is used as the detector to track objects. In Faster R-CNN, ResNet-50 is used as the backbone.}
   \label{fig:MOT}
\end{figure}
\begin{figure}[t]
  \centering
   \includegraphics[width=0.45\linewidth]{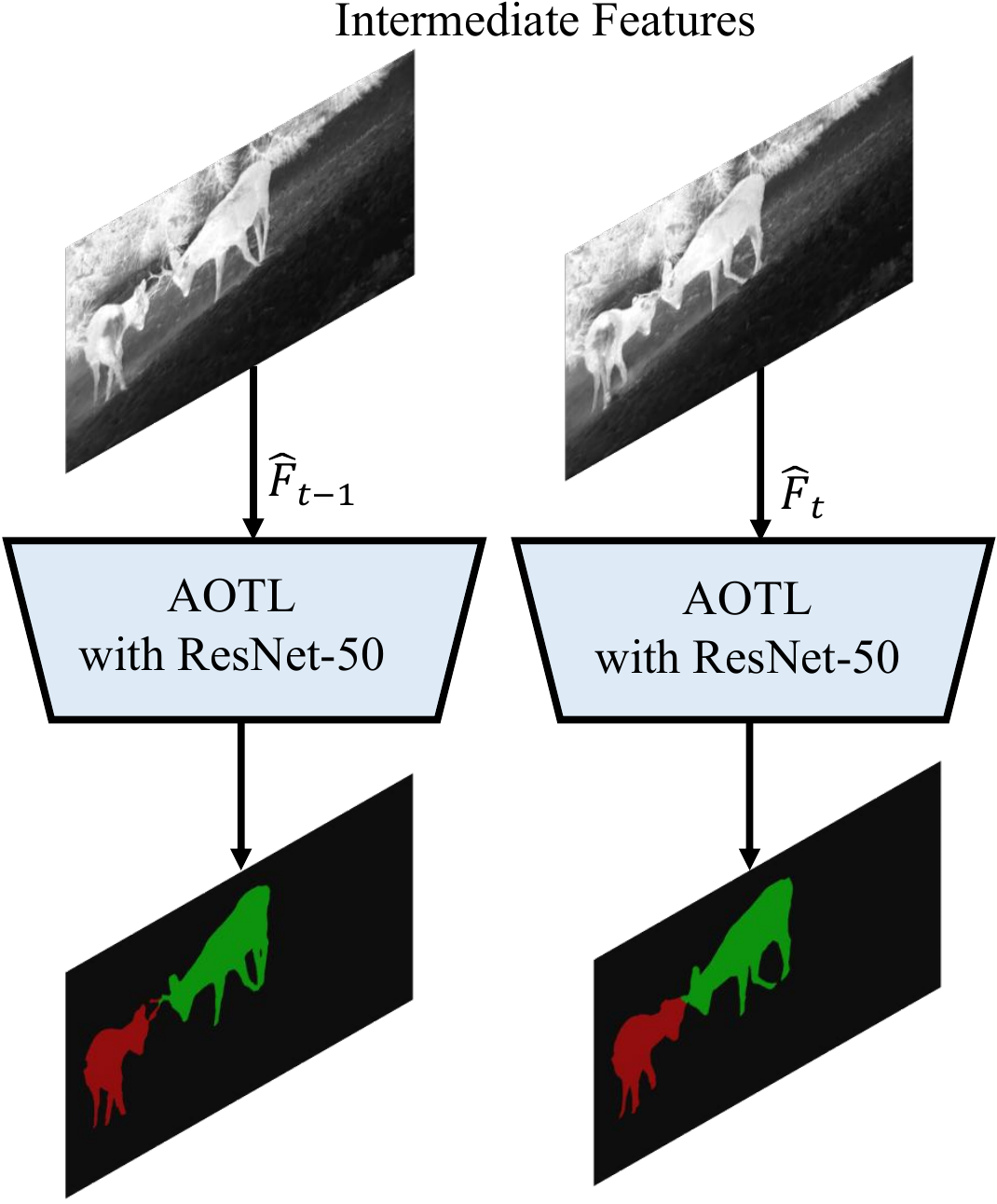}
   \caption{Architecture for video object segmentation. Associating Objects with Transformers (AOT) large version (AOTL) with ResNet-50 backbone is used for segmentation.}
   \label{fig:VOS}
\end{figure}
Then we fix the cross-domain motion encoder-decoder and train the contextual encoder-decoder. The target of contextual encoder-decoder is to compress the input frame $x_t$ into a compact coded representation $\hat{y}_t$ with a smaller size and decompress $\hat{y}_t$ to a partially decoded intermediate feature $\hat{F}_t$. However, we cannot obtain the ground truth of the feature. Therefore, to generate an intermediate feature that can be applied to different human and machine vision tasks, we train the contextual encoder-decoder with the help of the video reconstruction network. Specifically, we convert the partially decoded intermediate feature $\hat{F}_t$ to a reconstructed frame with the video reconstruction network and train the contextual encoder-decoder and the video reconstruction network jointly using the following loss function, 
\begin{equation}
    L_y= \lambda d(x_t, \hat{x}_t) + R_y,
\label{equ:contextual_RD}
\end{equation}
where $d(x_t, \hat{x}_t)$ denotes the distortion between the input frame $x_t$ and the reconstructed frame $\hat{x}_t$. $R_y$ represents the bit rate used for encoding the quantized contextual compact representation $\hat{y}_t$ and its hyper-prior. \par
Finally, we jointly optimize the rate-distortion cost of the whole feature-based compression loop,
\begin{equation}
    L=\lambda d(x_t, \hat{x}_t) + R_m + R_y. 
\label{equ:contextual_RD}
\end{equation}
The distortion is measured by MSE when PSNR is used as the distortion metric and is measured by 1$-$MS-SSIM when MS-SSIM is used as the distortion metric. Following existing neural video compression schemes, the distortion is calculated in RGB color space.

\subsection{Training for Task Networks}
After finishing the training procedure of the feature-based compression loop, we continue to train task networks.\par

\subsubsection{Video Reconstruction} As described in Section~\ref{sec:opti_loop}, the video reconstruction network is part of the training process of the feature-based compression loop, users can directly use the trained video reconstruction network to generate fully decoded frame $\hat{x}_t$ in the pixel domain. However, users are free to use other video reconstruction networks different from the one that helps train the feature-based compression loop. When training a new video reconstruction network, we fix the feature-based compression loop and use the MSE or 1--MS-SSIM between the input frame $x_t$ and the fully decoded frame $\hat{x}_t$ to train the network.\par

\subsubsection{Video Enhancement} Towards video enhancement, our framework targets to improve the quality of reconstructed videos while not harming the video reconstruction performance. Therefore, we fix the feature-based compression loop and only train the video enhancement networks with their corresponding loss functions. For example, we calculate the MSE between clean frames and denoised frames to train the video denoising network and calculate the MSE between ground truth high-resolution frames and super-resolved frames to train the video super-resolution network. \par

\subsubsection{Video Analysis} Towards video analysis, we also fix the feature-based compression loop and only train the video analysis networks. We take video action recognition, video object detection, video multiple object tracking, and video object segmentation as examples. For video action recognition, we follow~\cite{lin2020tsm} and use the standard categorical cross-entropy loss to train the network. For video object detection, we use the same multi-task loss in Faster R-CNN~\cite{7485869}. For video multiple object tracking, we use the joint detection loss and quasi-dense similarity learning loss. For video object segmentation, we use a combination of bootstrapped cross-entropy loss and soft Jaccard loss. More details can be found in ~\cite{lin2020tsm},\cite{7485869}, \cite{qdtrack_conf},\cite{yang2021aot}.

\section{Experimental Setup}\label{sec:setup}
To demonstrate the effectiveness of our proposed framework, we conduct extensive experiments on video reconstruction, video enhancement, and video analysis.\par

\subsection{Training and Testing Datasets}
\subsubsection{Video Reconstruction} We use the training split of the Vimeo-90k~\cite{xue2019video} dataset to train the feature-based compression loop and video reconstruction network. We use HEVC~\cite{sullivan2012overview}, UVG~\cite{mercat2020uvg}, and MCL-JCV~\cite{wang2016mcl} datasets to evaluate the reconstruction performance.
HEVC dataset contains 16 commonly used videos with different resolutions, including Class B, C, D, and E.  Following the setting in~\cite{sheng2022temporal}, we add an additional HEVC Class RGB defined in the common test conditions for HEVC range
extensions~\cite{david2013common}, which contains 6 1080p videos in RGB444 format. UVG and MCL-JCV datasets are also widely used. The UVG dataset has 7 1080p videos and the MCL-JCV dataset has 30 1080p videos.\par
\subsubsection{Video Enhancement}
The training and testing datasets for video enhancement tasks are listed as follows.\par 
\textbf{Blind/Non-Blind Video Denoising.} We train video denoising networks using the DAVIS-2017 training split~\cite{perazzi2016benchmark}. We evaluate the denoising performance using the DAVIS-2017 test split and Set8 dataset. Following~\cite{tassano2020fastdvdnet}, the sequences of Set8 are downscaled to a resolution of 960 $\times$ 540. \par

\textbf{BI-Degradation/Real-World Video SR.} For BI-degradation video SR, we train video SR networks using the Vimeo-90k and REDS~\cite{nah2019ntire} datasets, respectively. When the video SR network is trained on the Vimeo-90K dataset, we use the Vimeo-90K test dataset and Vid4 dataset~\cite{liu2013bayesian} for evaluation. When the network is trained on the REDS dataset, we use the REDS4 test dataset to evaluate the performance.
For real-world video SR, we train video SR networks with the RealVSR dataset~\cite{yang2021real} and use its test split to evaluate the performance.\par
\subsubsection{Video Analysis}
The training and testing datasets for video analysis tasks are listed as follows.\par 
\textbf{Video Action Recognition.} We train video action recognition networks on the UCF101 dataset~\cite{soomro2012ucf101} and Something-something-V2 (SthV2) dataset~\cite{goyal2017something}. For the UCF101 dataset, its first split of the official training protocol~\cite{idrees2017thumos} is adopted for training and its first testing split is used for evaluation. For the SthV2 dataset, its training split is used for training and its validation set is used for evaluation.\par

\textbf{Video Object Detection.} We train video object detection networks with a mixture of the ImageNet VID and DET datasets~\cite{russakovsky2015imagenet} with the split provided in FGFA~\cite{zhu2017flow}. We use the ImageNet VID validation set for evaluation.\par

\textbf{Video Multiple Object Tracking.} We train video multiple object tracking networks on the MOT17 dataset~\cite{milan2016mot16}. Following the setting in QDTrack~\cite{qdtrack_conf}, we use half of the training dataset to train the network and use the remaining half of the training dataset for evaluation.\par

\textbf{Video Object Segmentation.} We train video object segmentation networks on the DAVIS-2017 training split~\cite{perazzi2016benchmark} and evaluate the segmentation performance on the DAVIS-2017 testing split.

\begin{figure*}[t]
  \centering
  \begin{minipage}[c]{0.22\linewidth}
  \centering
    \includegraphics[width=\linewidth]{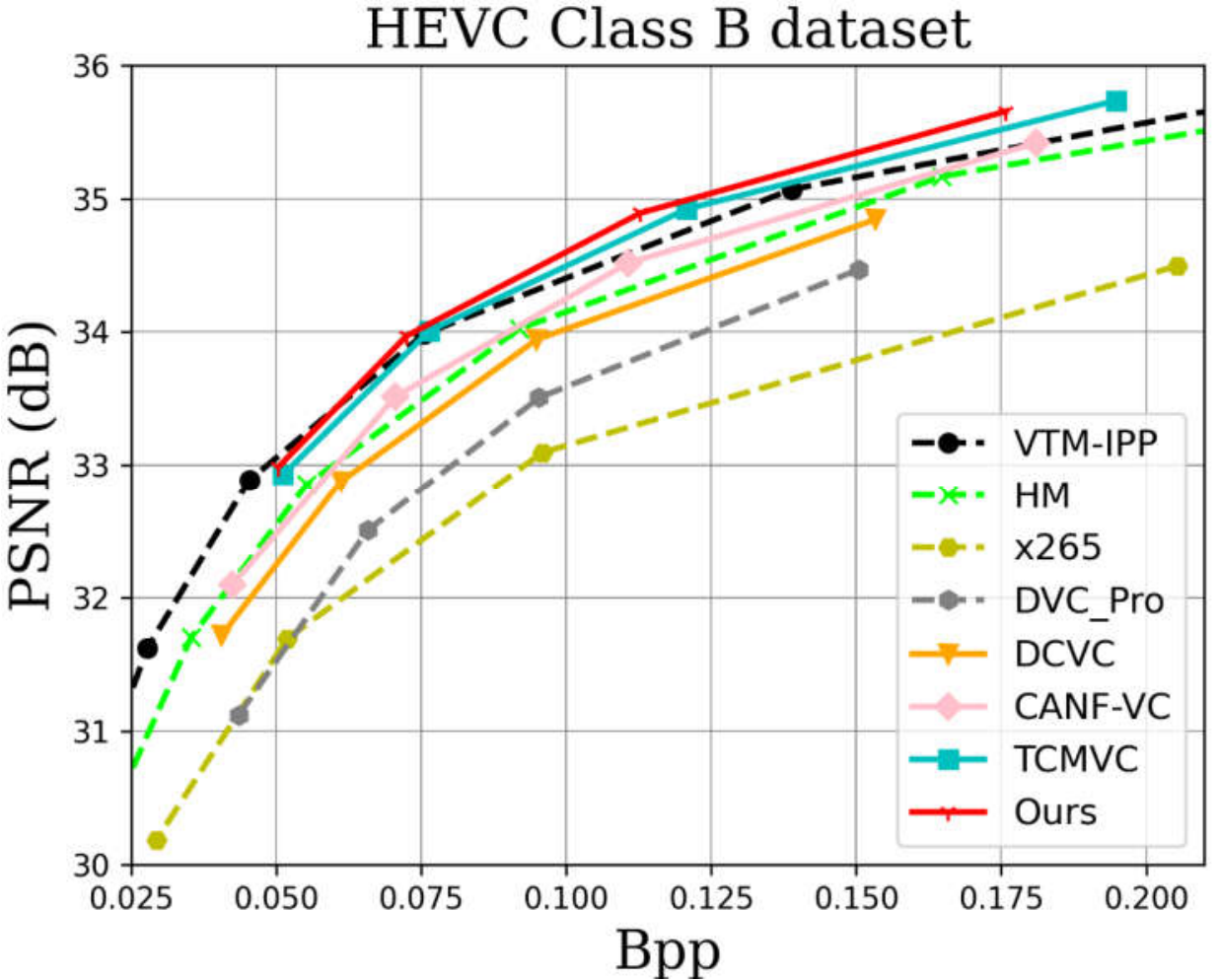}
 \end{minipage}%
   \begin{minipage}[c]{0.22\linewidth}
  \centering
    \includegraphics[width=\linewidth]{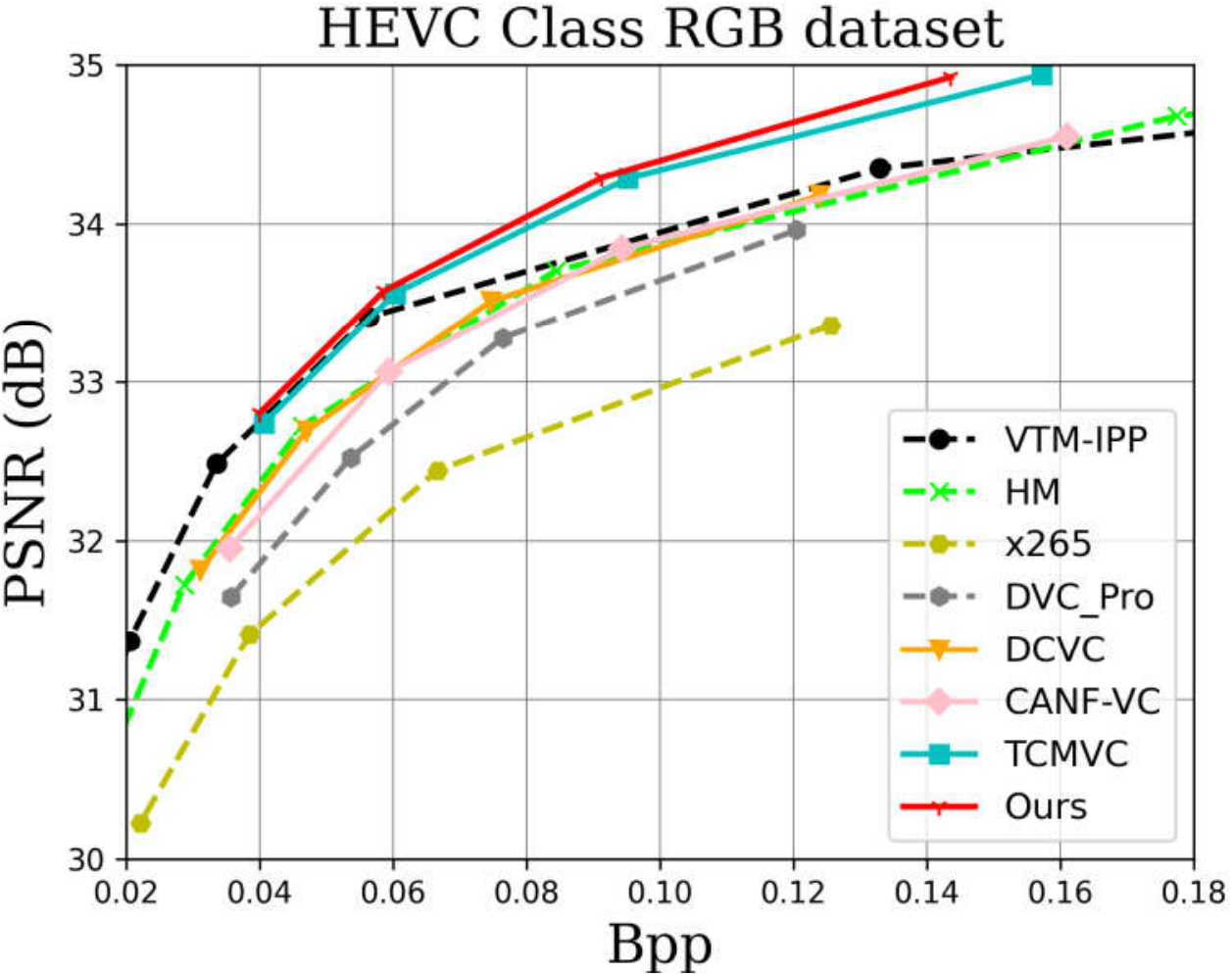}
 \end{minipage}%
  \begin{minipage}[c]{0.22\linewidth}
  \centering
    \includegraphics[width=\linewidth]{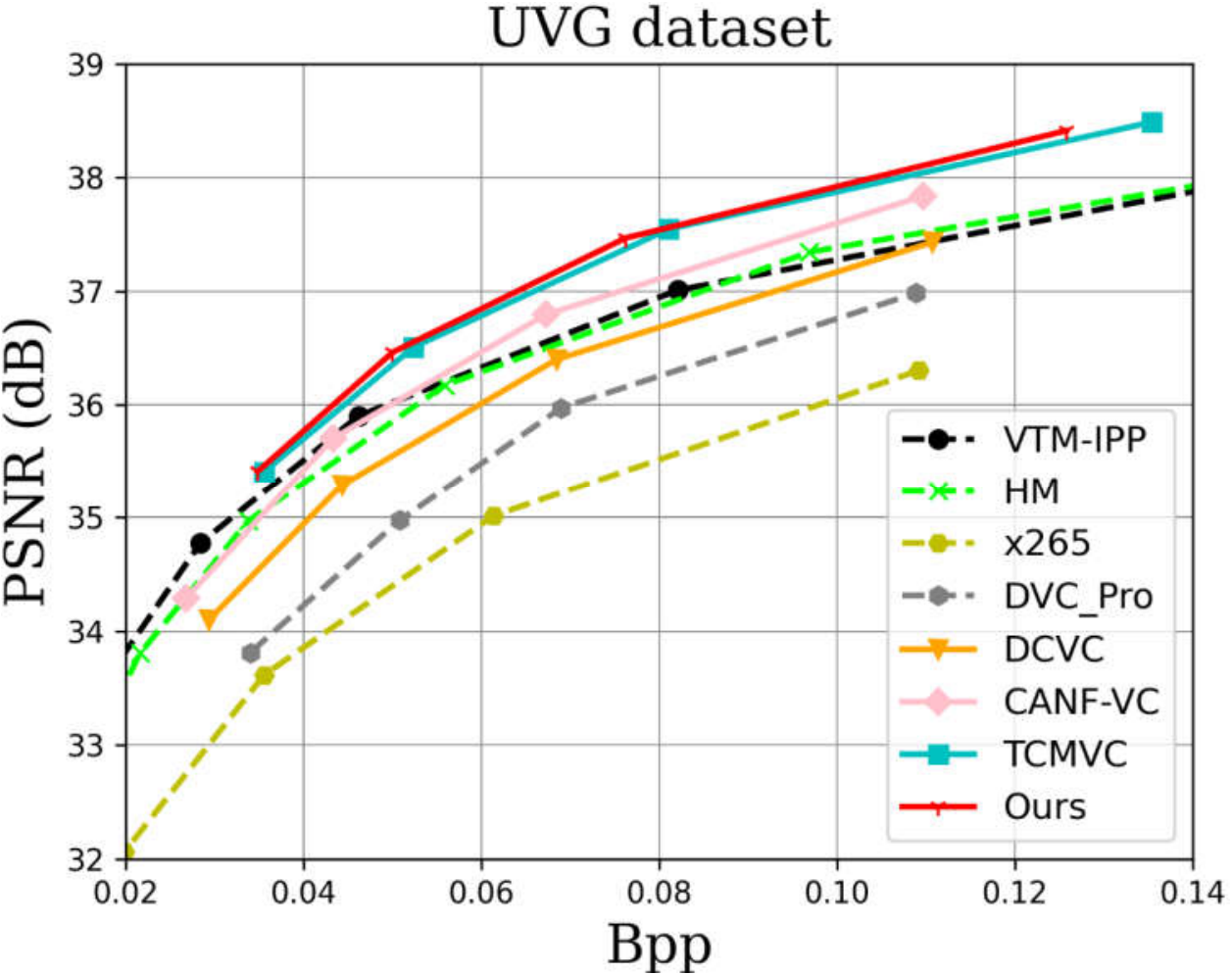}
  \end{minipage}%
  \begin{minipage}[c]{0.22\linewidth}
  \centering
    \includegraphics[width=\linewidth]{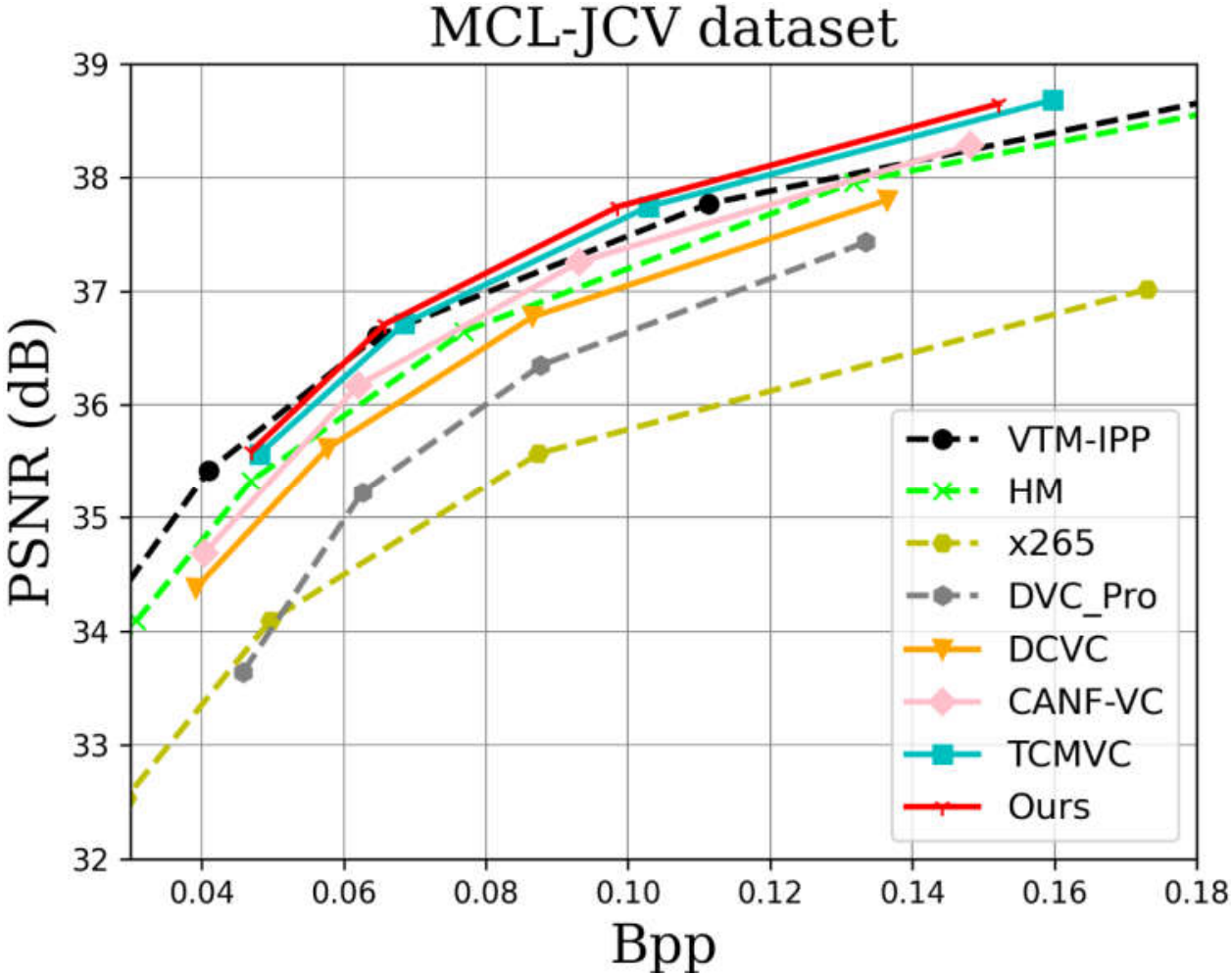}
  \end{minipage}%
  
  \begin{minipage}[c]{0.22\linewidth}
  \centering
    \includegraphics[width=\linewidth]{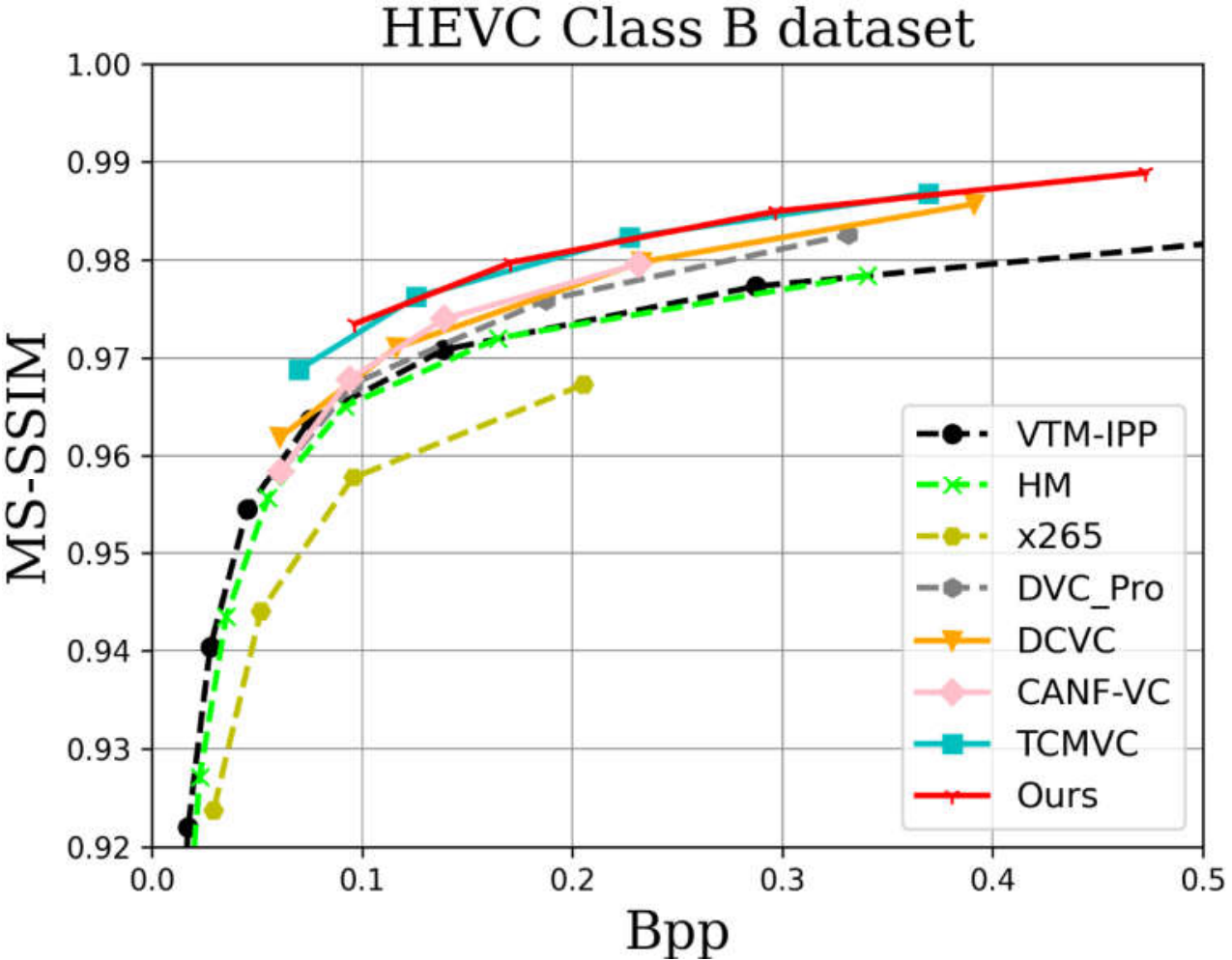}
  \end{minipage}%
  \begin{minipage}[c]{0.22\linewidth}
  \centering
    \includegraphics[width=\linewidth]{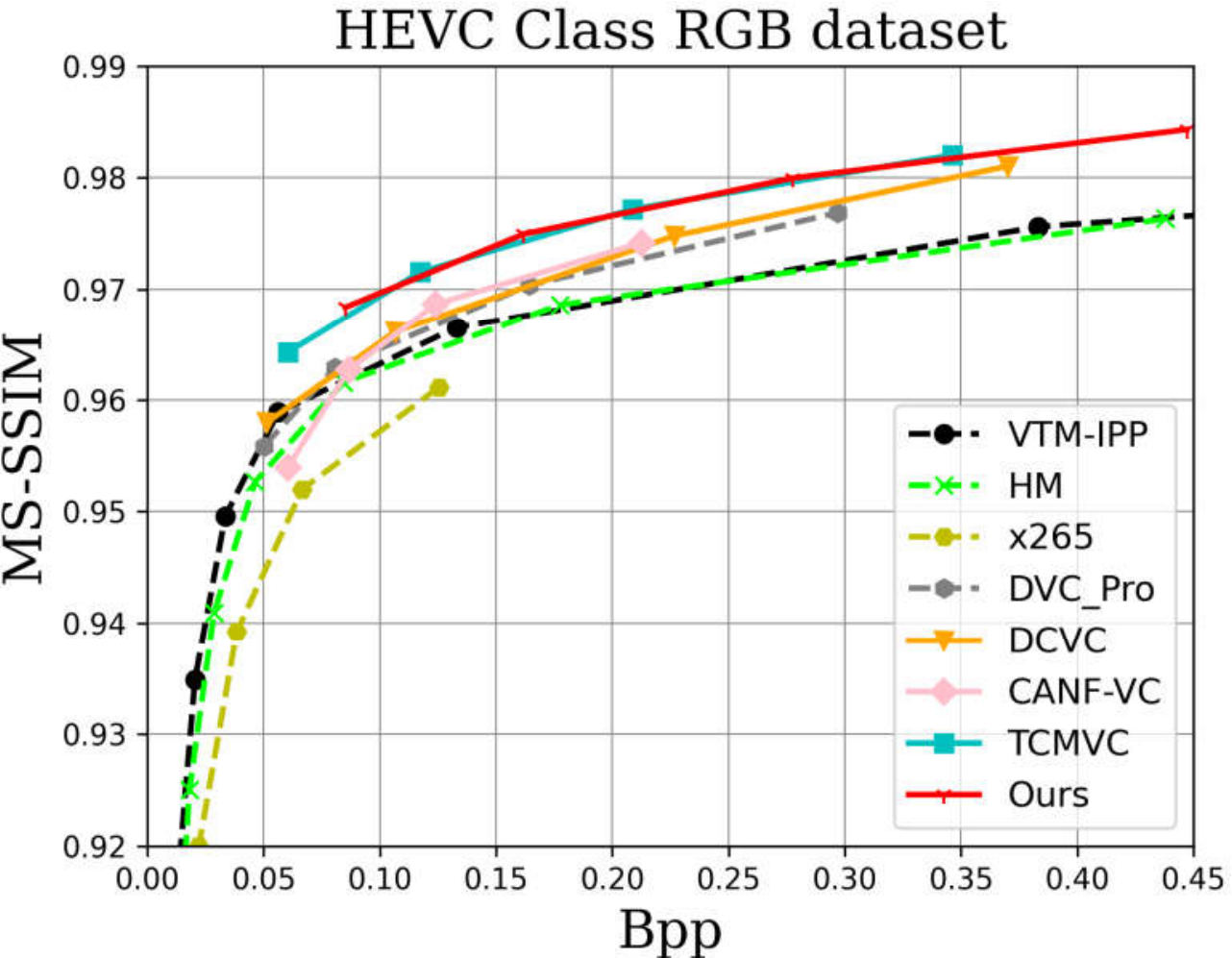}
  \end{minipage}%
  \begin{minipage}[c]{0.22\linewidth}
  \centering
    \includegraphics[width=\linewidth]{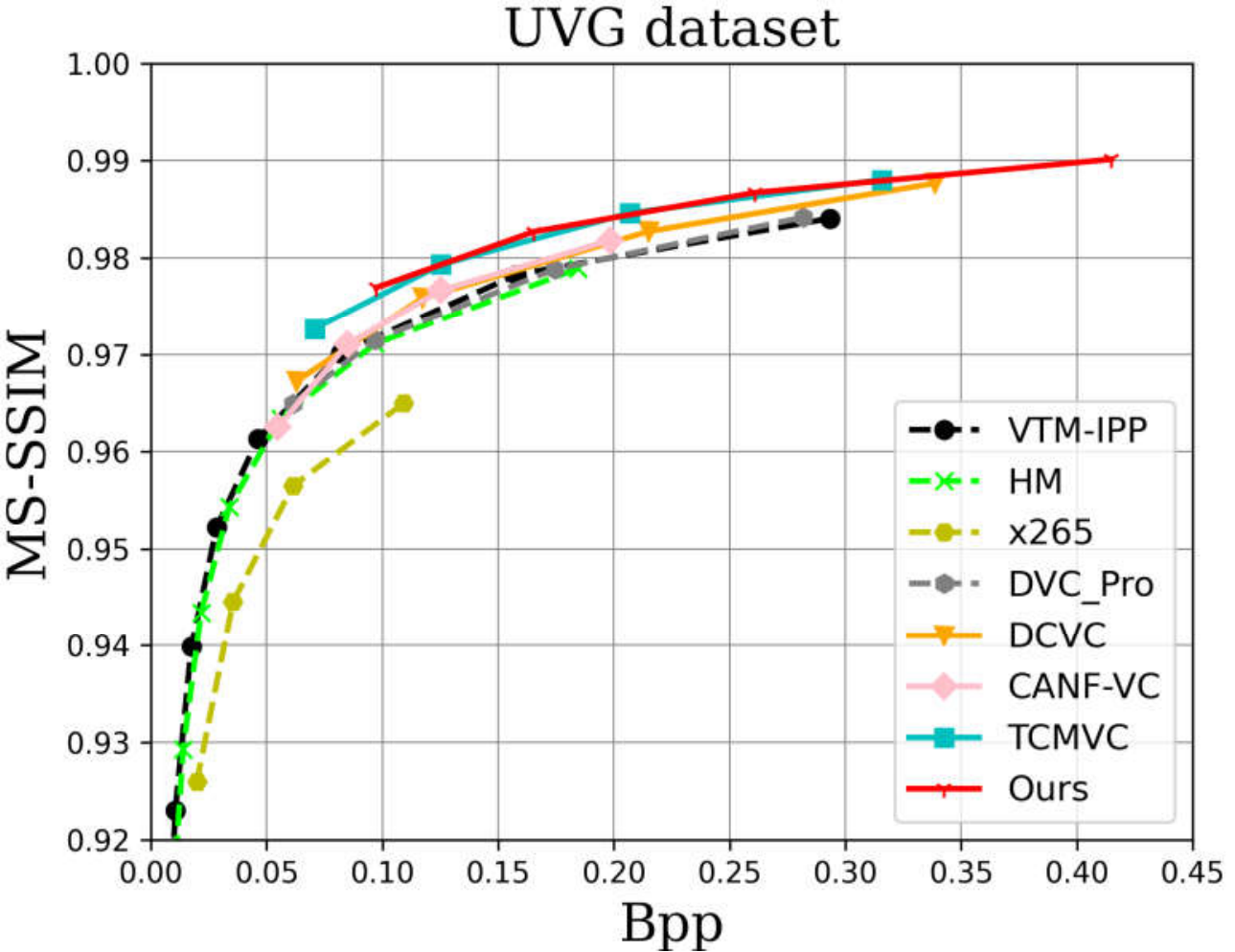}
  \end{minipage}%
  \begin{minipage}[c]{0.22\linewidth}
  \centering
    \includegraphics[width=\linewidth]{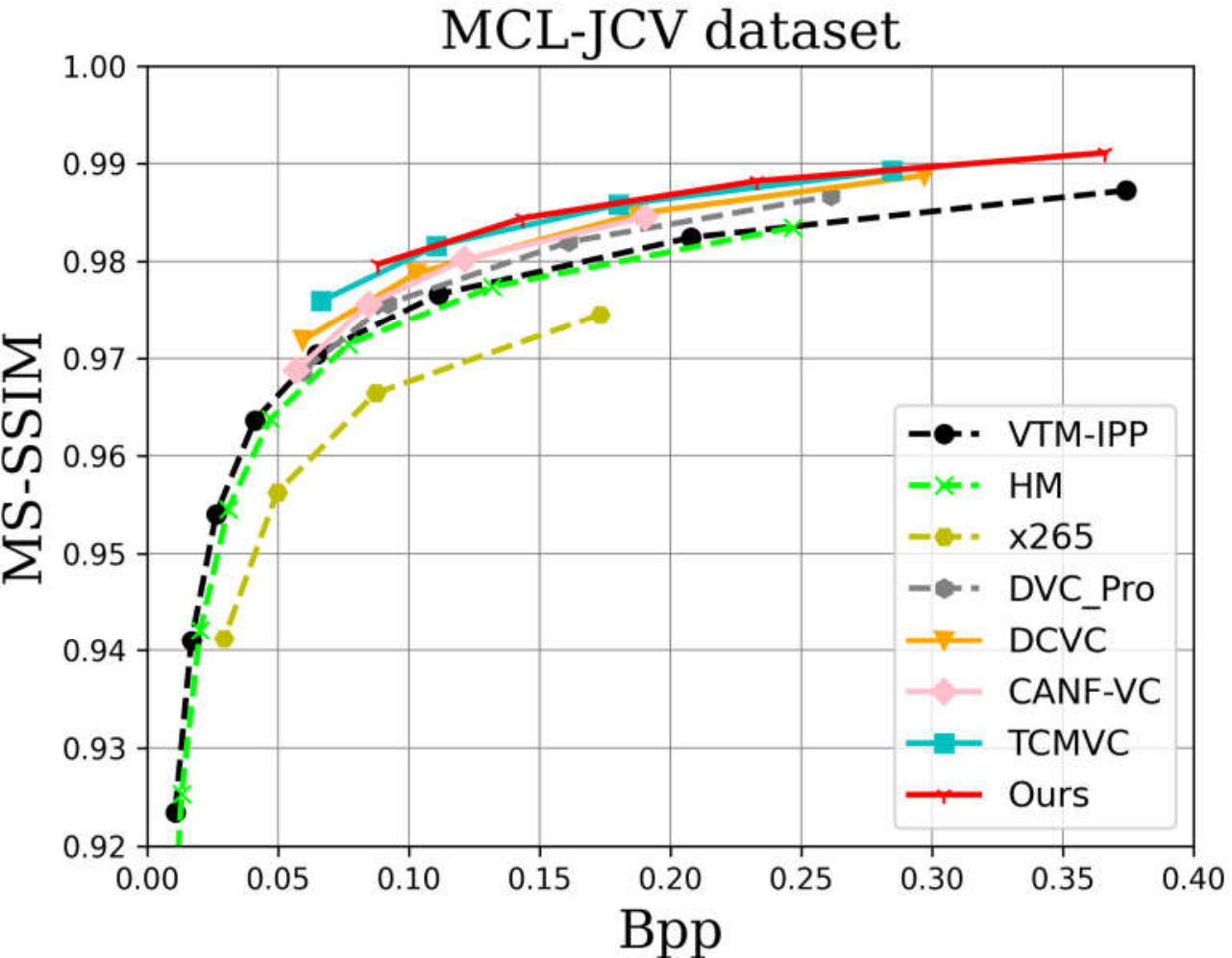}
  \end{minipage}%
    \caption{Rate-distortion curves on the HEVC Class B, HEVC Class RGB, UVG, and MCL-JCV datasets. Intra period is set to 12.}
  \label{fig:result_ip12}
\end{figure*}
\begin{table*}[t]
\caption{BD-rate (\%) for PSNR (intra period 12). The anchor is VTM-IPP. Negative values indicate bits saving while positive values indicate more bit costs.}
  \centering
\scalebox{1}{
\begin{threeparttable}
\begin{tabular}{l|c|c|c|c|c|c|c|c}
\toprule[1.5pt]
               & VTM-IPP & HM     & x265 &DVC\_Pro &DCVC & CANF-VC &TCMVC  &Ours  \\ \hline
HEVC Class B   &0.0        &20.1  &89.1  &67.3     &32.5 &15.6     &--1.0  &--5.2 \\ \hline
HEVC Class C   &0.0        &13.3  &49.5  &70.5     &44.8 &22.2     &15.7   &7.3   \\ \hline
HEVC Class D   &0.0        &11.6  &44.2  &47.2     &25.2 &8.2      &--3.4  &--7.7  \\ \hline
HEVC Class E   &0.0        &22.4  &79.5  &124.2    &66.8 &26.7     &7.8    &1.9    \\ \hline
HEVC Class RGB &0.0        &13.2  &94.5  &50.8     &22.9 &22.9     &--13.7 &--17.9  \\ \hline
MCL-JCV        &0.0        &15.3  &99.2  &63.0     &28.1 &12.5     &--1.6  &--5.3  \\ \hline
UVG            &0.0        &8.3   &97.4  &54.8     &21.0 &--1.6    &--20.4 &--22.3\\ \hline
Average        &0.0        &14.9  &79.1  &68.3     &34.5 &15.21    &--2.4  &--7.0\\ \bottomrule[1.5pt]
\end{tabular}
\end{threeparttable}}
\label{table:ip12_psnr}
\end{table*}
\begin{table*}[t]
\caption{BD-rate (\%) for MS-SSIM (intra period 12). The anchor is VTM-IPP.}
  \centering
\scalebox{1}{
\begin{threeparttable}
\begin{tabular}{l|c|c|c|c|c|c|c|c}
\toprule[1.5pt]
               & VTM-IPP & HM  & x265 & DVC\_Pro & DCVC   & CANF-VC &TCMVC  & Ours  \\ \hline
HEVC Class B   &0.0      &14.9 &72.3  &--12.9    &--25.2  &--15.5   &--49.5 &--53.6\\ \hline
HEVC Class C   &0.0      &13.5 &53.4  &--11.8    &--26.2  &--21.3   &--46.8 &--50.2\\ \hline
HEVC Class D   &0.0      &10.6 &46.0  &--30.8    &--40.1  &--31.8   &--56.5 &--58.4\\ \hline
HEVC Class E   &0.0      &17.0 &60.8  &17.3      &--8.0   &1.3      &--50.1 &--56.4\\ \hline
HEVC Class RGB &0.0      &14.7 &68.6  &--13.1    &--24.3  &--4.3    &--50.9 &--53.6\\ \hline
MCL-JCV        &0.0      &14.1 &75.4  &--12.0    &--26.8  &--17.6   &--40.6 &--44.7\\ \hline
UVG            &0.0      &10.2 &73.5  &4.7       &--9.2   &--3.4    &--29.1 &--32.6\\ \hline
Average        &0.0      &13.6 &64.3  &--8.4     &--22.8  &--13.2   &--46.2 &--49.9\\ \bottomrule[1.5pt]
\end{tabular}
\end{threeparttable}}
\label{table:ip12_msssim}
\end{table*}
\begin{figure*}[t]
  \centering
  \begin{minipage}[c]{0.22\linewidth}
  \centering
    \includegraphics[width=\linewidth]{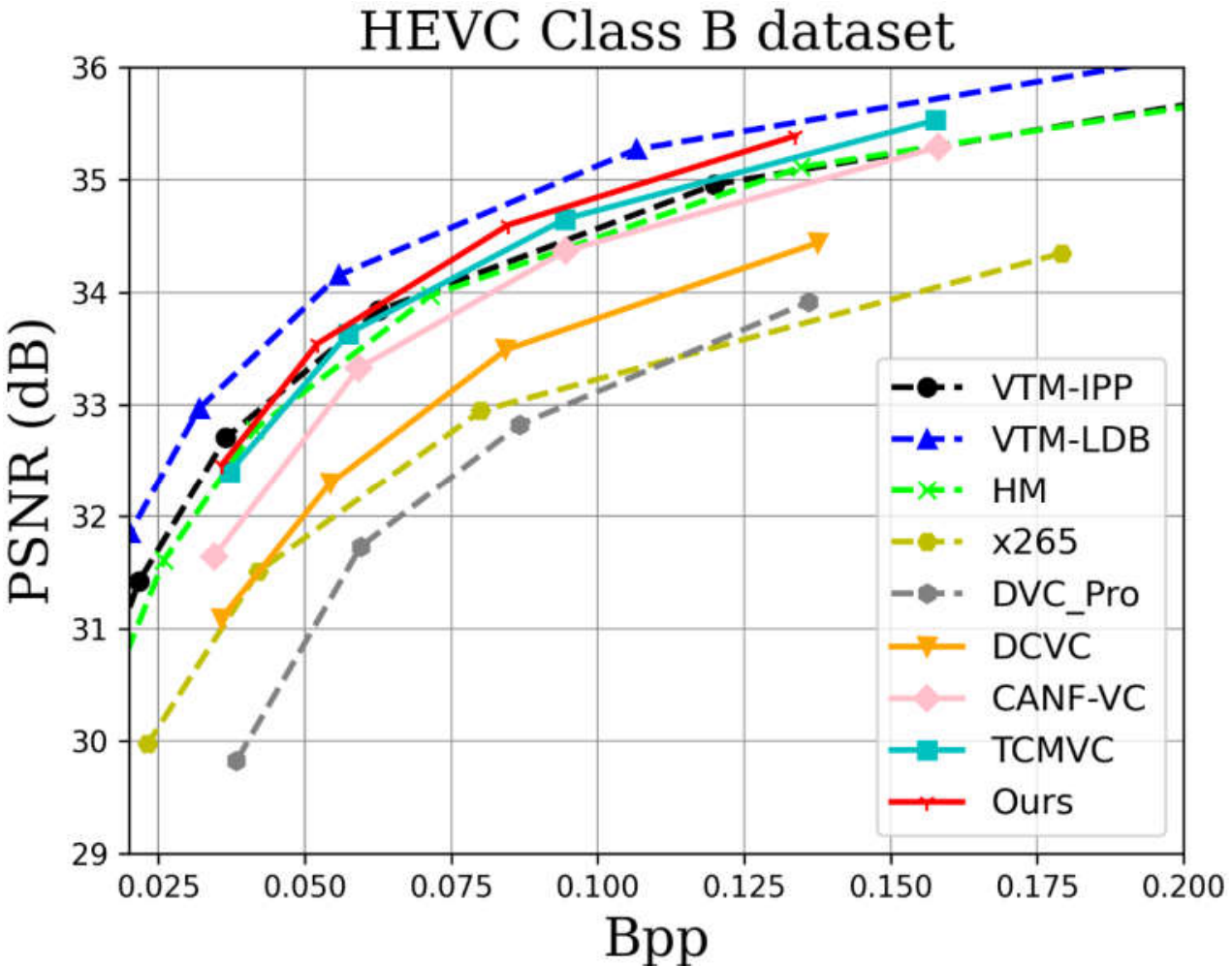}
 \end{minipage}%
  \begin{minipage}[c]{0.22\linewidth}
  \centering
    \includegraphics[width=\linewidth]{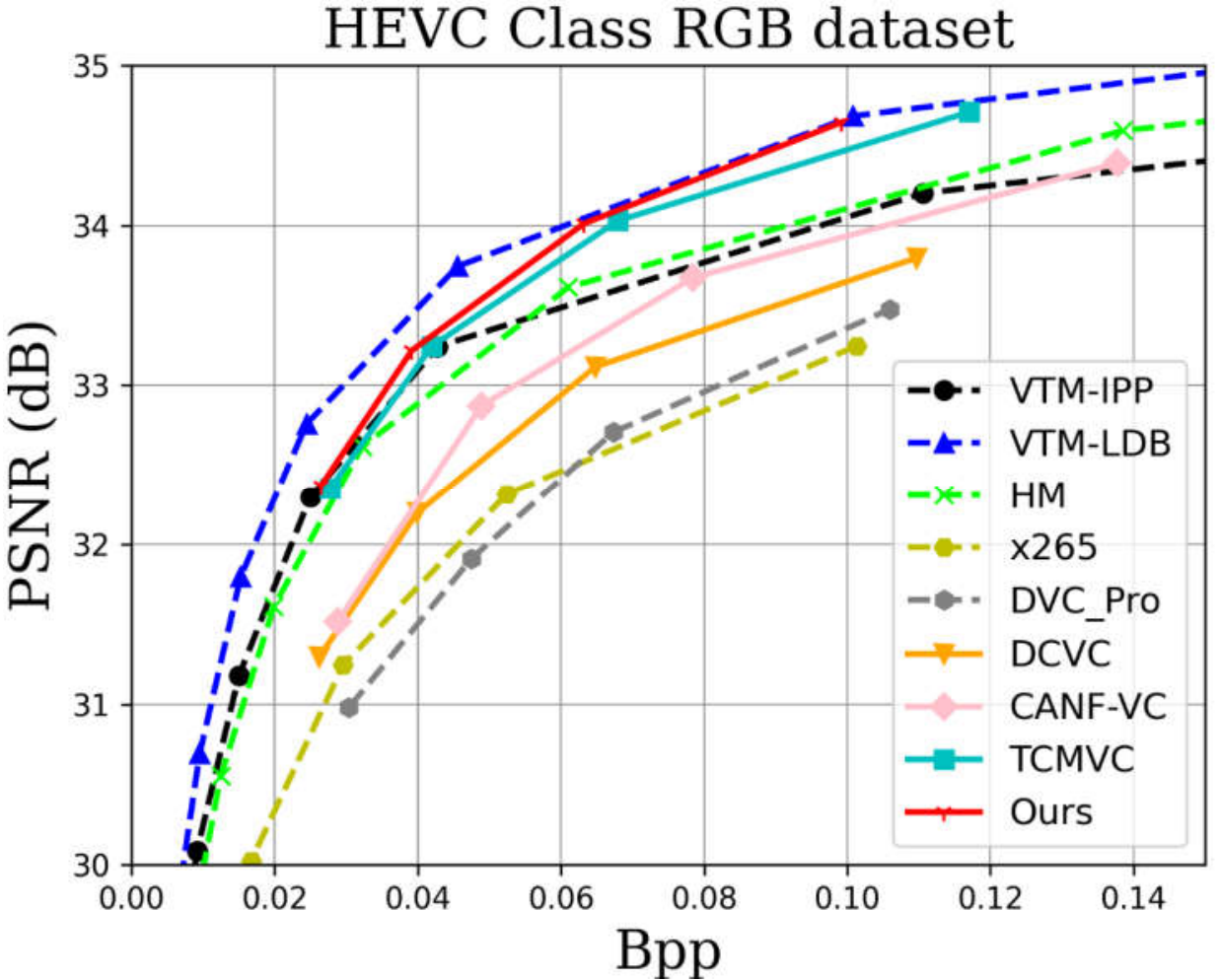}
 \end{minipage}%
  \begin{minipage}[c]{0.22\linewidth}
  \centering
    \includegraphics[width=\linewidth]{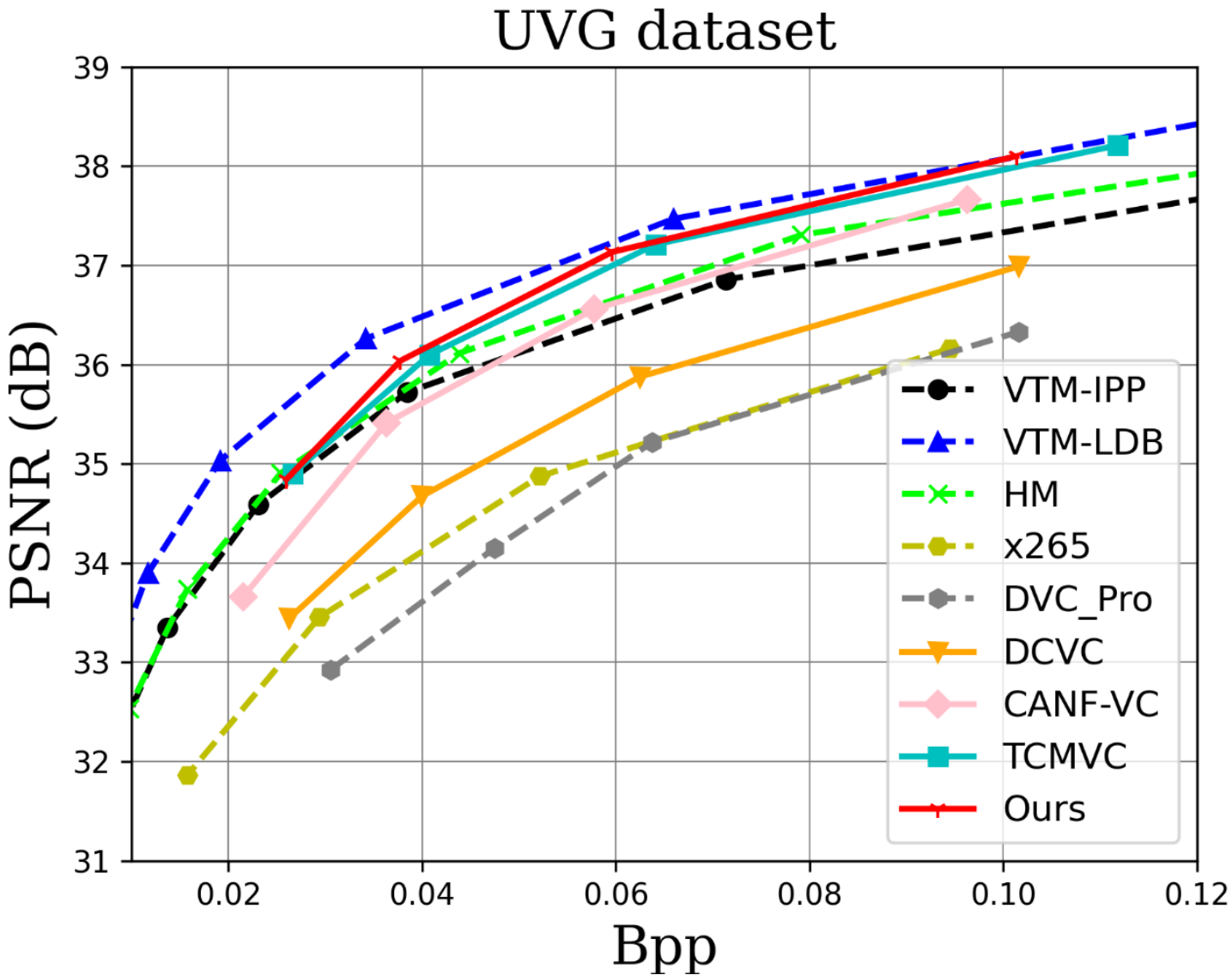}
  \end{minipage}%
  \begin{minipage}[c]{0.22\linewidth}
  \centering
    \includegraphics[width=\linewidth]{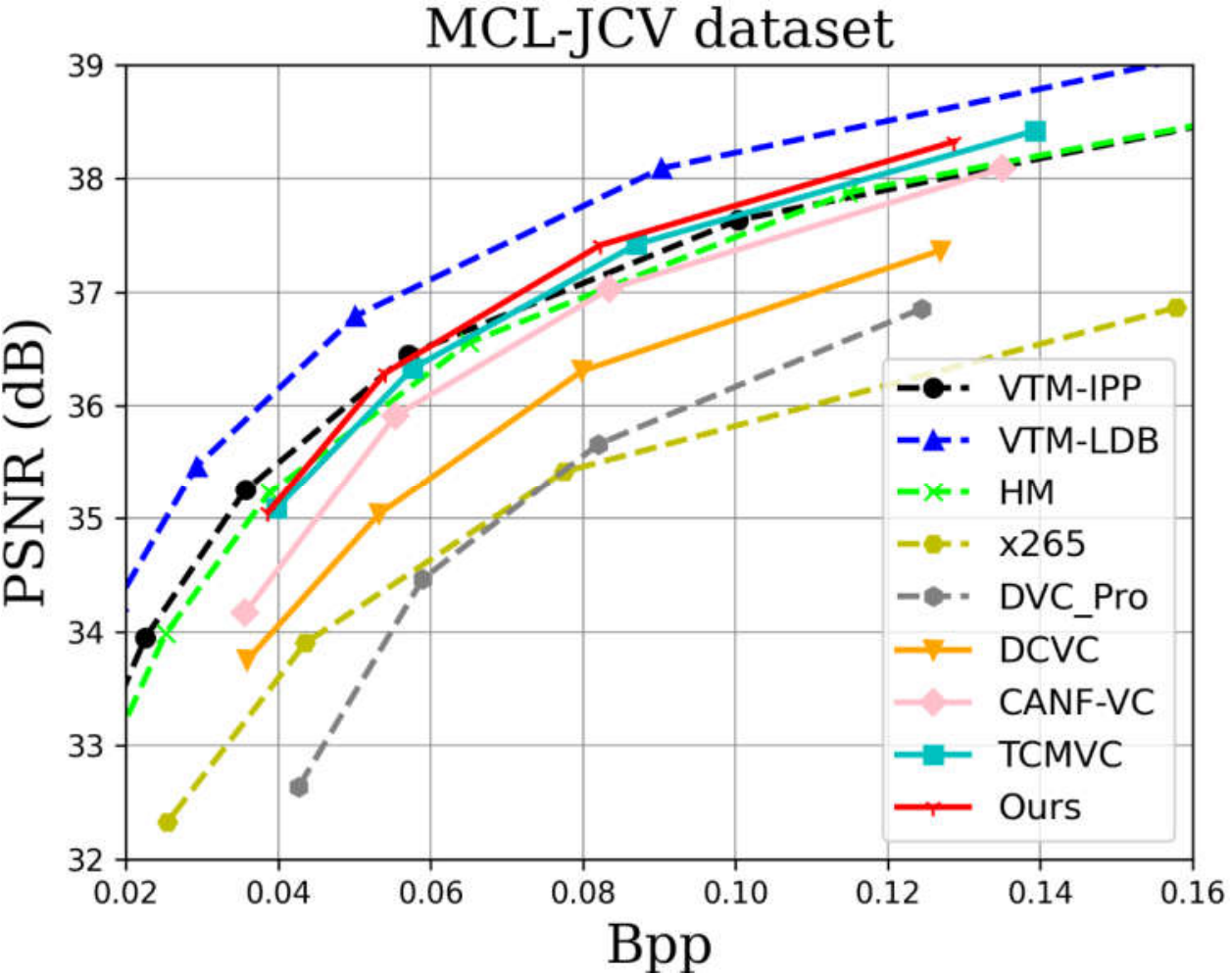}
  \end{minipage}%
  
  \begin{minipage}[c]{0.22\linewidth}
  \centering
    \includegraphics[width=\linewidth]{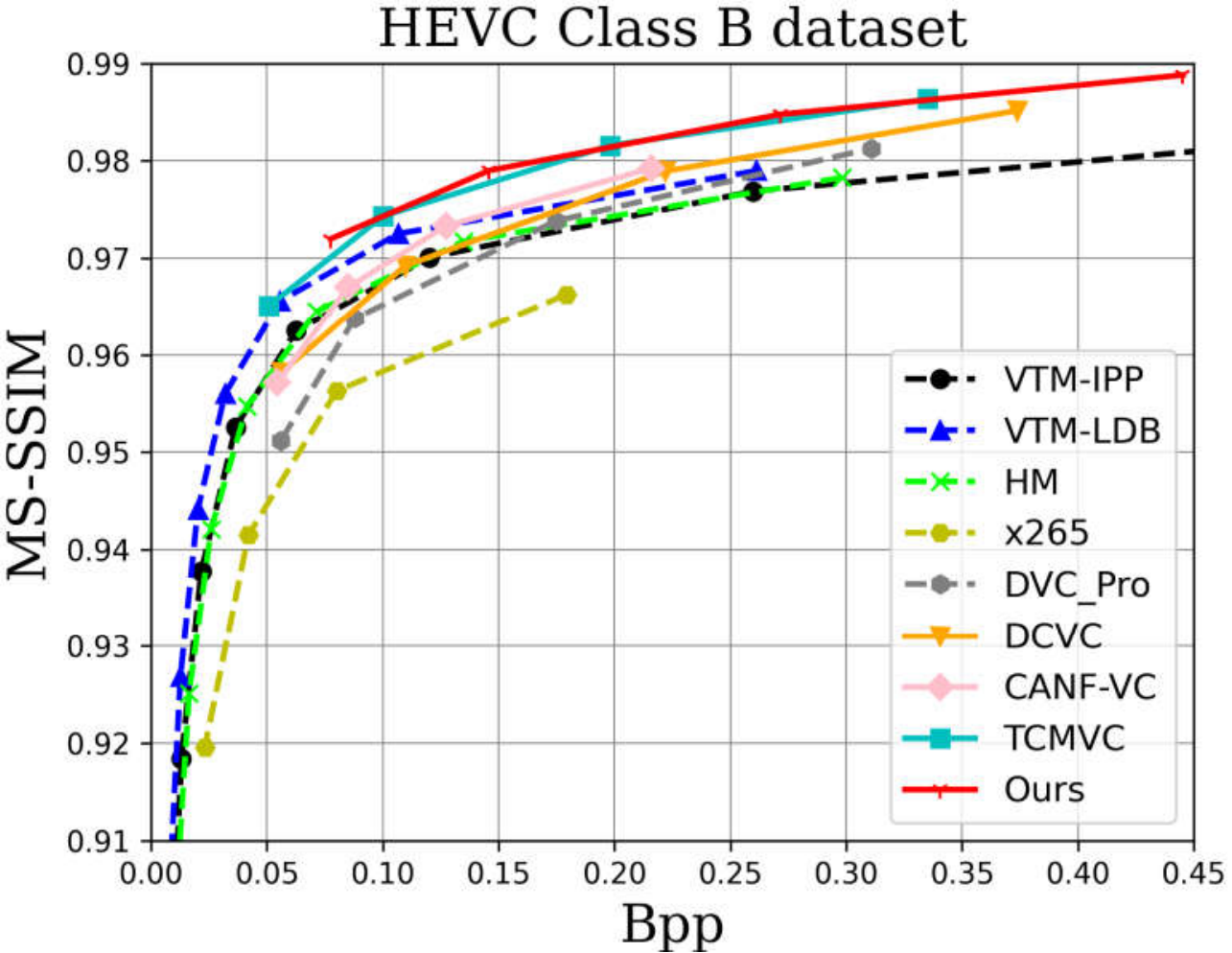}
  \end{minipage}%
  \begin{minipage}[c]{0.22\linewidth}
  \centering
    \includegraphics[width=\linewidth]{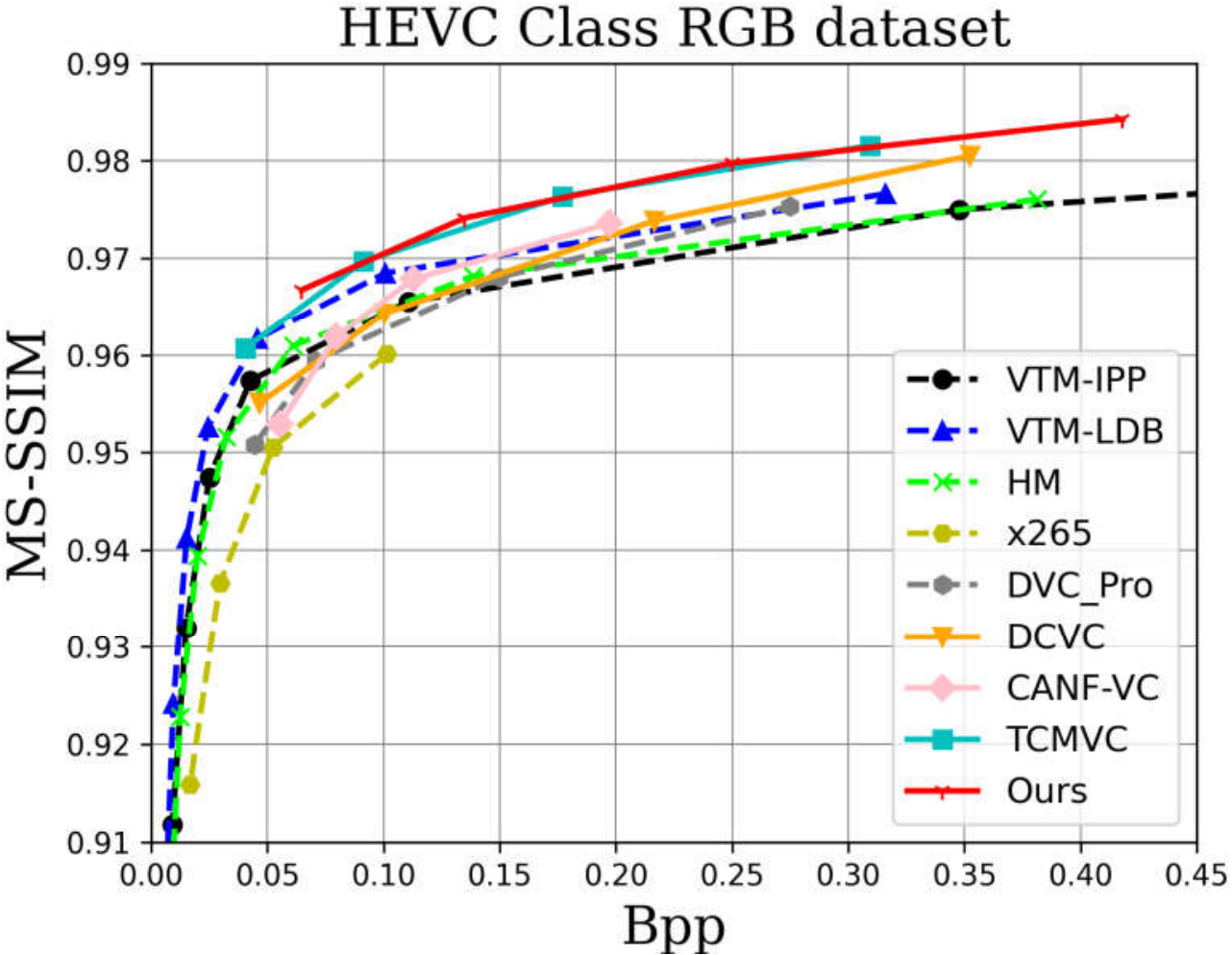}
  \end{minipage}%
  \begin{minipage}[c]{0.22\linewidth}
  \centering
    \includegraphics[width=\linewidth]{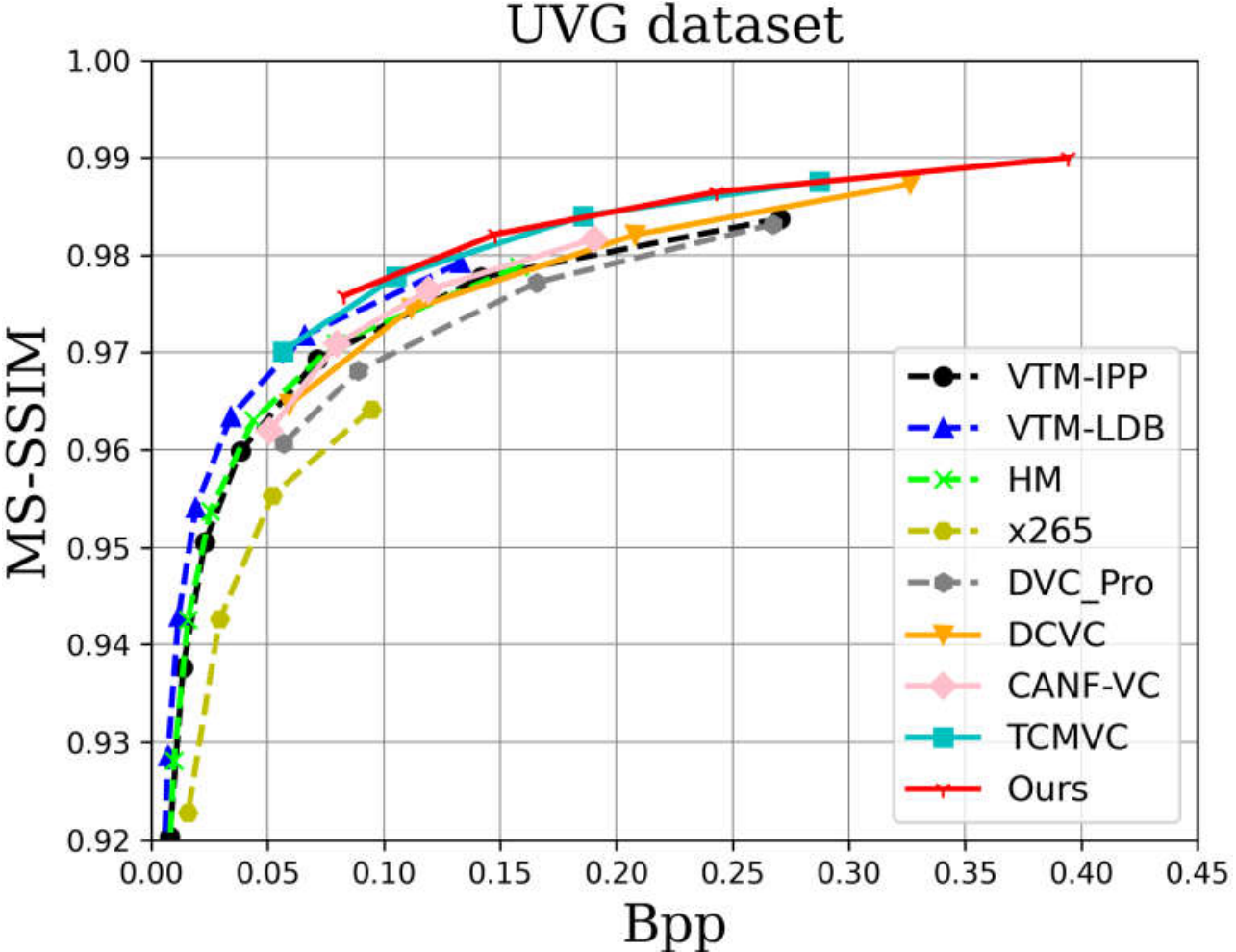}
  \end{minipage}%
  \begin{minipage}[c]{0.22\linewidth}
  \centering
    \includegraphics[width=\linewidth]{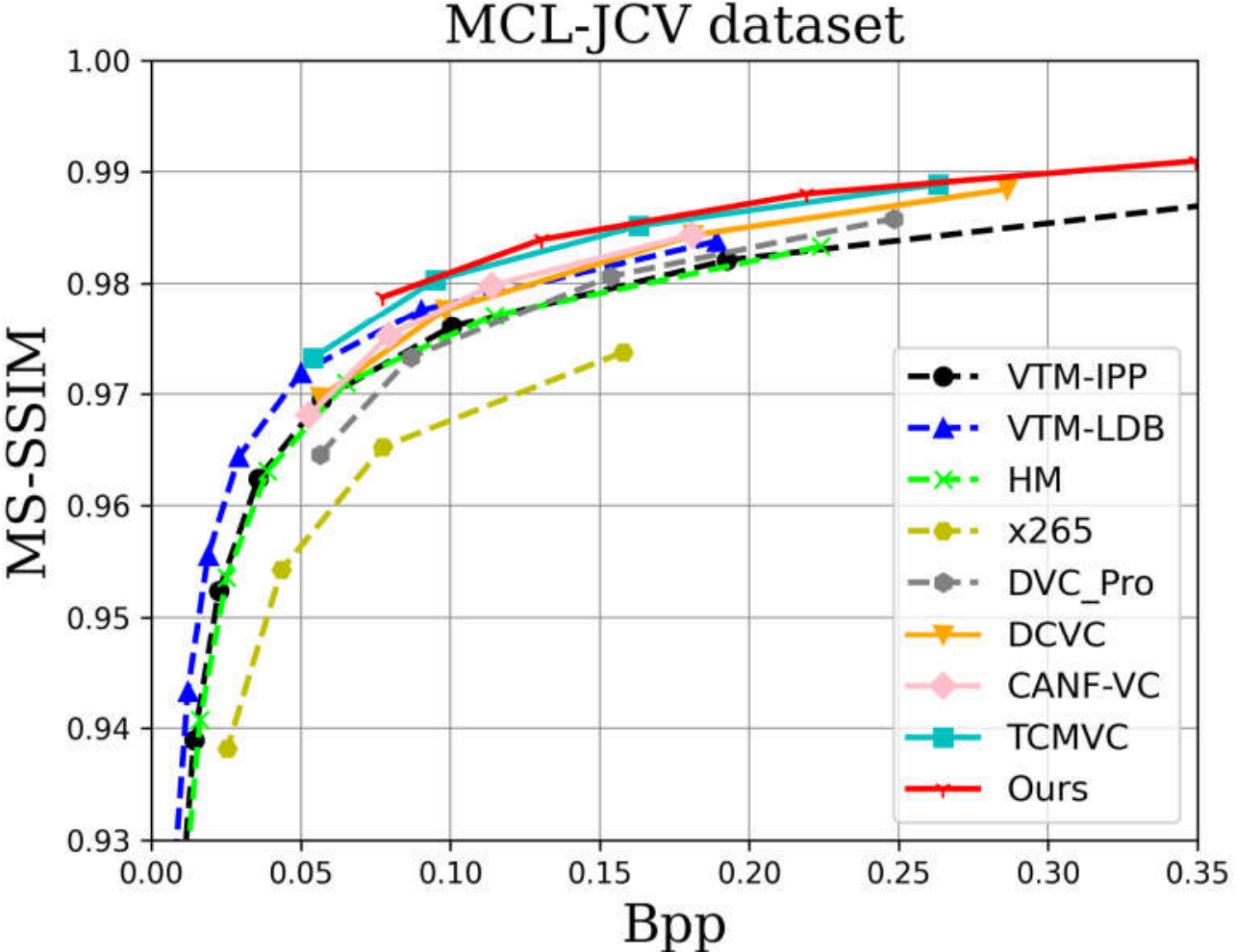}
  \end{minipage}%
    \caption{Rate-distortion curves on the HEVC Class B, HEVC Class RGB, UVG, and MCL-JCV datasets. Intra period is set to 32.}
  \label{fig:result_ip32}
\end{figure*}
\begin{table*}[t]
\caption{BD-rate (\%) for PSNR (intra period 32). The anchor is VTM-IPP. Negative values indicate bits saving while positive values indicate more bits costs.}
  \centering
\scalebox{1}{
\begin{threeparttable}
\begin{tabular}{l|c|c|c|c|c|c|c|c|c}
\toprule[1.5pt]
\diagbox{Datasets}{Schemes}& VTM-IPP & VTM-LDB & HM  & x265 & DVC\_Pro & DCVC  & CANF-VC &TCMVC & Ours  \\ \hline
HEVC Class B   &0.0      &--22.2   &9.0  &94.8  &149.2     &70.2   &21.7     &0.3   &--6.1\\ \hline
HEVC Class C   &0.0      &--25.7   &4.0  &51.6  &144.7     &87.5   &25.0     &16.6  &6.5\\ \hline
HEVC Class D   &0.0      &--23.7   &3.6  &47.7  &104.9     &58.0   &8.9      &--5.7 &--10.6\\ \hline
HEVC Class E   &0.0      &--24.0   &11.1 &91.5  &352.1     &202.5  &67.4     &24.8  &19.0\\ \hline
HEVC Class RGB &0.0      &--28.6   &1.7  &101.0 &121.5     &65.0   &37.0     &--14.2&--18.8\\ \hline
MCL-JCV        &0.0      &--25.5   &7.7  &103.6 &119.8     &55.9   &20.7     &2.7   &--1.3\\ \hline
UVG            &0.0      &--31.9   &--4.4 &104.4 &128.8     &58.1   &6.4      &--18.3&--20.7\\ \hline
Average        &0.0      &--25.9   &4.7  &84.9  &160.1     &85.3   &26.7     &0.9   &--4.6\\ \bottomrule[1.5pt]
\end{tabular}
\end{threeparttable}}
\label{table:ip32_psnr}
\end{table*}
\begin{table*}[t]
\caption{BD-rate (\%) for MS-SSIM (intra period 32). The anchor is VTM-IPP.}
  \centering
\scalebox{1}{
\begin{threeparttable}
\begin{tabular}{l|c|c|c|c|c|c|c|c|c}
\toprule[1.5pt]
\diagbox{Datasets}{Schemes}& VTM-IPP & VTM-LDB & HM  & x265 &DVC\_Pro & DCVC   & CANF-VC &TCMVC  & Ours  \\ \hline
HEVC Class B   &0.0      &--23.5   &4.4  &78.1  &15.8     &--10.9  & --11.2  &--47.3 &--54.4\\ \hline
HEVC Class C   &0.0      &--25.1   &3.5  &56.0  &15.4     &--12.9  & --18.2  &--46.5 &--52.2\\ \hline
HEVC Class D   &0.0      &--23.8   &3.1  &49.6  &--12.9   &--30.4  & --31.2  &--57.4 &--60.9\\ \hline
HEVC Class E   &0.0      &--23.1   &8.5  &71.8  &112.6    &52.2    & 40.4    &--41.2 &--51.1\\ \hline
HEVC Class RGB &0.0      &--24.3   &4.1  &75.1  &15.7     &--9.3   & 40.8    &--50.3 &--56.0\\ \hline
MCL-JCV        &0.0      &--25.5   &6.2  &81.4  &7.0      &--17.9  & --16.0  &--38.2 &--44.7\\ \hline
UVG            &0.0      &--28.4   &--2.2&79.6  &28.4     &3.9     & --0.1   &--27.4 &--32.7\\ \hline
Average        &0.0      &--24.8   &3.9  &70.2  &26.0     &--3.6   &0.6      &--44.0 &--50.3\\ \bottomrule[1.5pt]
\end{tabular}
\end{threeparttable}}
\label{table:ip32_msssim}
\end{table*}
\begin{table}[t]
 \centering
 \caption{Average encoding/decoding time for a 1080p frame. The time for entropy encoding/decoding is included.}
\scalebox{1}{
\begin{tabular}{c|c|c}
\toprule[1.5pt]
Schemes  & Enc T (s) & Dec T (s)            \\ \hline
DVC\_Pro &7.075&30.422         \\ \hline
DCVC     &7.580& 31.746        \\ \hline
CANF-VC  &2.160&1.087              \\ \hline
TCMVC    &1.593&0.937              \\ \hline
Ours     &1.336&0.961                 \\ \bottomrule[1.5pt]
\end{tabular}}
\label{time}
\end{table}

\subsection{Implementation Details}
\subsubsection{Video Reconstruction} Towards video reconstruction, following~\cite{sheng2022temporal,li2021deep,canfvc}, four models with different $\lambda$ values ($\lambda = 256, 512, 1024, 2048$) are trained for different coding rates. We use AdamW~\cite{Loshchilov2019DecoupledWD} optimizer and set the batch size to 4. When using 1$-$MS-SSIM~\cite{wang2003multiscale} as the distortion, each $\lambda$ value is divided by 50.\par
\subsubsection{Video Enhancement}
The implementation details for video enhancement tasks are listed as follows.\par 
\textbf{Blind/Non-Blind Video Denoising.} Towards non-blind video denoising, for each bit rate, we train two models for noisy videos with different white Gaussian noise standard deviations ($\delta$=20 and 40). The larger $\delta$ is, the more noise exists. Towards blind video denoising, for each bit rate, we train a single model that can be applied to the noisy videos with any $\delta \in [0~55]$.
\par
\textbf{BI-Degradation/Real-World Video SR.}
Towards BI-degradation video SR, we use the Bicubic to obtain downsampled training patches. The corresponding high-resolution patches are regarded as the ground truth. For real-world video SR, we use real-world low/high-resolution videos as training patches. \par
\subsubsection{Video Analysis}
The implementation details for video analysis tasks are listed as follows.\par 
\textbf{Video Action Recognition.} We use MMaction2~\cite{2020mmaction2} to implement video action recognition networks. \emph{tsm\_k400\_pretrained\_r50\_1$\times$1$\times$8\_25e\_ucf101\_rgb.py} and \emph{tsm\_r50\_1$\times$1$\times$8\_50e\_sthv2\_rgb.py} are used as configuration files for training and evaluation. \par
\textbf{Video Object Detection.} 
We use MMtracking~\cite{mmtrack2020} to implement video object detection networks. The configuration file \emph{selsa\_faster\_rcnn\_r101\_dc5\_1$\times$\_imagenetvid.py} is used for training and evaluation. Different from the original SELAS model, we only aggregate the temporal information of a single previous frame for online detection.\par
\textbf{Video Multiple Object Tracking.} 
We use MMtracking to implement video multiple object tracking networks. The configuration file \emph{qdtrack\_faster-rcnn\_r50\_fpn\_4e\_mot-private-half.py} is used for training and evaluation. \par
\textbf{Video Object Segmentation.} 
We use AOT benchmark~\cite{yang2021aot} to implement video object segmentation networks. We only use the ``PRE\_DAV" configuration without the static ``PRE" for training and evaluation.
\begin{figure*}[t]
  \centering
  \begin{minipage}[c]{0.22\linewidth}
  \centering
    \includegraphics[width=\linewidth]{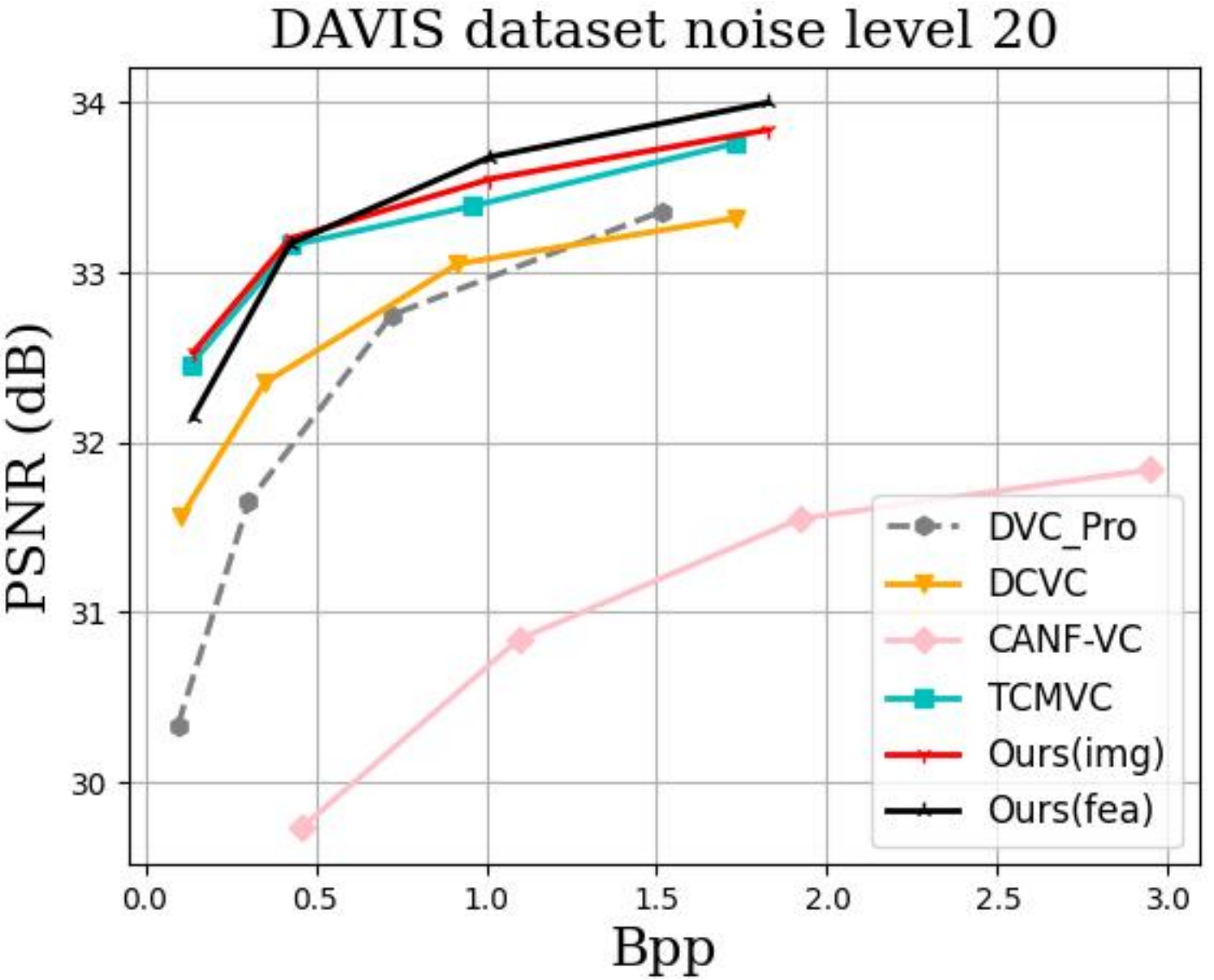}
 \end{minipage}%
  \begin{minipage}[c]{0.22\linewidth}
  \centering
    \includegraphics[width=\linewidth]{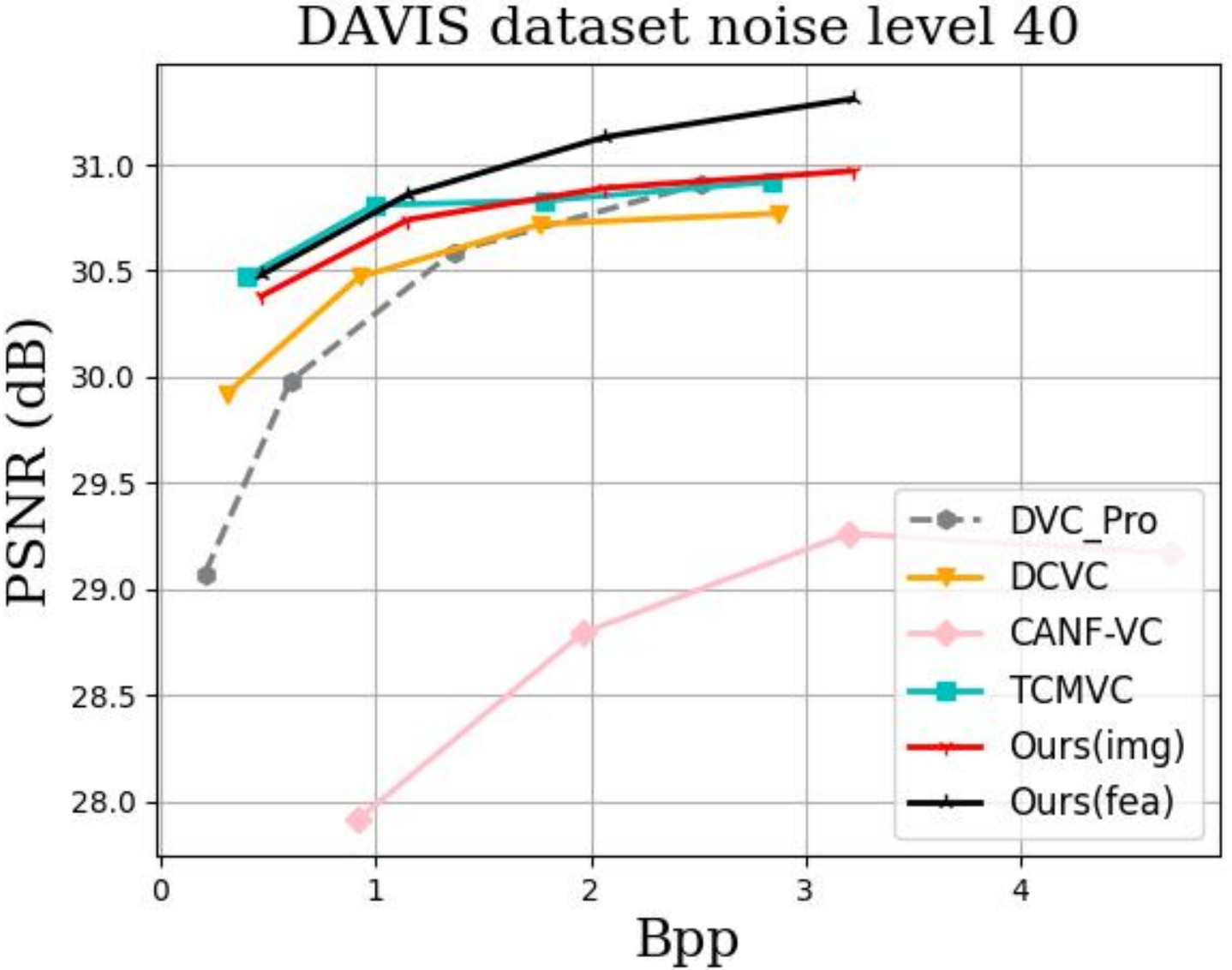}
 \end{minipage}%
  \begin{minipage}[c]{0.22\linewidth}
  \centering
    \includegraphics[width=\linewidth]{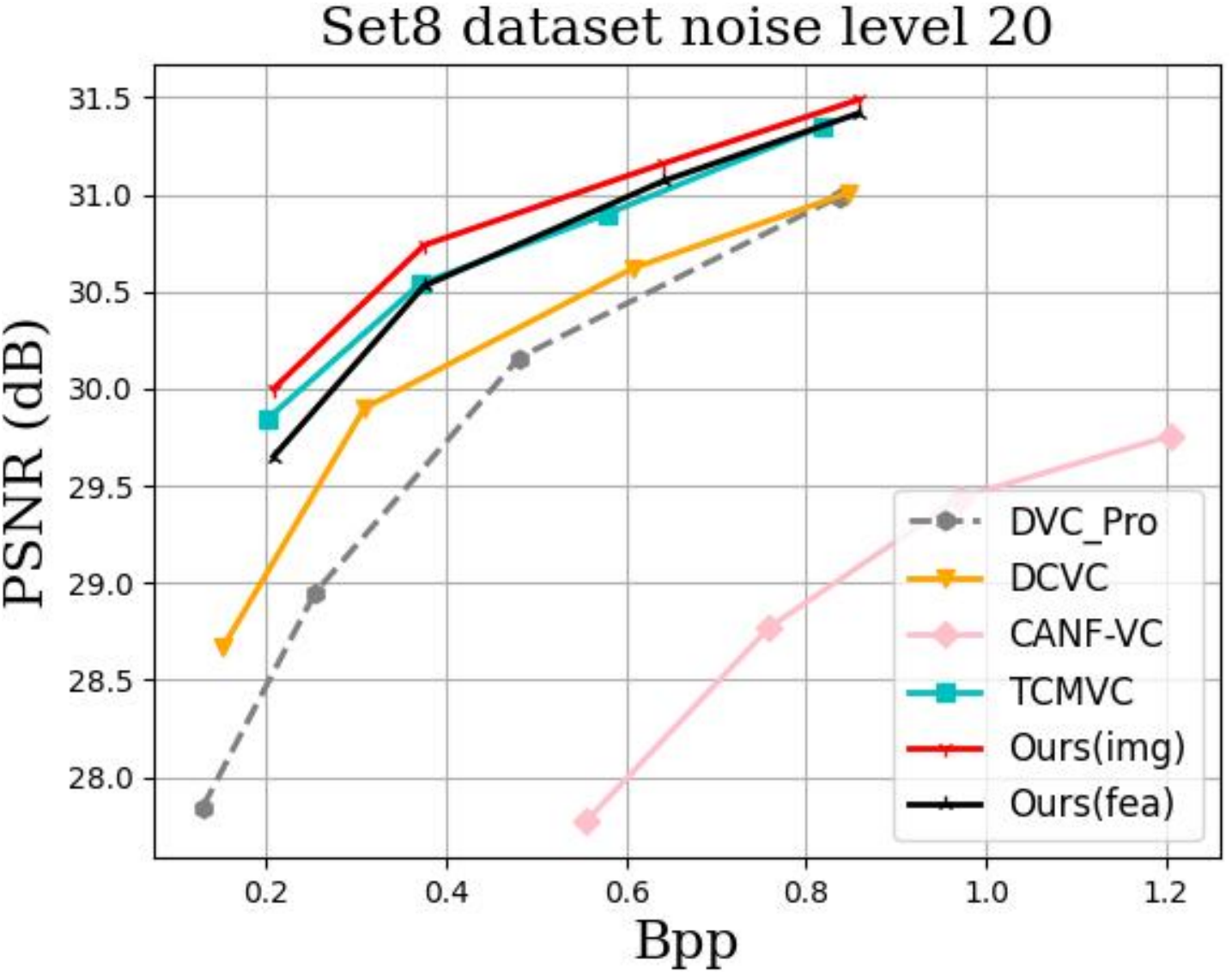}
  \end{minipage}%
  \begin{minipage}[c]{0.22\linewidth}
  \centering
    \includegraphics[width=\linewidth]{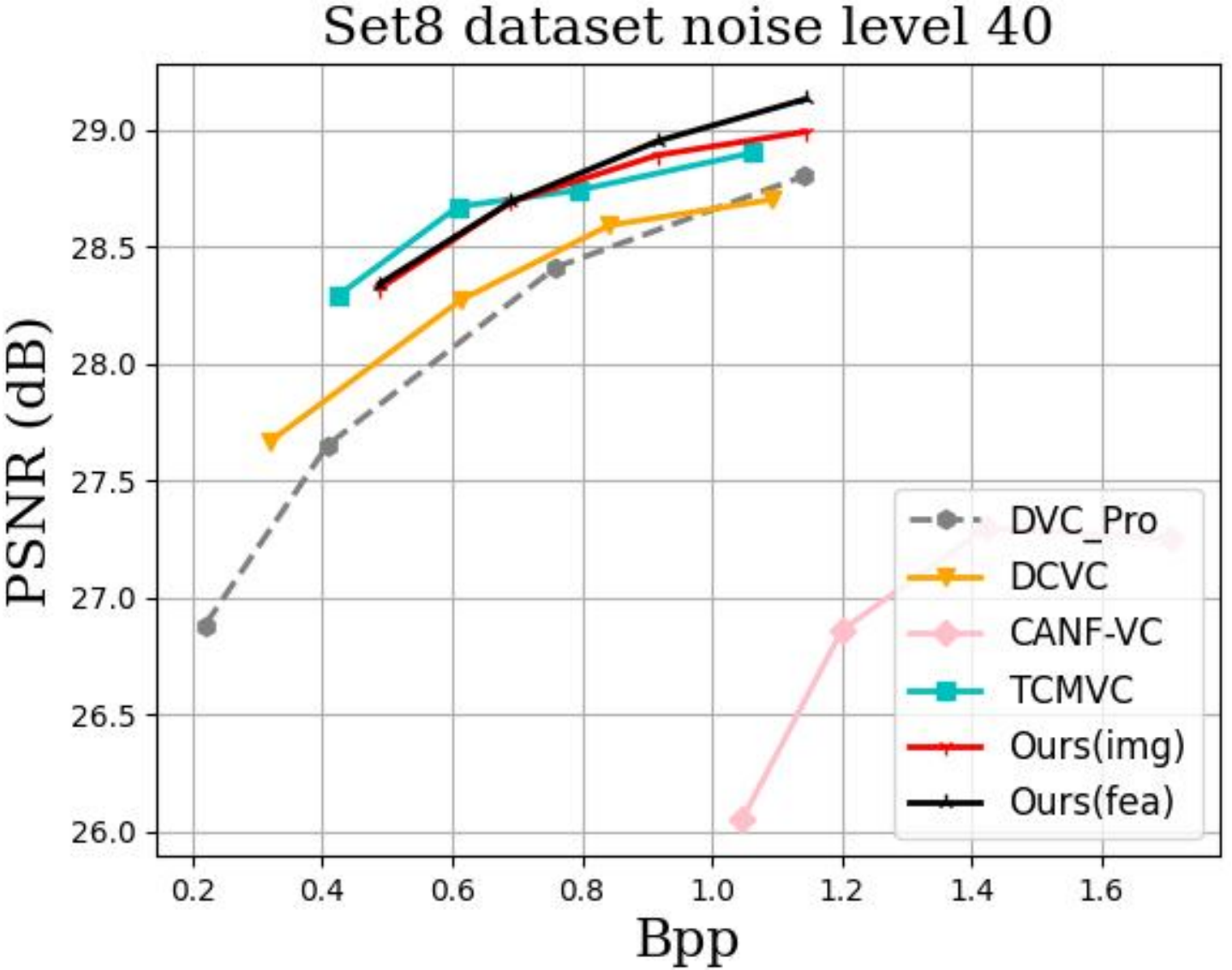}
  \end{minipage}%
  \\
  \begin{minipage}[c]{0.22\linewidth}
  \centering
    \includegraphics[width=\linewidth]{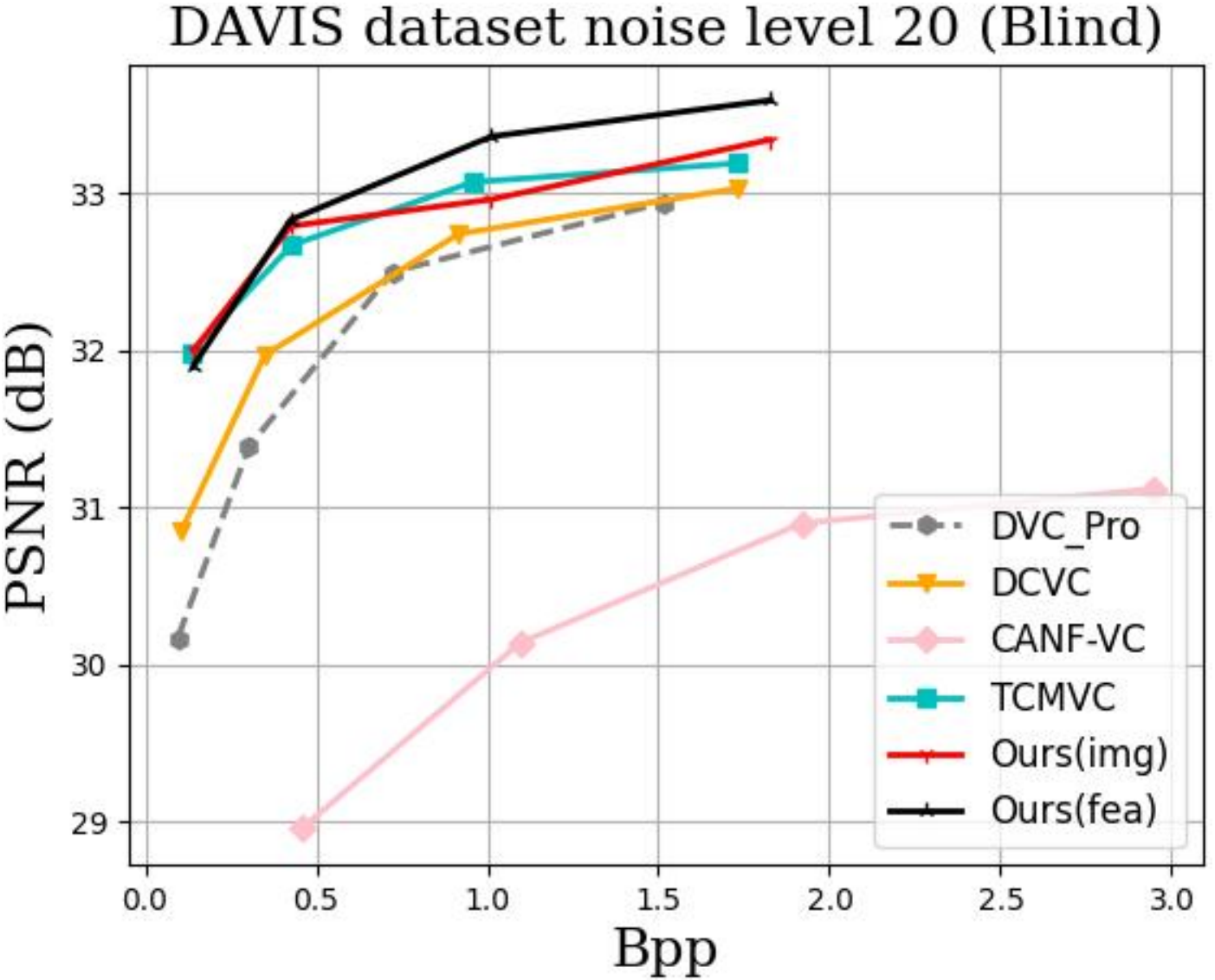}
 \end{minipage}%
  \begin{minipage}[c]{0.22\linewidth}
  \centering
    \includegraphics[width=\linewidth]{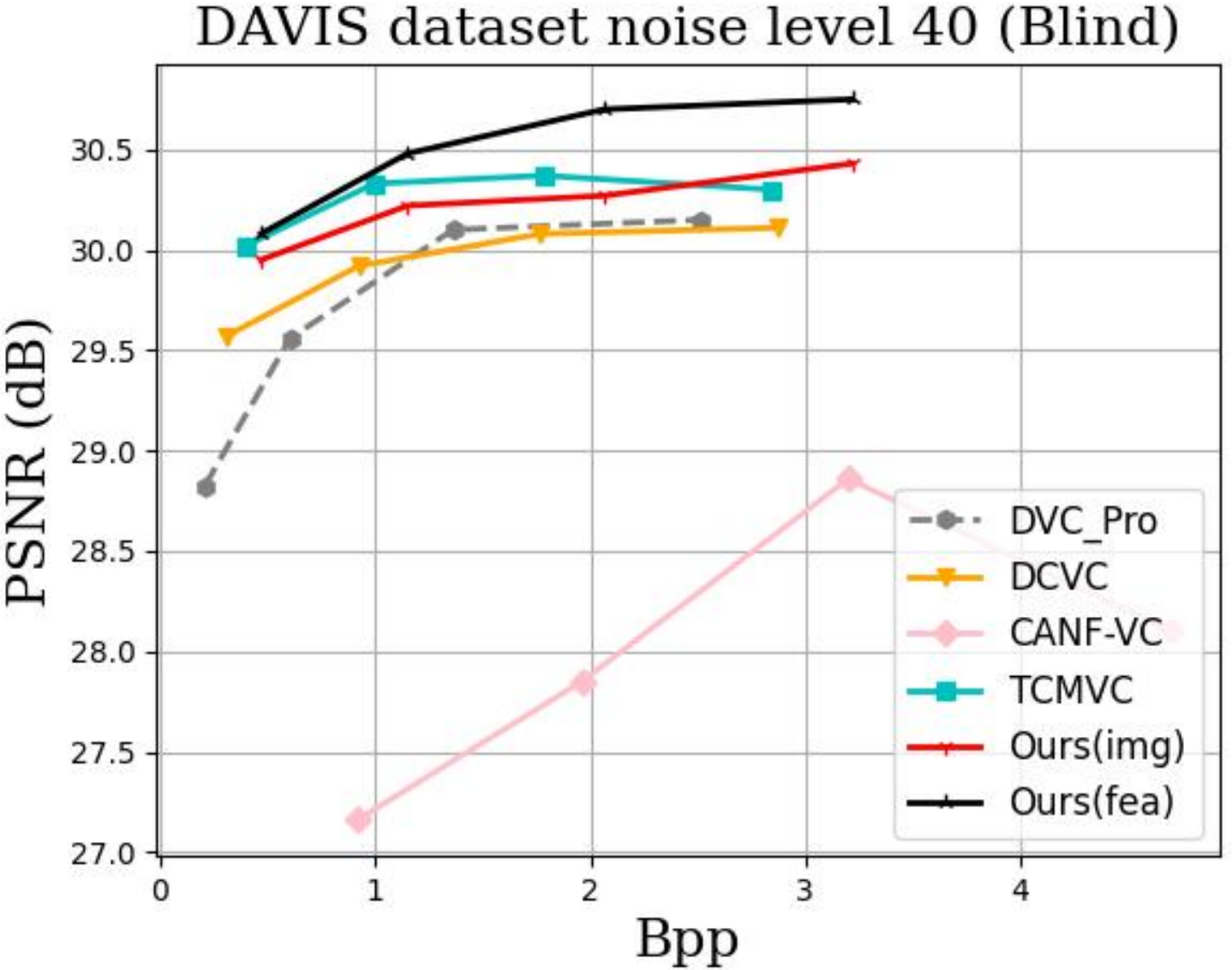}
 \end{minipage}%
  \begin{minipage}[c]{0.22\linewidth}
  \centering
    \includegraphics[width=\linewidth]{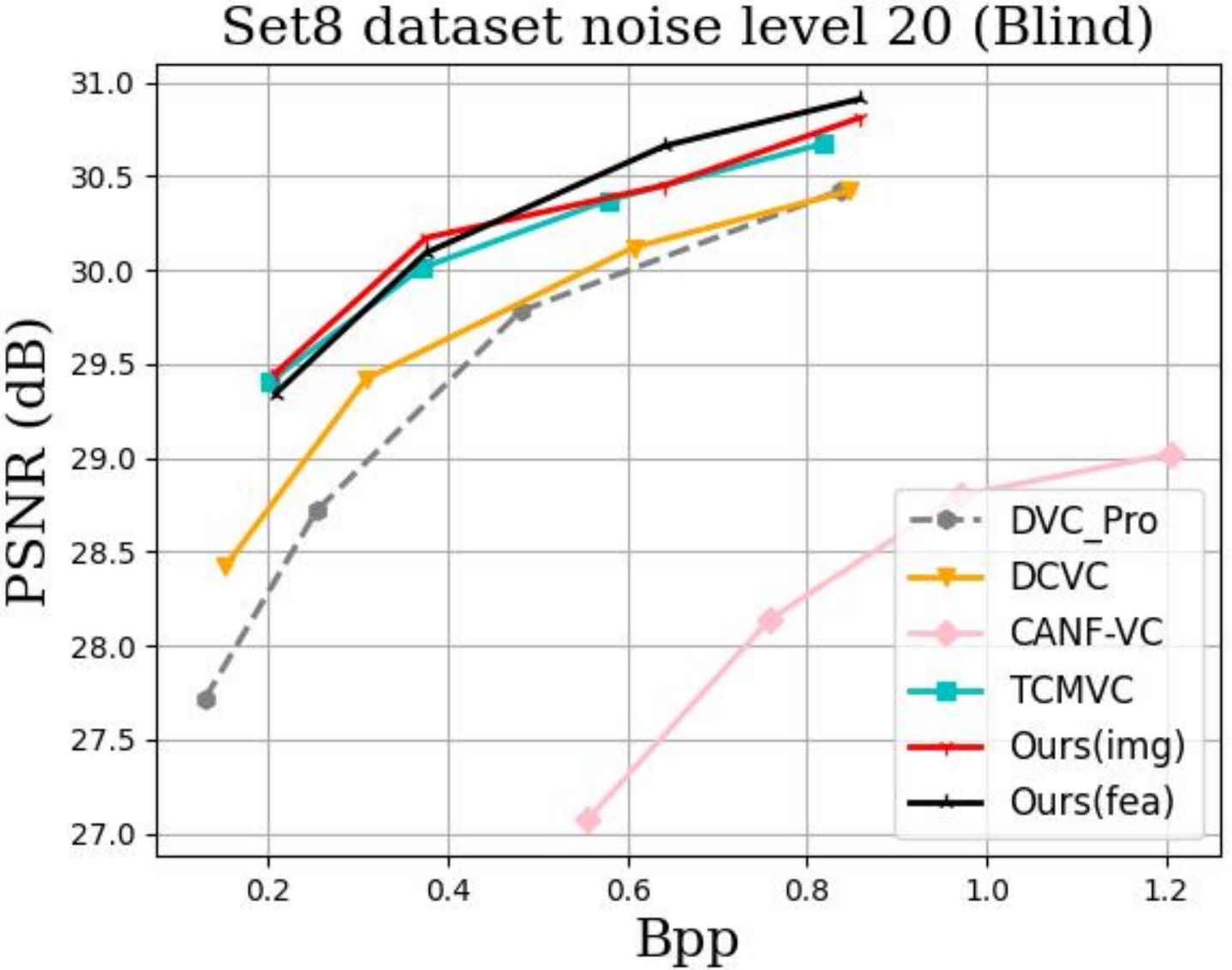}
  \end{minipage}%
  \begin{minipage}[c]{0.22\linewidth}
  \centering
    \includegraphics[width=\linewidth]{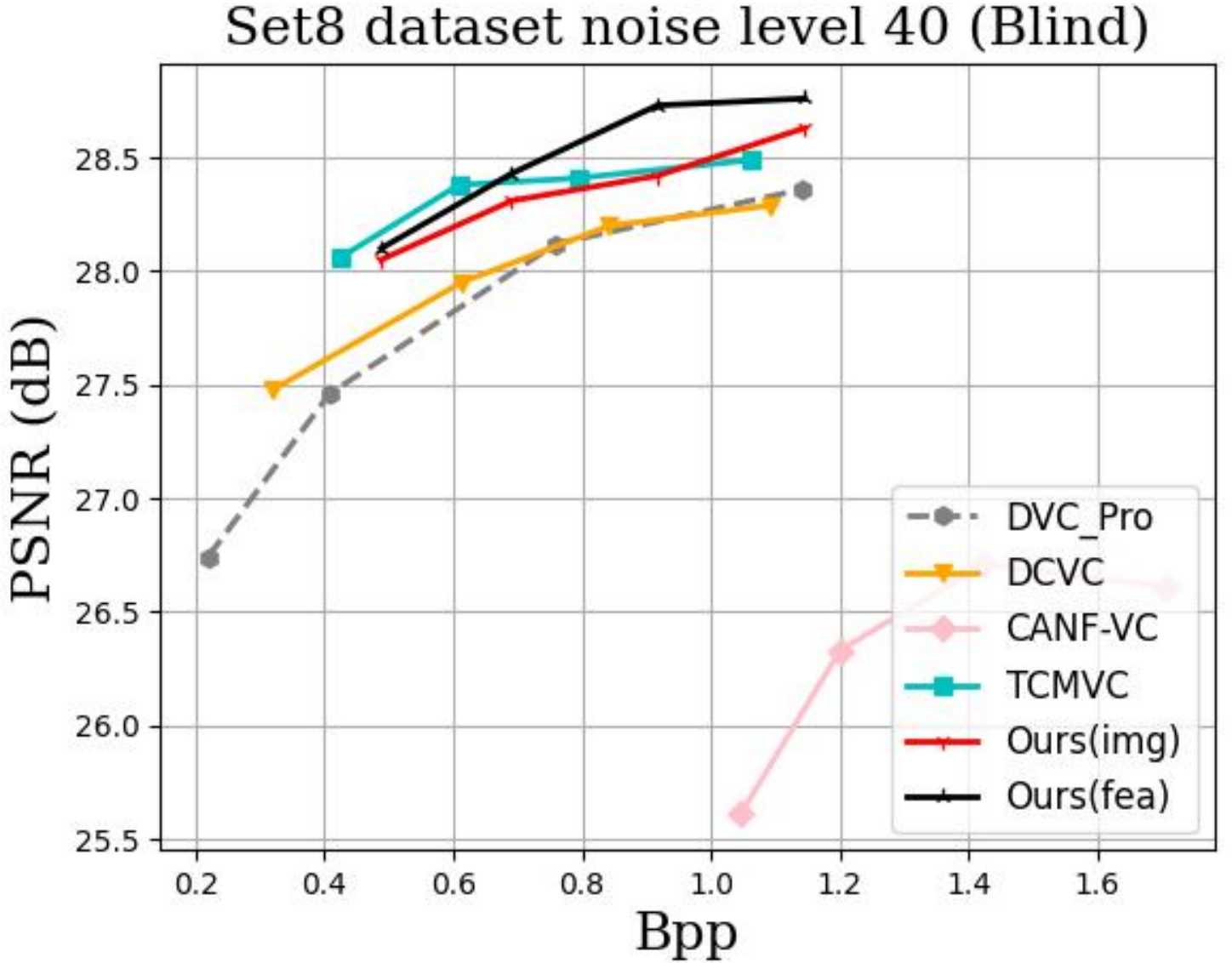}
  \end{minipage}%
    \caption{Rate-performance (PSNR between denoised videos and clean videos) curves of non-blind and blind video denoising task on the DAVIS and Set8 dataset. For DAVIS, the average PSNR of noisy uncompressed videos when $\delta$ is set to 20 and 40 are 22.1104 dB and 16.0898 dB, respectively. For Set8, the average PSNR of noisy uncompressed videos when $\delta$ is set to 20 and 40 are also 22.1101 dB and 16.0895 dB, respectively.}
  \label{fig:non_blind_all_codecs_denoising}
\end{figure*}

\section{Experimental Results}\label{sec:experiment}
For video reconstruction, we use rate-distortion curves and BD-rate to compare the video reconstruction performance of different video codecs. For video enhancement and video analysis tasks, we use rate-performance curves and BD-performance to compare the task performances of different video codecs. When generating rate-performance curves, for each sequence in a dataset, we first compress it and calculate its average compression rate (bits per pixel). Then, we use the decoded intermediate features or decoded frames to perform downstream video enhancement or video analysis tasks and obtain task performance evaluation metrics. Then, we calculate the average compression rate and average task performance evaluation metrics of all sequences in the dataset. We regard the average compression rate as the x-axis and the average task performance evaluation metrics as the y-axis to generate the rate-performance curves.

\subsection{Results for Video Reconstruction}
To verify the compression performance of our 
framework, we compare it with traditional 
 video codecs of H.265~\cite{sullivan2012overview} and H.266~\cite{bross2021overview}. Specifically, for H.265,  we compare with its industrial software x265 under \emph{veryslow} preset and official reference software HM-16.20 under \emph{encoder\_lowdelay\_main\_rext} configuration. For H.266, we 
 compare with its official reference software VTM-13.2. Following~\cite{sheng2022temporal}, we use VTM-LDB (\emph{encoder\_lowdelay\_vtm}) and VTM-IPP (one reference frame and flat QP) configurations. Furthermore,  previous state-of-the-art neural video compression schemes are also included for comparison, including DVC\_Pro~\cite{lu2020end}, DCVC~\cite{li2021deep}, CANF-VC~\cite{canfvc}, TCMVC~\cite{sheng2022temporal}. For CANF-VC and TCMVC, we use their models with error propagation training~\cite{sheng2022temporal}. For all the neural video codecs, we use the unified \emph{cheng2020anchor} model without the auto-regressive entropy model as the I-frame codec. We set the intra period to 12 and 32 and encode 96 frames for each video in all testing datasets. \par
For intra period 12, we illustrate the rate-distortion curves in Fig.~\ref{fig:result_ip12} and list the detailed BD-rate results in Table~\ref{table:ip12_psnr} and  Table~\ref{table:ip12_msssim}. When calculating the BD-rate, we regard VTM-IPP as the anchor. Experimental results show that our proposed framework achieves an average 7.0\% BD-rate reduction against VTM-IPP in terms of PSNR and achieves about an average 49.9\% BD-rate reduction in terms of MS-SSIM. For intra period 32, we illustrate the RD-curves in Fig.~\ref{fig:result_ip32} and list the BD-rate results in Table~\ref{table:ip32_psnr} and Table~\ref{table:ip32_msssim}. Experimental results show that the compression performance of most neural video codecs is greatly reduced while our framework still achieves an average 4.6\% BD-rate reduction against VTM-IPP in terms of PSNR and achieves about 50.3\% BD-rate reduction in terms of MS-SSIM. \par
\begin{table}[t]
\caption{BD-performmance for non-blind video denoising. The performance is measured by PSNR (dB). The anchor is Ours(fea). Negative values indicate performance loss while positive values indicate higher performance.}
  \centering
\scalebox{1}{
\begin{threeparttable}
\begin{tabular}{l|c|c|c|c}
\toprule[1.5pt]
\multicolumn{1}{l|}{\multirow{2}{*}{\diagbox{Schemes}{Datasets}}} & \multicolumn{2}{c|}{DAVS}                         & \multicolumn{2}{c}{Set8}      \\ \cline{2-5} 
\multicolumn{1}{l|}{}    & \multicolumn{1}{c|}{$\delta$=20} & \multicolumn{1}{c|}{$\delta$=40} & \multicolumn{1}{c|}{$\delta$=20} & $\delta$=40  \\ \hline
Ours(fea)  &0.00      &0.00  &0.00 &0.00\\ \hline
Ours(img)  &0.03      &--0.16   &0.19  &-0.03\\ \hline
TCMVC    &--0.02      &--0.07   &0.03  &--0.01\\\hline
CANF-VC    &--3.03      &--2.44   &--2.69  &--2.74\\\hline
DCVC      &--0.62      &--0.34   &--0.39  &--0.31\\ \hline
DVC\_Pro   &--1.01      &--0.43   &--0.74  &--0.38\\ \bottomrule[1.5pt]
\end{tabular}
\end{threeparttable}}
\label{table:non_blindvd}
\end{table}
\begin{table}[t]
\caption{BD-performmance for blind video denoising. The anchor is Ours(fea). Negative values indicate performance loss while positive values indicate higher performance.}
  \centering
\scalebox{1}{
\begin{threeparttable}
\begin{tabular}{l|c|c|c|c}
\toprule[1.5pt]
\multicolumn{1}{l|}{\multirow{2}{*}{\diagbox{Schemes}{Datasets}}} & \multicolumn{2}{c|}{DAVS}                         & \multicolumn{2}{c}{Set8}      \\ \cline{2-5} 
\multicolumn{1}{l|}{}    & \multicolumn{1}{c|}{$\delta$=20} & \multicolumn{1}{c|}{$\delta$=40} & \multicolumn{1}{c|}{$\delta$=20} & $\delta$=40  \\ \hline
Ours(fea)  &0.00      &0.00  &0.00 &0.00\\ \hline
Ours(img)  &--0.11      &--0.27   &0.00  &-0.16\\ \hline
TCMVC    &--0.17      &--0.16   &--0.08  &--0.08\\\hline
CANF-VC    &--3.41 &--2.96      &--2.91   &--2.89 \\\hline
DCVC      &--0.66 &--0.50      &--0.46   &--0.41\\ \hline
DVC\_Pro   &--0.96      &--0.53   &--0.67  &--0.42\\ \bottomrule[1.5pt]
\end{tabular}
\end{threeparttable}}
\label{table:blind_vd}
\end{table}
 \begin{table}[t]
 \caption{Average decoding time (entropy decoding time is excluded), task time, decoding GFLOPs, and task GFLOPs for each frame of the DAVIS and Set8 datasets.}
   \centering
\scalebox{1}{
\begin{threeparttable}
\begin{tabular}{lcccc}
\toprule[1.5pt]
\multicolumn{1}{l|}{Schemes}          & \multicolumn{1}{c|}{Dec T (s)} & \multicolumn{1}{c|}{Task T (s)} & \multicolumn{1}{c|}{Dec G} & Task G \\ \hline
\multicolumn{5}{c}{DAVIS}                                                                                                                                \\ \hline
\multicolumn{1}{l|}{Ours(fea)} & \multicolumn{1}{c|}{0.013}             & \multicolumn{1}{c|}{0.008}         & \multicolumn{1}{c|}{1034.17}         & 529.07      \\ \hline
\multicolumn{1}{c|}{Ours(img)} & \multicolumn{1}{c|}{0.021}             & \multicolumn{1}{c|}{0.007}         & \multicolumn{1}{c|}{1482.97}         & 464.59      \\ \hline
\multicolumn{5}{c}{Set8}                                                                                                                                         \\ \hline
\multicolumn{1}{l|}{Ours(fea)} & \multicolumn{1}{c|}{0.014}             & \multicolumn{1}{c|}{0.011}         & \multicolumn{1}{c|}{1246.55}         & 637.72      \\ \hline
\multicolumn{1}{l|}{Ours(img)} & \multicolumn{1}{c|}{0.022}             & \multicolumn{1}{c|}{0.010}         & \multicolumn{1}{c|}{1787.51}         & 560.00      \\ 
\bottomrule[1.5pt]
\end{tabular}
  \begin{tablenotes}
   \item \footnotesize \dag To show the decoding complexity of the video reconstruction network is large, we only compare the network inference time of the decoder and task network. i.e., the entropy decoding time is excluded. If included, the decoding time is longer. The same setting is used for other vision tasks.
  \end{tablenotes}
\end{threeparttable}}
\label{table:denoising_time}
\end{table}
 
 \begin{figure*}[t]
  \centering
  \begin{minipage}[c]{0.22\linewidth}
  \centering
    \includegraphics[width=\linewidth]{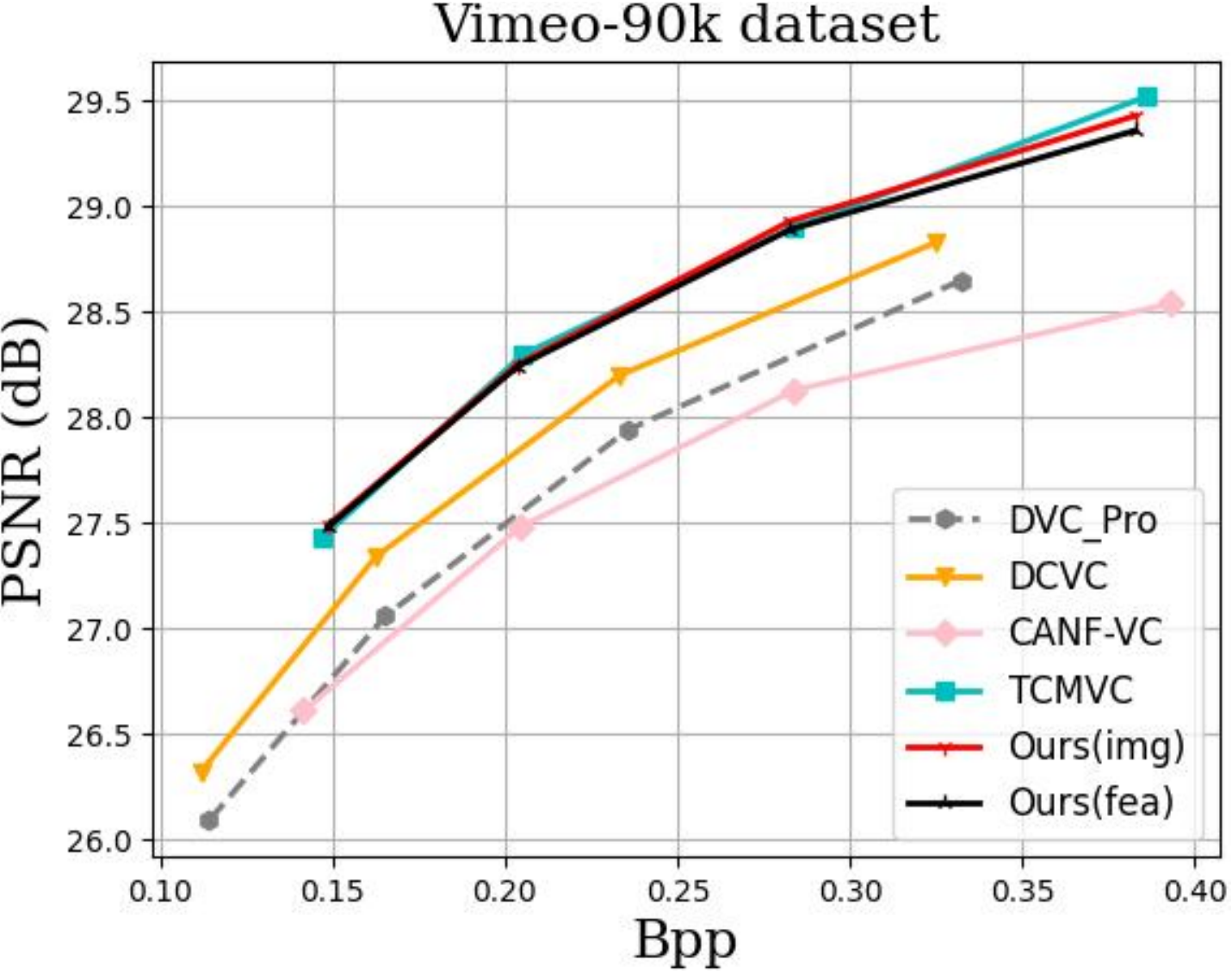}
  \end{minipage}%
  \begin{minipage}[c]{0.22\linewidth}
  \centering
    \includegraphics[width=\linewidth]{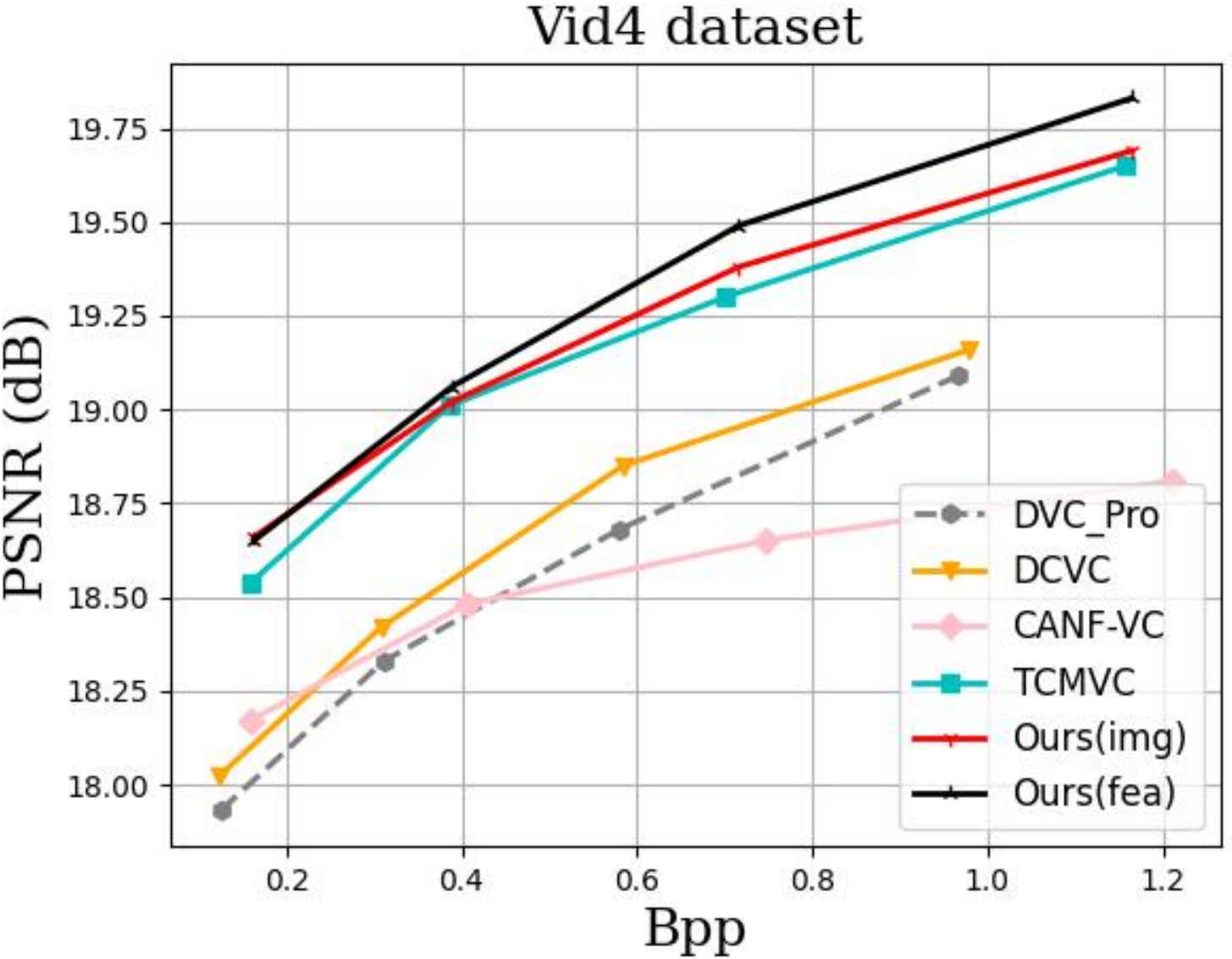}
  \end{minipage}
  \begin{minipage}[c]{0.22\linewidth}
  \centering
    \includegraphics[width=\linewidth]{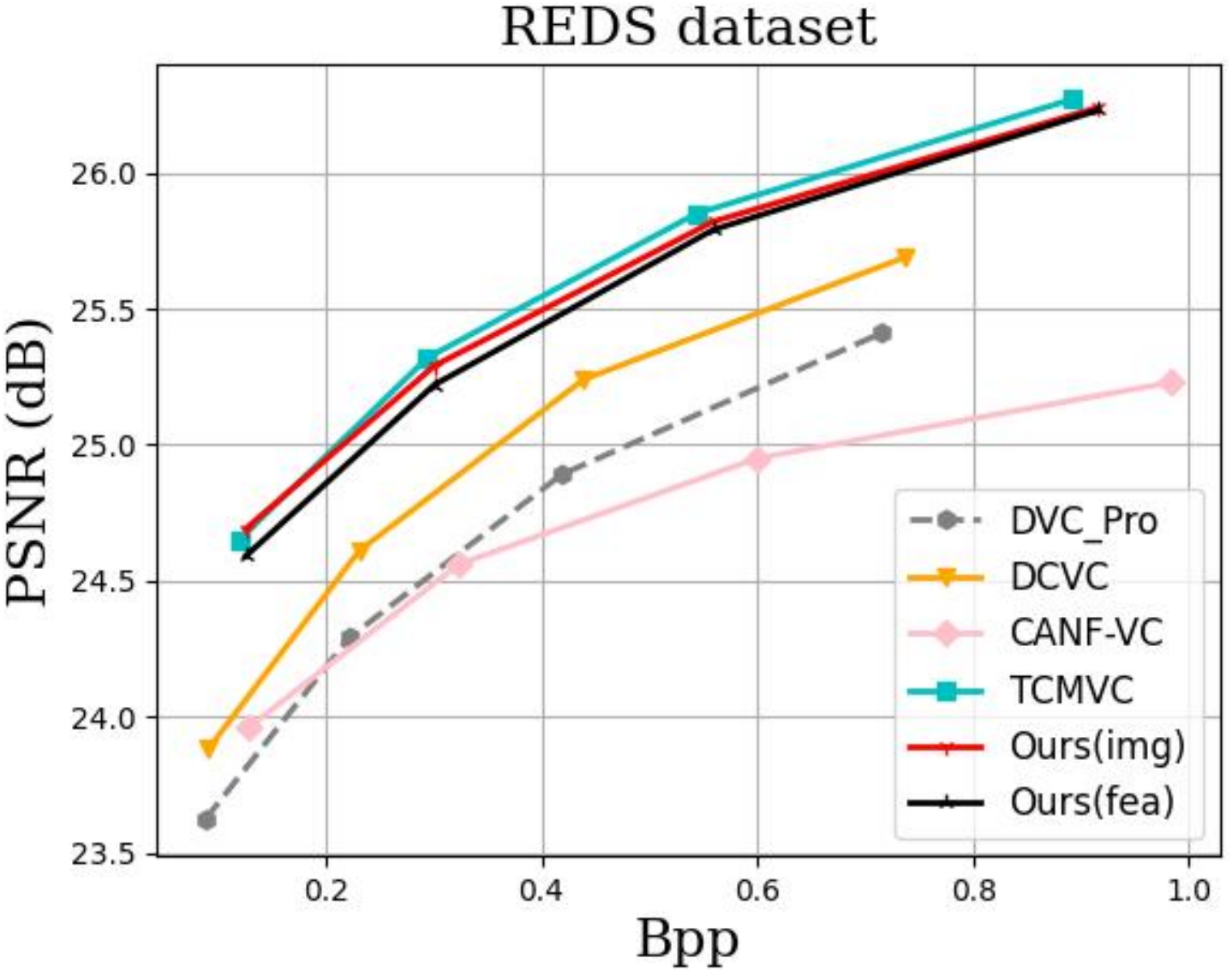}
 \end{minipage}
  \begin{minipage}[c]{0.22\linewidth}
  \centering
    \includegraphics[width=\linewidth]{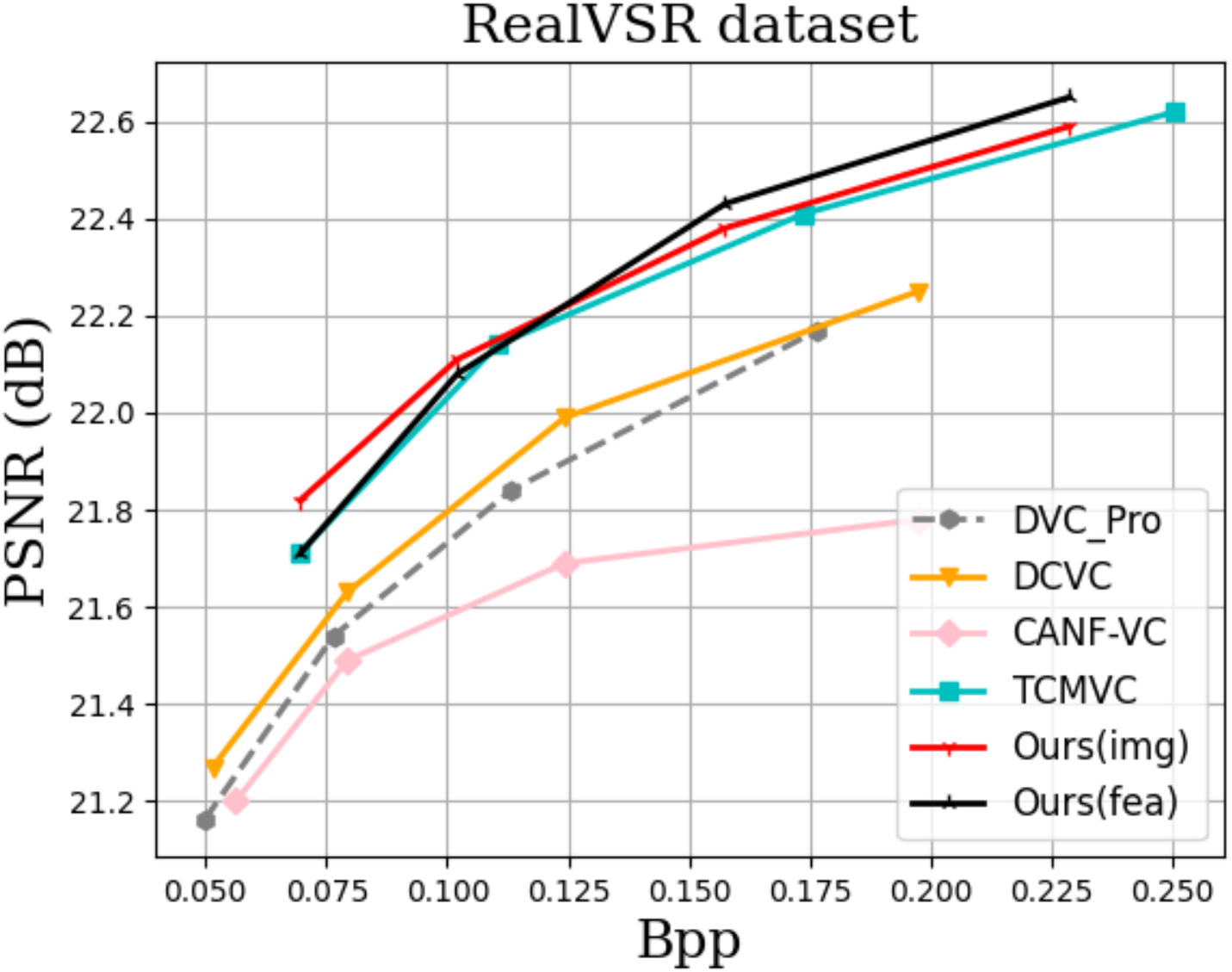}
  \end{minipage}%
    \caption{Rate-performance (PSNR between the output super-resolved videos and the ground truth high-resolution videos) curves of BI-degradation and real-world video super-resolution tasks on the Vimeo-90k, Vid4, REDS4, and RealVSR datasets.}
  \label{fig:all_codecs_sr}
\end{figure*}

 We also list the average encoding and decoding time for one 1080p frame of different neural video codecs in Table~\ref{time}. Following~\cite{sheng2022temporal}, the time for entropy encoding/decoding time is included. The comparison results show that the encoding time of our proposed framework is the shortest since its encoder removes the separate motion estimation network and does not need to obtain the full reconstructed videos in the pixel domain. The decoding time of our framework is longer than TCMVC since it uses a more complex video reconstruction network on the decoder side for higher compression efficiency. However, when oriented to video enhancement and analysis tasks, the video reconstruction network can be removed. 

\subsection{Results for Video Enhancement}
To explore the effectiveness of using decoded intermediate features to perform video enhancement,  we design two contrasting models for comparison, i.e., Ours(fea) and Ours(img). For Ours(fea), we feed the partially decoded intermediate features into the video enhancement networks shown in Fig.~\ref{fig:denoise_sr} to perform video enhancement directly. For Ours(img), we use the fully decoded pixel-domain frames to perform video enhancement. We change the input of the video enhancement shown in Fig.~\ref{fig:denoise_sr} from $\hat{F}_{t}$ to $\hat{x}_t$. To exploit the temporal correlation, we also change the other input, the temporal context $\bar{C}_{t}^{0}$, to the pixel-domain predicted frame $\tilde{x}_t$ described in Eq.(\ref{equ:mv_RD}). The detailed architectural difference between Ours(fea) and Ours(img) can be found in the supplemental material. We further compare with DVC\_Pro, DCVC, CANF-VC, and TCMVC, which also use the fully decoded pixel-domain frames to perform enhancement using the networks with the same architecture as Ours(img). \par
\begin{table}[t]
\caption{BD-performmance for BI-degradation and real-world video super-resolution. The anchor is Ours(fea). Negative values indicate performance loss while positive values indicate higher performance.}
  \centering
\scalebox{1}{
\begin{threeparttable}
\begin{tabular}{l|c|c|c|c}
\toprule[1.5pt]
\multicolumn{1}{l|}{\multirow{2}{*}{\diagbox{Schemes}{Datasets}}} & \multicolumn{1}{l|}{\multirow{2}{*}{Viemo}} & \multicolumn{1}{l|}{\multirow{2}{*}{Vid4}} & \multicolumn{1}{l|}{\multirow{2}{*}{REDS4}} & \multirow{2}{*}{RealVSR} \\
\multicolumn{1}{l|}{}                         & \multicolumn{1}{l|}{}                       & \multicolumn{1}{l|}{}                      & \multicolumn{1}{l|}{}                      &                       \\ \hline
Ours(fea)  &0.00      &0.00  &0.00 &0.00\\ \hline
Ours(img)  &0.02      &--0.6   &0.06  &0.00\\ \hline
TCMVC    &0.03      &--0.09   &0.10 &--0.04\\\hline
CANF-VC    &--0.78 &--0.67      &--0.77   &--0.54 \\\hline
DCVC      &--0.34 &--0.51      &--0.37   &--0.26\\ \hline
DVC\_Pro   &--0.62      &--0.62   &--0.66  &--0.32\\ \bottomrule[1.5pt]
\end{tabular}
\end{threeparttable}}
\label{table:sr}
\end{table}
\subsubsection{Results for Blind/Non-Blind Video Denoising}
For blind and non-blind video denoising, the standard deviation $\delta$ of white Gaussian noise of testing noisy videos are both set to 20 and 40. We illustrate the rate-performance curves on DAVIS and Set8 datasets in Fig.~\ref{fig:non_blind_all_codecs_denoising} and list the detailed BD-performance in Table~\ref{table:non_blindvd} and Table~\ref{table:blind_vd}. We also present the subjective comparison in the supplemental material.
The experimental results show that Ours(fea) achieves an average better denoising performance than Ours(img) when the bit rate is larger. It may be because, for a higher bit rate, the video reconstruction network focuses on reconstructing noisy frames for higher signal fidelity. Therefore, the fully decoded frames contain more noise. However, the partially decoded intermediate feature before the video reconstruction network of Ours(fea) has more feature channels than the fully decoded frame, which leads to the intermediate feature retaining more information on video content to restore clean videos. However, for a lower bit rate and smaller noise level, the performance gain of Ours(fea) is limited and even worse than that of Ours(img). This may be owing to that when the bit rate is small, the fully decoded frame already contains enough information to restore clean videos and its video reconstruction network and video denoising network form a two-stage denoiser~\cite{huang2022neural}. \par
To demonstrate that using the decoded intermediate feature to perform video denoising is more efficient in terms of computational complexity, we compare the decoding time, denoising task time, decoding GFLOPs, task GFLOPs of Ours(fea) and Ours(img) in Table~\ref{table:denoising_time}. The comparison results show that although feeding the intermediate feature to the denoising network leads to a complexity increase, it can reduce large decoding complexity. Taking the DAVIS dataset as an example, the runtime (0.021s-0.013s) of converting the intermediate feature to pixel domain is even more than that of video denoising. \par
 
\subsubsection{Results for BI-Degradation/Real-World Video SR}
For BI-degradation/real-world video SR, we illustrate the rate-performance curves on the Vimeo, Vid4, REDS4, and RealVSR datasets in Fig.~\ref{fig:all_codecs_sr} and list the detailed BD-performance in Table~\ref{table:sr}. We also show the subjective comparison in the supplemental material. The comparison results show that Ours(fea) achieves a similar performance compared with Ours(img), which means using decoded intermediate features or decoded pixel-domain frames to perform video SR does not have obvious performance differences. However, in terms of computational complexity, as listed in Table~\ref{table:sr_time}, Ours(fea) is more efficient for video SR since it can save the large complexity of converting the partially decoded intermediate features to fully decoded pixel-domain frames. The runtime used to convert the intermediate feature to the fully decoded frame is even more than that used for video SR. \par
  \begin{table}[t]
 \caption{Average decoding time (entropy decoding time is excluded), task time, decoding GFLOPs, and task GFLOPs for each frame of Viemo, Vid4, REDS4, and RealVSR datasets.}
   \centering
\scalebox{1}{
\begin{threeparttable}
\begin{tabular}{lcccc}
\toprule[1.5pt]
\multicolumn{1}{l|}{Schemes}          & \multicolumn{1}{c|}{Dec T (s)} & \multicolumn{1}{c|}{Task T (s)} & \multicolumn{1}{c|}{Dec G} & Task G \\ \hline
\multicolumn{5}{c}{Viemo}                                                                                                                                \\ \hline
\multicolumn{1}{l|}{Ours(fea)} & \multicolumn{1}{c|}{0.010}             & \multicolumn{1}{c|}{0.007}         & \multicolumn{1}{c|}{18.468}         & 24.972      \\ \hline
\multicolumn{1}{c|}{Ours(img)} & \multicolumn{1}{c|}{0.020}             & \multicolumn{1}{c|}{0.006}         & \multicolumn{1}{c|}{26.482}         & 23.82      \\ \hline
\multicolumn{5}{c}{Vid4}                                                                                                                                         \\ \hline
\multicolumn{1}{l|}{Ours(fea)} & \multicolumn{1}{c|}{0.011}             & \multicolumn{1}{c|}{0.008}         & \multicolumn{1}{c|}{83.10}         & 112.37      \\ \hline
\multicolumn{1}{l|}{Ours(img)} & \multicolumn{1}{c|}{0.021}             & \multicolumn{1}{c|}{0.007}         & \multicolumn{1}{c|}{119.17}         & 107.19      \\\hline 
\multicolumn{5}{c}{REDS}                                                                                                                                \\ \hline
\multicolumn{1}{l|}{Ours(fea)} & \multicolumn{1}{c|}{0.012}             & \multicolumn{1}{c|}{0.008}         & \multicolumn{1}{c|}{138.51}         & 187.29      \\ \hline
\multicolumn{1}{c|}{Ours(img)} & \multicolumn{1}{c|}{0.022}             & \multicolumn{1}{c|}{0.007}         & \multicolumn{1}{c|}{198.61}         & 179.86      \\ \hline
\multicolumn{5}{c}{RealVSR}                                                                                                                                \\ \hline
\multicolumn{1}{l|}{Ours(fea)} & \multicolumn{1}{c|}{0.011}             & \multicolumn{1}{c|}{0.008}         & \multicolumn{1}{c|}{73.87}         & 99.89      \\ \hline
\multicolumn{1}{c|}{Ours(img)} & \multicolumn{1}{c|}{0.021}             & \multicolumn{1}{c|}{0.007}         & \multicolumn{1}{c|}{105.93}         & 95.28      \\
\bottomrule[1.5pt]
\end{tabular}
\end{threeparttable}}
\label{table:sr_time}
\end{table}
\subsection{Results for Video Analysis}
In addition to human vision tasks,  we further explore the effectiveness of using decoded intermediate features to perform machine vision tasks. Similarly, we design two models for comparison, i.e., Ours(fea) and Ours(img). For Ours(fea), we directly feed the decoded intermediate features into the video analysis models that allow features as inputs, as shown in Fig.~\ref{fig:action_recognition}, Fig.~\ref{fig:VID}, Fig.~\ref{fig:MOT}, and Fig.~\ref{fig:VOS}. For Ours(img), we use the decoded frames to perform video analysis using the models that allow pixel-domain frames as inputs. The difference between the video analysis models of Ours(fea) and Ours(img) can be found in Section~\ref{task_network}. We further compare with DVC\_Pro, DCVC, CANF-VC, and TCMVC, which use decoded frames to perform video analysis using the networks with the same architecture as Ours(img).
\begin{figure}[t]
  \centering
  \begin{minipage}[c]{0.48\linewidth}
  \centering
    \includegraphics[width=\linewidth]{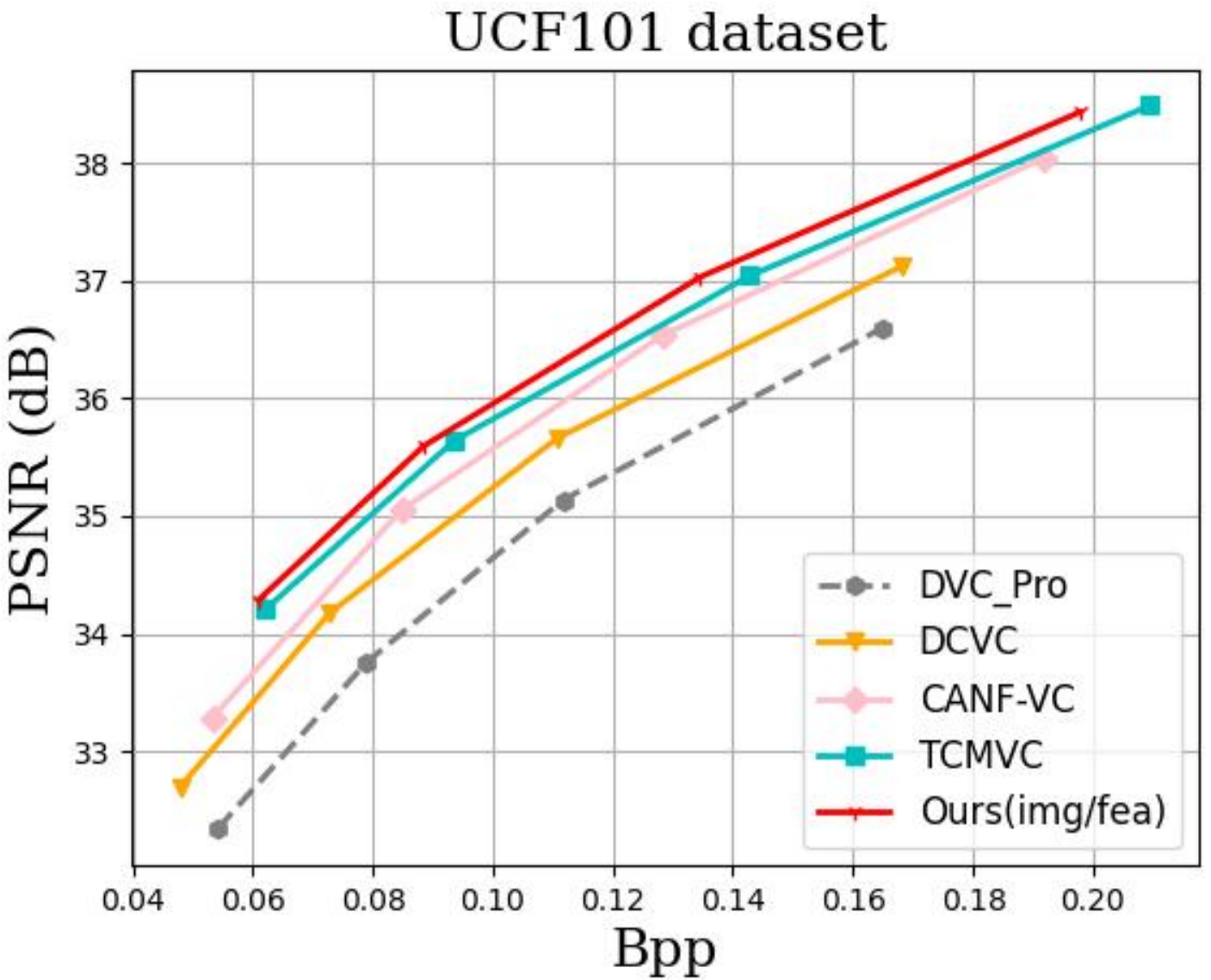}
 \end{minipage}%
  \begin{minipage}[c]{0.48\linewidth}
  \centering
    \includegraphics[width=\linewidth]{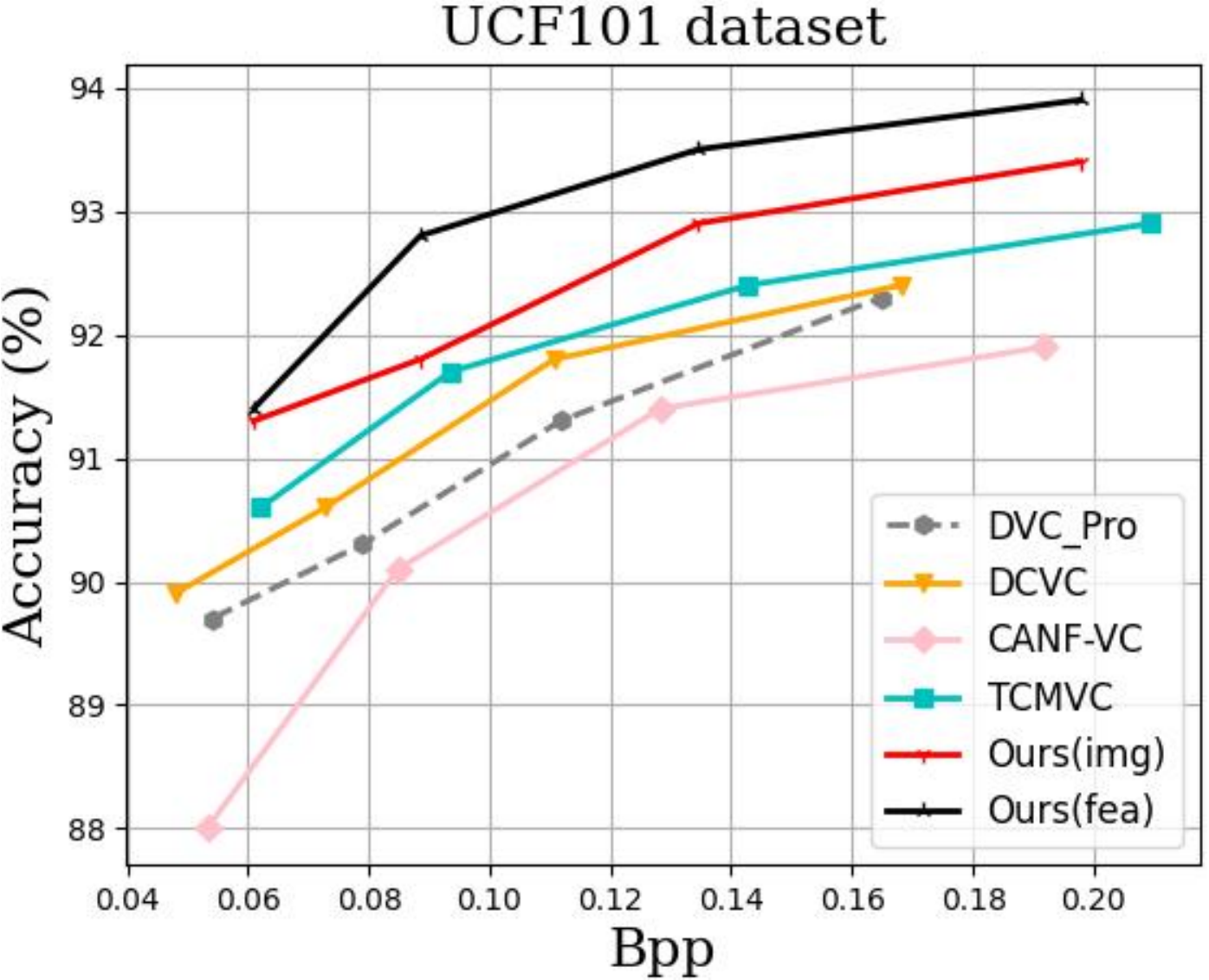}
  \end{minipage}\\
  \begin{minipage}[c]{0.48\linewidth}
  \centering
    \includegraphics[width=\linewidth]{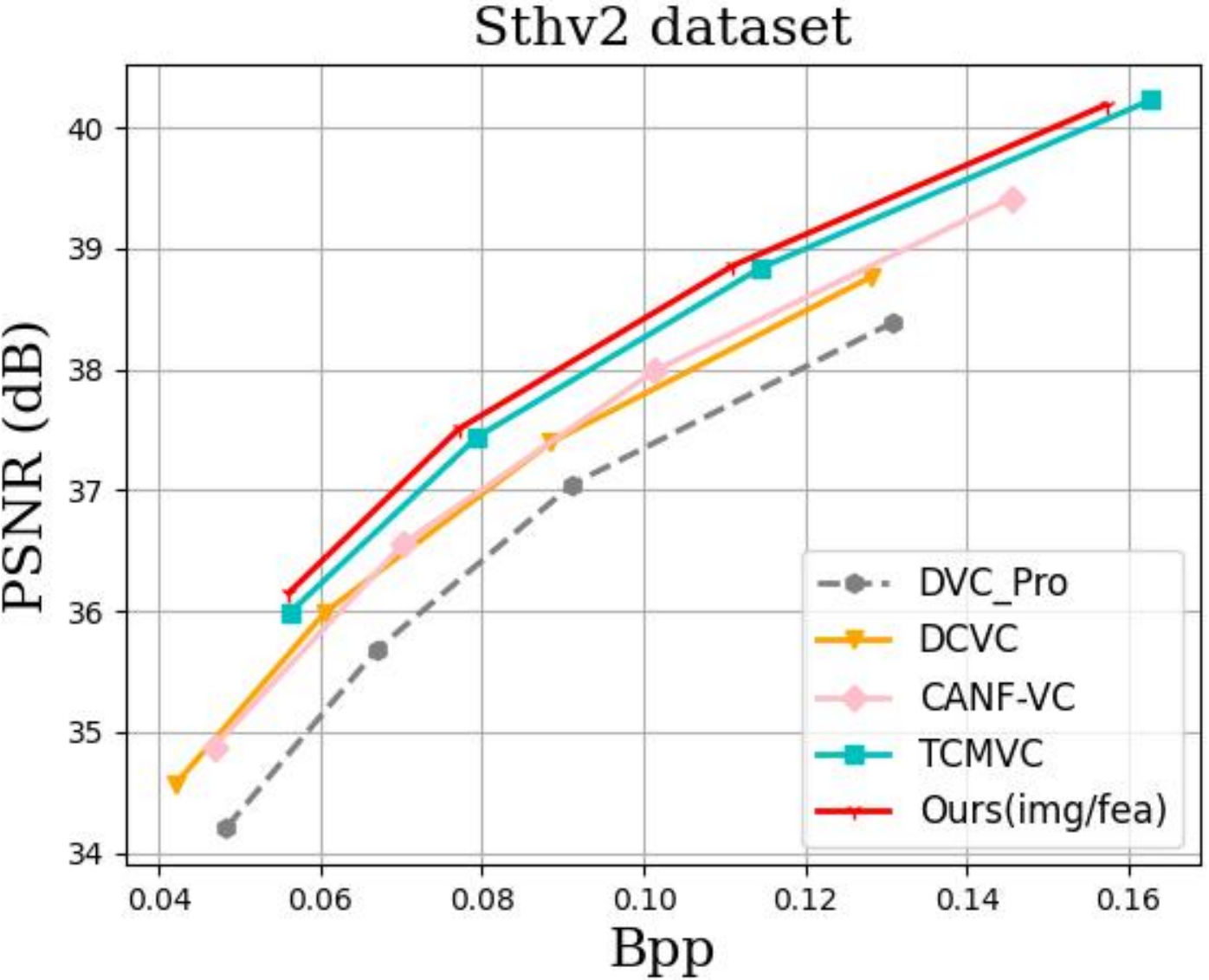}
 \end{minipage}%
  \begin{minipage}[c]{0.48\linewidth}
  \centering
    \includegraphics[width=\linewidth]{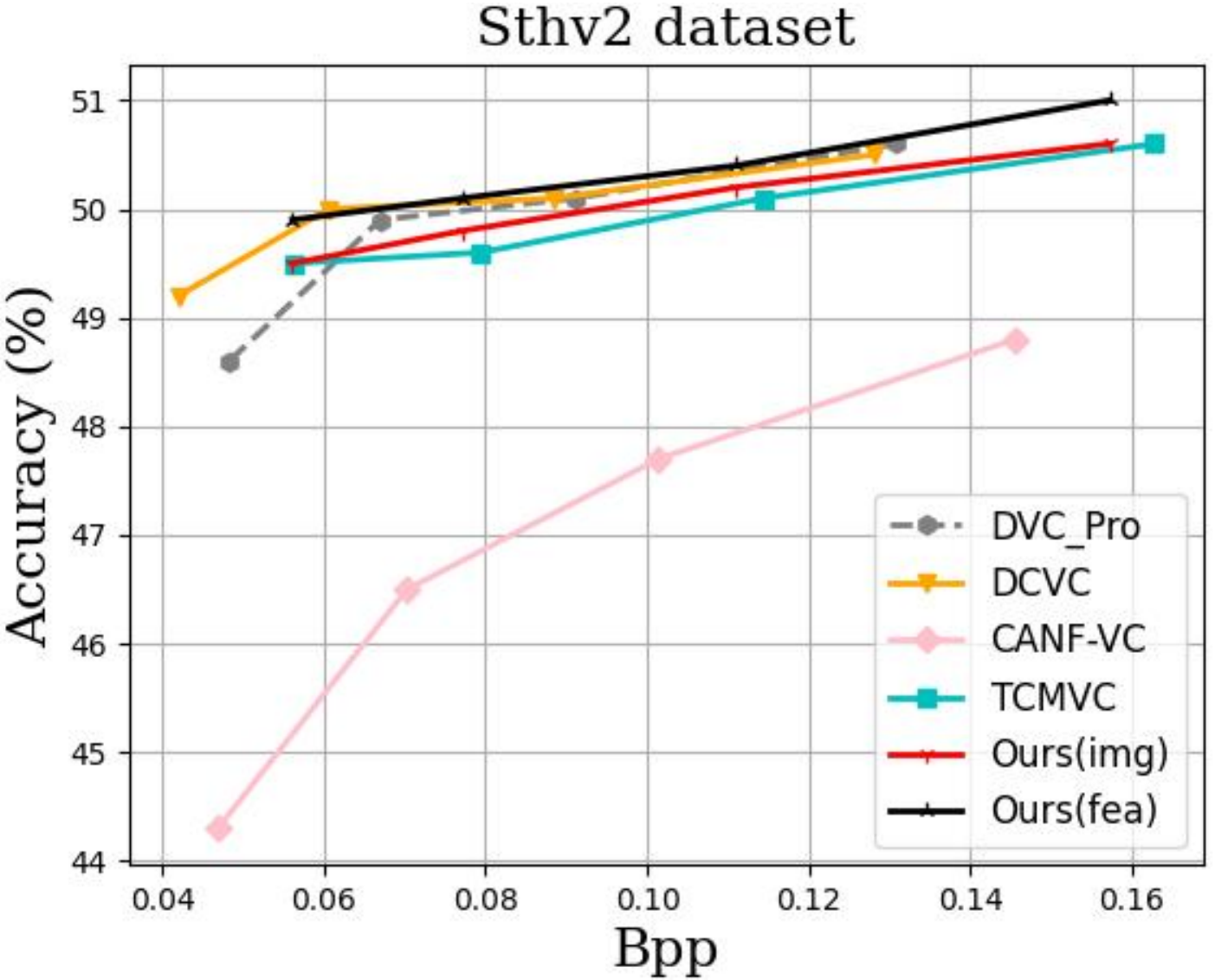}
  \end{minipage}%
    \caption{Rate-distortion curves and rate-performance (video action recognition top1-accuracy) curves on the UCF101 and Something-something-V2 datasets.}
  \label{fig:all_codecs_ar}
\end{figure}
\begin{table}[!t]
\caption{BD-performmance for video action recognition. The anchor is Ours(fea). Negative values indicate performance loss while positive values indicate higher performance.}
  \centering
\scalebox{1}{
\begin{threeparttable}
\begin{tabular}{l|c|c}
\toprule[1.5pt]
\multicolumn{1}{l|}{\multirow{2}{*}{\diagbox{Schemes}{Datasets}}} & \multicolumn{1}{l|}{\multirow{2}{*}{UCF101}} & \multirow{2}{*}{Sthv2} \\
\multicolumn{1}{l|}{}                         & \multicolumn{1}{l|}{}                       &                     \\ \hline
Ours(fea)  &0.00      &0.00 \\ \hline
Ours(img)  &--0.68      &--0.31\\ \hline
TCMVC    &--1.15      &--0.46\\\hline
CANF-VC    &--2.29 &--3.01 \\\hline
DCVC      &--1.45 &--0.14\\ \hline
DVC\_Pro   &--1.90      &--0.21\\ \bottomrule[1.5pt]
\end{tabular}
\end{threeparttable}}
\label{table:ar}
\end{table}
 \begin{table}[!t]
 \caption{Average decoding time (entropy decoding time is excluded), task time, decoding GFLOPs, and task GFLOPs for sampled frames in each video of the UCF101 and Something-something-V2 datasets.}
   \centering
\scalebox{1}{
\begin{threeparttable}
\begin{tabular}{lcccc}
\toprule[1.5pt]
\multicolumn{1}{l|}{Schemes}          & \multicolumn{1}{c|}{Dec T (s)} & \multicolumn{1}{c|}{Task T (s)} & \multicolumn{1}{c|}{Dec G} & Task G \\ \hline
\multicolumn{5}{c}{UCF101}                                                                                                                                \\ \hline
\multicolumn{1}{l|}{Ours(fea)} & \multicolumn{1}{c|}{0.072}             & \multicolumn{1}{c|}{0.010}         & \multicolumn{1}{c|}{1181.92}         & 68.12      \\ \hline
\multicolumn{1}{c|}{Ours(img)} & \multicolumn{1}{c|}{0.112}             & \multicolumn{1}{c|}{0.009}         & \multicolumn{1}{c|}{1694.83}         & 43.05      \\ \hline
\multicolumn{5}{c}{Sthv2}                                                                                                                                         \\ \hline
\multicolumn{1}{l|}{Ours(fea)} & \multicolumn{1}{c|}{0.432}             & \multicolumn{1}{c|}{0.010}         & \multicolumn{1}{c|}{7091.52}         & 408.75      \\ \hline
\multicolumn{1}{l|}{Ours(img)} & \multicolumn{1}{c|}{0.72}             & \multicolumn{1}{c|}{0.009}         & \multicolumn{1}{c|}{10168.99}         & 258.31      \\ 
\bottomrule[1.5pt]
\end{tabular}
  \begin{tablenotes}
   \item \footnotesize \dag For the UCF101 dataset, 8 frames are sampled. For the Sthv2 dataset, 48 frames are sampled.
   \item \footnotesize \dag\dag Sample frames are stacked along the batch size dimension for parallel recognition.
  \end{tablenotes}
\end{threeparttable}}
\label{table:ar_time}
\end{table}
\subsubsection{Results for Video Action Recognition}
For video action recognition, we illustrate the rate-performance (video action recognition top1-accuracy) curves on the UCF101 and SthV2 datasets in Fig.~\ref{fig:all_codecs_ar}. We also list the detailed BD-performance in Table~\ref{table:ar}. Experimental results show that, for video action recognition,  Ours(fea) achieves higher action recognition accuracy compared with Ours(img), which shows that the intermediate feature is more effective for the TSM model. It may be because if the intermediate feature is converted into pixel domain using the video reconstruction network, some information that is useful for action recognition is lost.  When humans need to further analyze the videos recognized by machines, video reconstruction is needed, i.e., converting the intermediate features to pixel-domain decoded frames. Therefore, we also illustrate the rate-distortion curves in Fig.~\ref{fig:all_codecs_ar}. Compared with other video codecs, Ours(fea) achieves the best video reconstruction performance. \par
\begin{figure}[t]
  \centering
  \begin{minipage}[c]{0.48\linewidth}
  \centering
    \includegraphics[width=\linewidth]{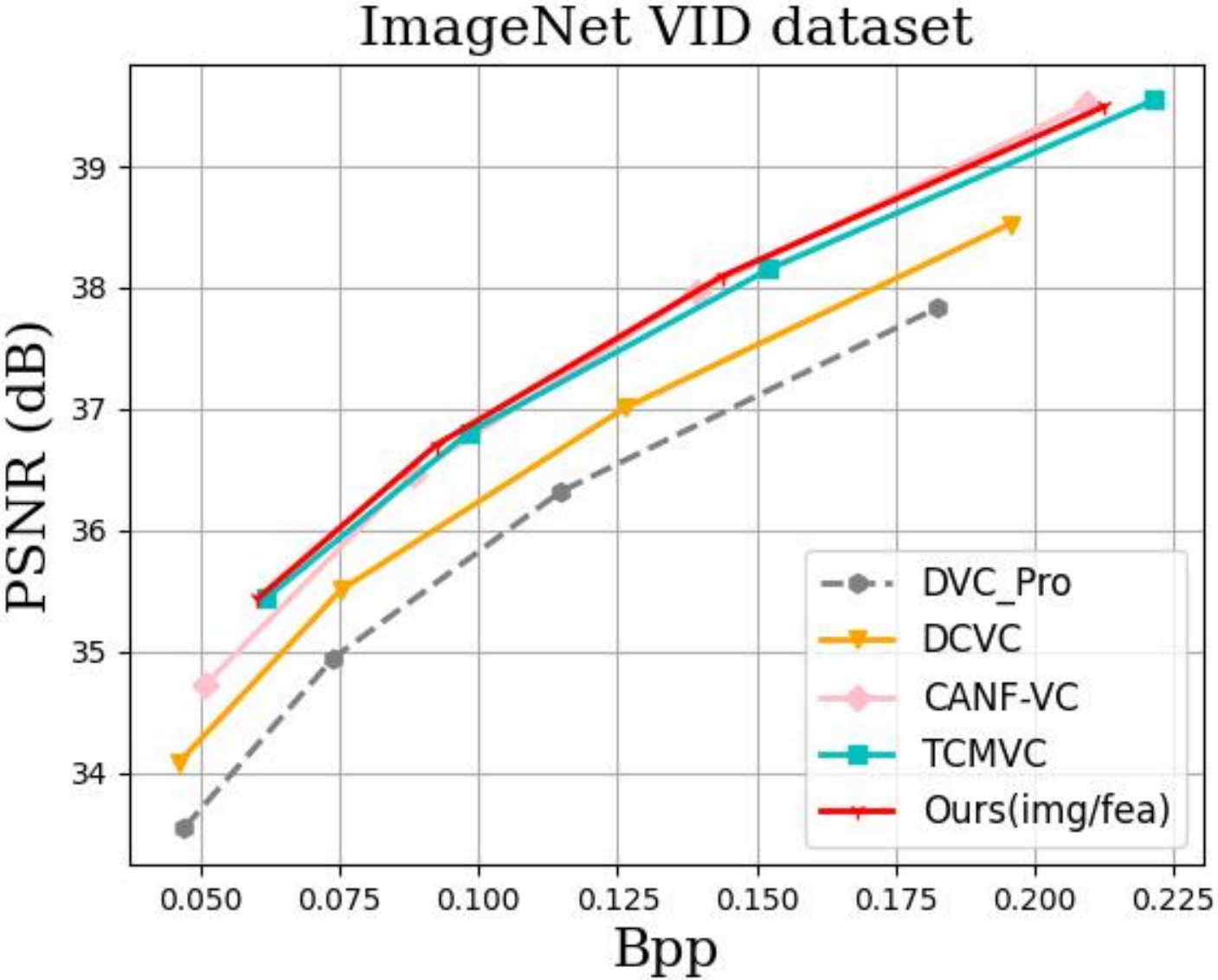}
 \end{minipage}%
  \begin{minipage}[c]{0.48\linewidth}
  \centering
    \includegraphics[width=\linewidth]{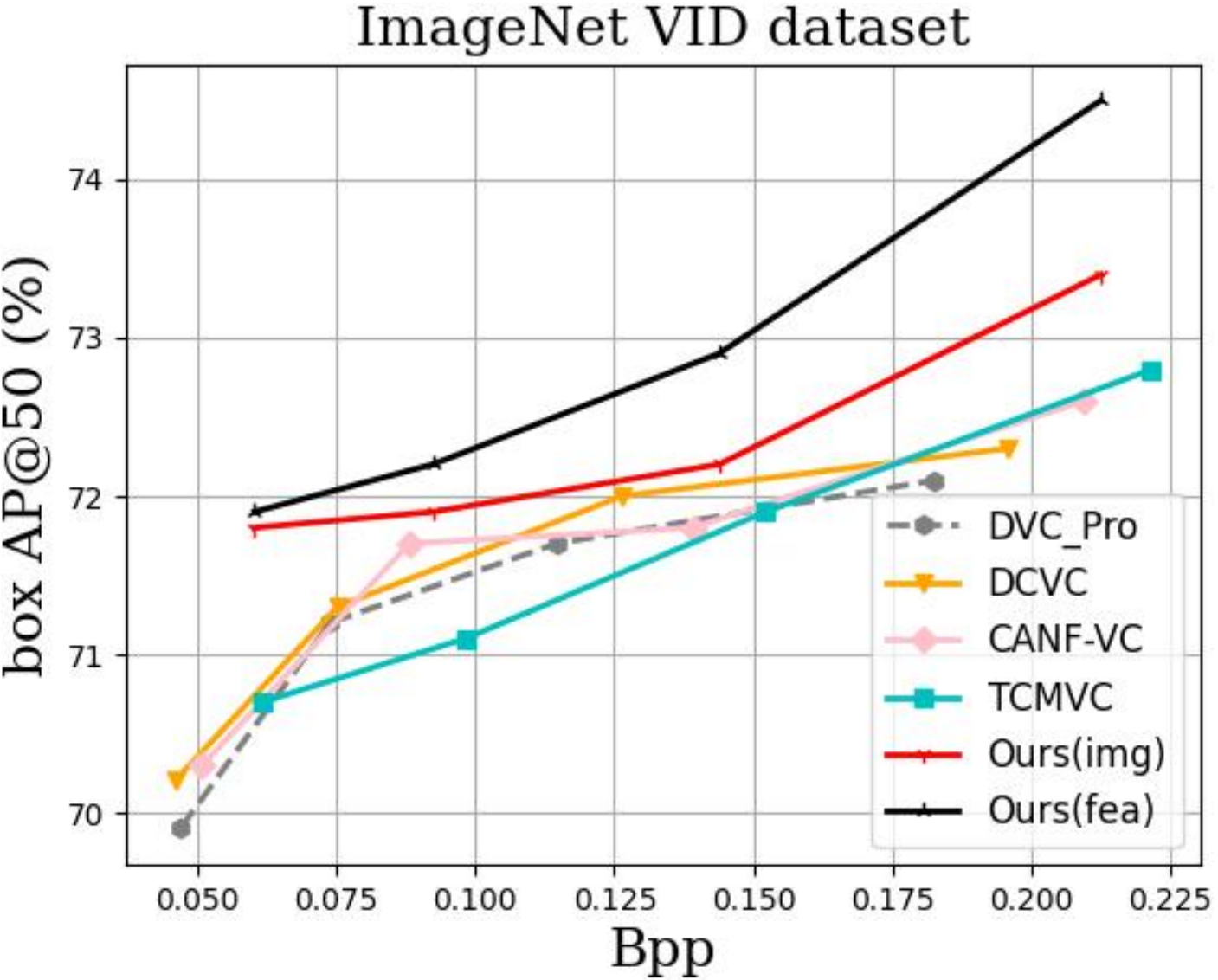}
  \end{minipage}%
    \caption{Rate-distortion and rate-performance (video object detection box AP@50) curves on the ImageNet VID dataset.}
  \label{fig:all_codecs_vid}
\end{figure}
\begin{table}[t]
\caption{BD-performmance for video object detection on the ImageNet VID dataset. The anchor is Ours(fea). Negative values indicate performance loss while positive values indicate higher performance.}
  \centering
\scalebox{1}{
\begin{threeparttable}
\begin{tabular}{l|c}
\toprule[1.5pt]

Ours(fea)  &0.00      \\ \hline
Ours(img)  &--0.51      \\ \hline
TCMVC    &--1.24     \\\hline
CANF-VC    &--0.89  \\\hline
DCVC      &--0.81\\ \hline
DVC\_Pro   &--0.92  \\ \bottomrule[1.5pt]
\end{tabular}
\end{threeparttable}}
\label{table:VID}
\end{table}
 \begin{table}[!t]
 \caption{Average decoding time (entropy decoding time is excluded), task time, decoding GFLOPs, and task GFLOPs for each frame of the ImageNet VID dataset.}
   \centering
\scalebox{1}{
\begin{threeparttable}
\begin{tabular}{lcccc}
\toprule[1.5pt]
\multicolumn{1}{l|}{Schemes}          & \multicolumn{1}{c|}{Dec T (s)} & \multicolumn{1}{c|}{Task T (s)} & \multicolumn{1}{c|}{Dec G} & Task G \\ \hline
\multicolumn{1}{l|}{Ours(fea)} & \multicolumn{1}{c|}{0.019}             & \multicolumn{1}{c|}{0.080}         & \multicolumn{1}{c|}{1329.65}         & 166.32      \\ \hline
\multicolumn{1}{l|}{Ours(img)} & \multicolumn{1}{c|}{0.034}             & \multicolumn{1}{c|}{0.078}         & \multicolumn{1}{c|}{1906.68}         & 136.35      \\ 
\bottomrule[1.5pt]
\end{tabular}
\end{threeparttable}}
\label{table:vid_time}
\end{table}
To demonstrate that using the decoded intermediate feature to perform video action recognition is more efficient in terms of computational complexity, we list the decoding time, action recognition time, decoding GFLOPs, and action recognition GFLOPs of Ours(fea) and Ours(img) in Table~\ref{table:ar_time}. Most existing action recognition networks need to sample a part of frames from an entire video and then perform action recognition. As listed in Table~\ref{table:ar_time}, decoding all the sampled frames takes much more time than action recognition. Therefore, decreasing the video decoding complexity is necessary for action recognition. Comparing the computational complexity of Ours(fea) and Ours(img), Ours(fea) is more efficient for video action recognition since it can save much decoding complexity than Ours(img).
 \par

\begin{figure}[t]
  \centering
  \begin{minipage}[c]{0.48\linewidth}
  \centering
    \includegraphics[width=\linewidth]{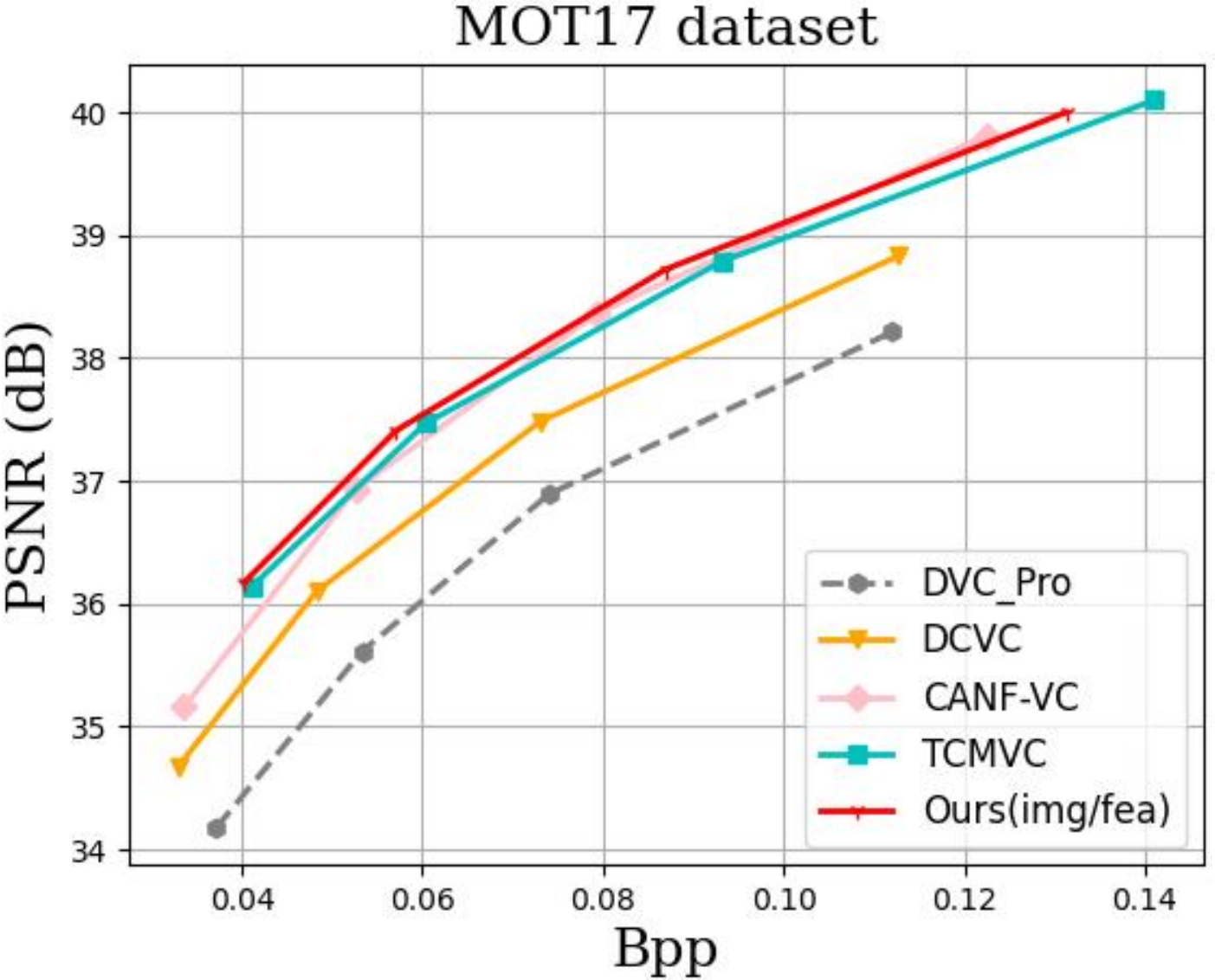}
 \end{minipage}%
  \begin{minipage}[c]{0.48\linewidth}
  \centering
    \includegraphics[width=\linewidth]{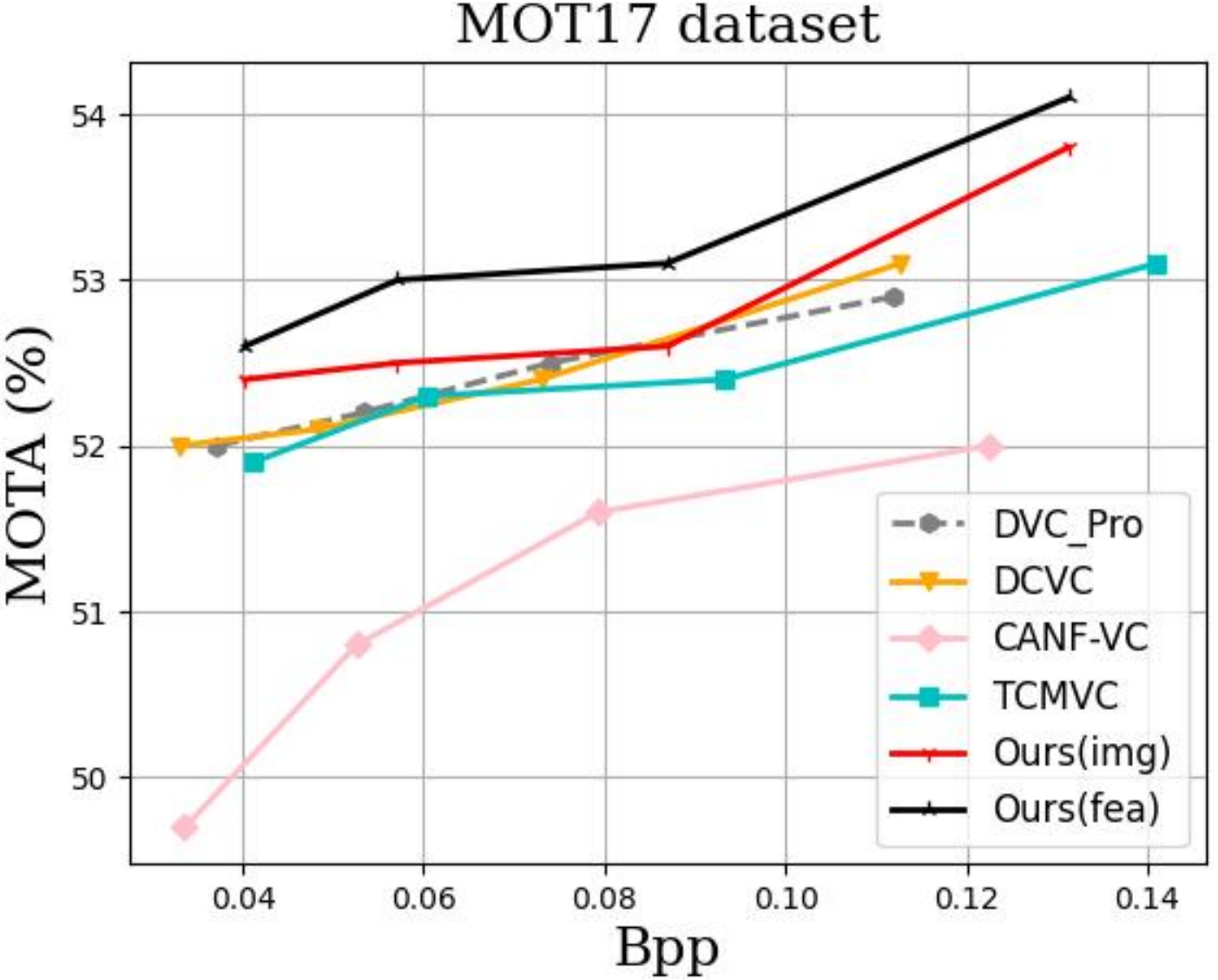}
  \end{minipage}%
    \caption{Rate-distortion and rate-performance (video multiple object tracking MOTA) curves on the MOT17 dataset.}
  \label{fig:all_codecs_mot}
\end{figure}
\begin{table}[t]
\caption{BD-performmance for video multiple object tracking on the MOT dataset. The anchor is Ours(fea). Negative values indicate performance loss while positive values indicate higher performance.}
  \centering
\scalebox{1}{
\begin{threeparttable}
\begin{tabular}{l|c}
\toprule[1.5pt]
Ours(fea)  &0.00      \\ \hline
Ours(img)  &--0.44      \\ \hline
TCMVC    &--0.79     \\\hline
CANF-VC    &--1.79  \\\hline
DCVC      &--0.63\\ \hline
DVC\_Pro   &--0.61  \\ \bottomrule[1.5pt]
\end{tabular}
\end{threeparttable}}
\label{table:MOT}
\end{table}
 \begin{table}[!t]
 \caption{Average decoding time (entropy decoding time is excluded), task time, decoding GFLOPs, and task GFLOPs for each frame of the  MOT17 dataset.}
   \centering
\scalebox{1}{
\begin{threeparttable}
\begin{tabular}{lcccc}
\toprule[1.5pt]
\multicolumn{1}{l|}{Schemes}          & \multicolumn{1}{c|}{Dec T (s)} & \multicolumn{1}{c|}{Task T (s)} & \multicolumn{1}{c|}{Dec G} & Task G \\ \hline
\multicolumn{1}{l|}{Ours(fea)} & \multicolumn{1}{c|}{0.020}             & \multicolumn{1}{c|}{0.065}         & \multicolumn{1}{c|}{2040.64}         & 242.52      \\ \hline
\multicolumn{1}{l|}{Ours(img)} & \multicolumn{1}{c|}{0.036}             & \multicolumn{1}{c|}{0.064}         & \multicolumn{1}{c|}{2926.22}         & 199.23      \\ 
\bottomrule[1.5pt]
\end{tabular}
\label{table:mot_time}
\end{threeparttable}}
\end{table}

\begin{figure}[t]
  \centering
  \begin{minipage}[c]{0.48\linewidth}
  \centering
    \includegraphics[width=\linewidth]{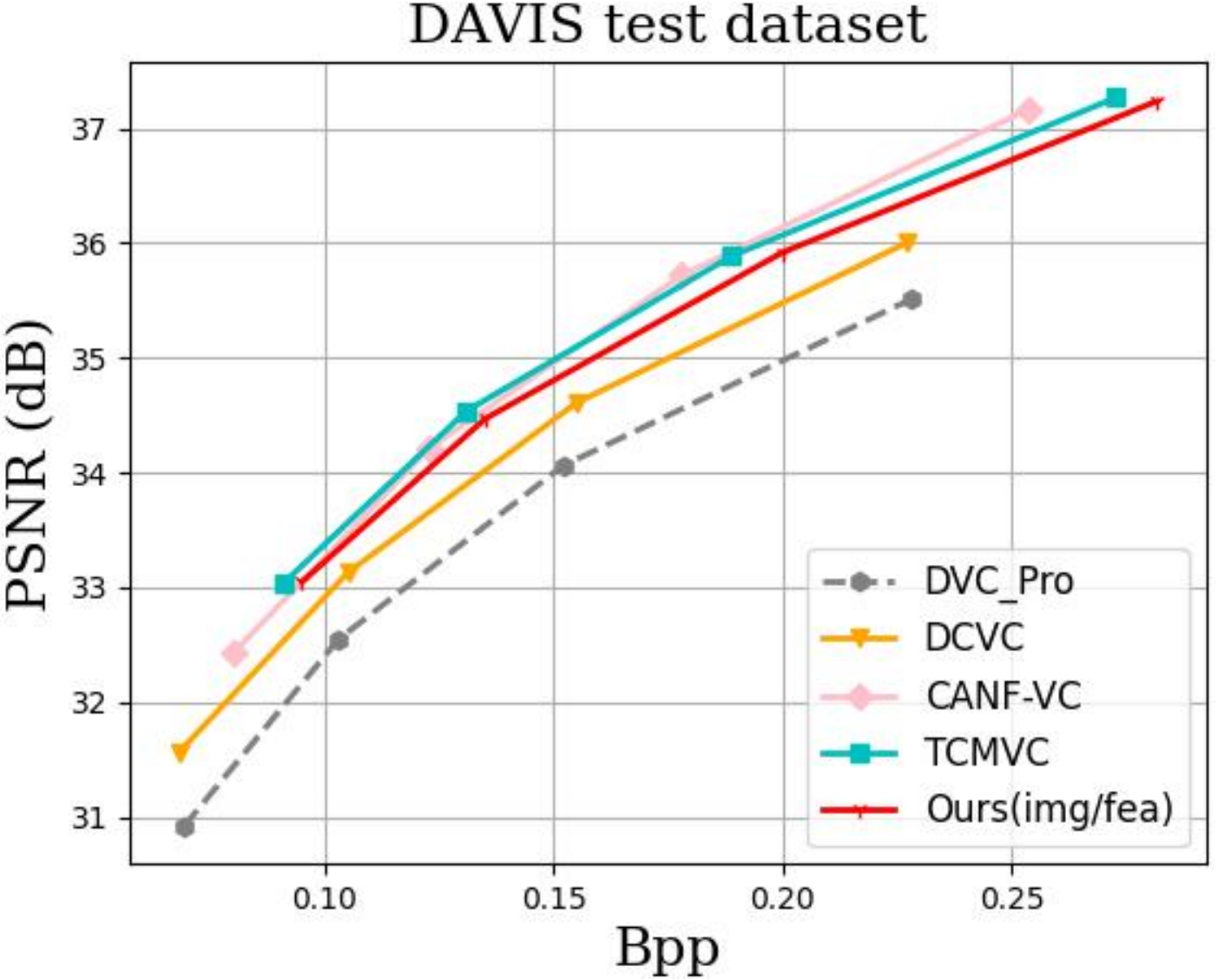}
 \end{minipage}%
  \begin{minipage}[c]{0.48\linewidth}
  \centering
    \includegraphics[width=\linewidth]{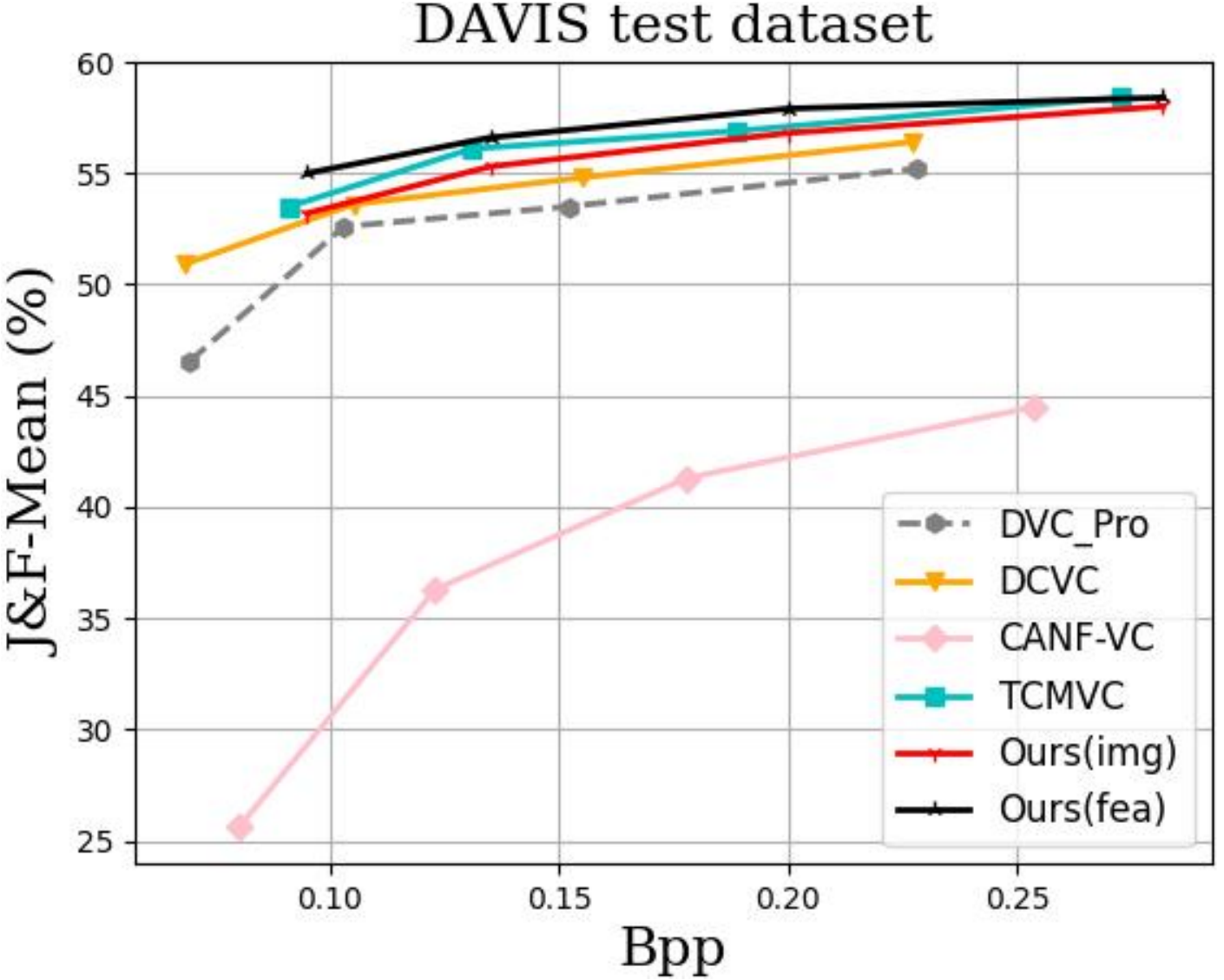}
  \end{minipage}%
    \caption{Rate-distortion and rate-performance (video object segmentation J\&F-Mean) curves on the DAVIS 2017 test dataset.}
  \label{fig:all_codecs_VOS}
\end{figure}
\begin{table}[t]
\caption{BD-performmance for video object segmentation on the DAVIS 2017 test dataset. The anchor is Ours(fea). Negative values indicate performance loss while positive values indicate higher performance.}
  \centering
\scalebox{1}{
\begin{threeparttable}
\begin{tabular}{l|c}
\toprule[1.5pt]
Ours(fea)  &0.00      \\ \hline
Ours(img)  &--1.18      \\ \hline
TCMVC    &--0.60     \\\hline
CANF-VC    &--17.94  \\\hline
DCVC      &--2.12\\ \hline
DVC\_Pro   &--3.31  \\ \bottomrule[1.5pt]
\end{tabular}
\end{threeparttable}}
\label{table:VOS}
\end{table}
 \begin{table}[!t]
 \caption{Average decoding time (entropy decoding time is excluded), task time, decoding GFLOPs, and task GFLOPs for each frame of the DAVIS 2017 test dataset.}
   \centering
\scalebox{1}{
\begin{threeparttable}
\begin{tabular}{lcccc}
\toprule[1.5pt]
\multicolumn{1}{l|}{Schemes}          & \multicolumn{1}{c|}{Dec T (s)} & \multicolumn{1}{c|}{Task T (s)} & \multicolumn{1}{c|}{Dec G} & Task G \\ \hline
\multicolumn{1}{l|}{Ours(fea)} & \multicolumn{1}{c|}{0.013}             & \multicolumn{1}{c|}{0.064}         & \multicolumn{1}{c|}{1034.17}         & 127.16      \\ \hline
\multicolumn{1}{l|}{Ours(img)} & \multicolumn{1}{c|}{0.021}             & \multicolumn{1}{c|}{0.063}         & \multicolumn{1}{c|}{1482.97}         & 87.97      \\ 
\bottomrule[1.5pt]
\end{tabular}
\label{table:vos_time}
\end{threeparttable}}
\end{table}
\subsubsection{Results for Video Object Detection}For video object detection, we illustrate the rate-performance (video object detection box AP@50) curves on the ImageNet VID dataset in Fig.~\ref{fig:all_codecs_vid}. We also list the detailed BD-performance in Table\ref{table:VID}. From the experimental results, we can observe that Ours(fea) achieves higher object detection accuracy compared with Ours(img), which shows that the intermediate feature is more effective for the SELAS model. This may be owing to that the intermediate feature contains more useful information for video object detection. When humans need to further analyze the videos detected by machines, video reconstruction is needed. Therefore, we also illustrate the rate-distortion curves in Fig.~\ref{fig:all_codecs_vid}. Compared with other video codecs, Ours(fea) achieves the best video reconstruction performance. \par
In terms of computational complexity, we list the complexity comparison of Ours(fea) and Our(img) in Table~\ref{table:vid_time}. The comparison results show that Ours(fea) can bring much decoding complexity saving with a little task complexity increase, which is more efficient for video object detection.

\subsubsection{Results for Video Multiple Object Tracking}
For video multiple object tracking, we present the rate-performance (video multiple object tracking MOTA) curves on the MOT17 dataset in Fig.~\ref{fig:all_codecs_mot}. We also list the detailed BD-performance in Table~\ref{table:MOT}. The experimental results show that Ours(fea) achieves higher video multiple object tracking accuracy compared with Ours(img). When humans need to further analyze the videos tracked by machines, video reconstruction is needed. We present the rate-distortion curves in Fig.~\ref{fig:all_codecs_mot}. Compared with other video codecs, Ours(fea) achieves the best video reconstruction performance.\par
We compare the computation complexity of Ours(fea) and Ours(img), as listed in Table~\ref{table:mot_time}. The comparison shows that Ours(fea) can save much decoding time and decoding GFLOPs for video reconstruction, which is more efficient for video multiple object tracking.\par
\subsubsection{Results for Video Object Segmentation}
For video object segmentation, we present the rate-performance (video object segmentation J\&F-Mean) curves on the DAVIS2017 test dataset in Fig.~\ref{fig:all_codecs_VOS}. We also list the detailed BD-performance in Table~\ref{table:VOS}. The experimental results show that Ours(fea) achieves higher video object segmentation performance compared with Ours(img). When humans need to further analyze the videos segmented by machines, we convert the intermediate feature to the pixel domain to perform video reconstruction. We present the rate-distortion curves in Fig.~\ref{fig:all_codecs_VOS}. Compared with other video codecs, the video reconstruction performance of Ours(fea) is a little worse than TCMVC and CANF-VC. However, the video segmentation performance of Ours(fea) is better than that of TCMVC and CANF-VC.\par
In terms of computation cost, we compare the decoding
time, segmentation time, decoding GFLOPs, and segmentation GFLOPs of Ours(fea) and Ours(img). As listed in Table~\ref{table:vos_time}, Ours(fea) can save much decoding complexity that is used for video reconstruction.
\begin{table}[t]
\caption{Influence of the complexity of the video reconstruction network on video compression efficiency of the  HEVC dataset. Compression efficiency difference is measured by BD-rate (\%).}
  \centering
\begin{tabular}{l|c|c|c|c|c}
\toprule[1.5pt]
\diagbox{Schemes}{Datasets} & B   & C  &  D & E &RGB \\ \hline
2 U-Net (Ours)    & 0.0 & 0.0& 0.0& 0.0& 0.0\\ \hline
1 U-Net     & 0.9 & 1.0& 0.2& 2.1& 1.5\\ \hline
2 residual blocks    & 3.1 & 2.2&1.2&5.7&3.7\\ \hline
1 residual blocks     &3.0&3.3&2.1&6.5&4.7\\
\bottomrule[1.5pt]
\end{tabular}
\label{ablation:frame_genrator}
\end{table}

\section{Ablation Study and Analysis}\label{sec:ablation}
\subsection{Influence of the Complexity of the Video Reconstruction Network}
The complexity of the decoder affects the compression performance~\cite{mallikarachchi2020decoding,li2022hybrid}. To explore the influence of the complexity of the video reconstruction network in our proposed framework on the performance of video compression, we perform an ablation study by replacing the video reconstruction network which is made up of two cascaded U-Nets with one U-Net, two residual blocks, and one residual block. The BD-rate comparison results on the HEVC dataset are shown in Table~\ref{ablation:frame_genrator}. The anchor is our framework (Ours) with the video reconstruction network consisting of two cascaded U-Nets. From this table, we can find that the compression performance increases a lot when the complexity of the video reconstruction network increases. This proves that when high-quality reconstructed videos are needed, a high-complexity video reconstruction network is needed to transform the intermediate feature $\hat{F}_t$ to a fully decoded frame $\hat{x}_t$ in the pixel domain. However, when oriented to video enhancement and video analysis tasks, the complex video reconstruction network takes a large complexity burden. If using the decoded pixel-domain frames to perform downstream tasks, the heavy decoding complexity is not efficient for real-time task execution.

\subsection{Analysis of Feature-based Temporal Context Mining}
How to maintain high video reconstruction quality and save decoding complexity for downstream vision tasks is the main problem that we target to address. To achieve this target, the choice of which features to use as the reference feature to learn temporal contexts becomes the key to the problem. As illustrated by Fig.~\ref{fig:feature_based_compression_loop}, there have been three kinds of ways to learn temporal contexts so far, i.e., the intermediate feature before being fed into the video reconstruction network (Ours) (Fig.~\ref{fig:TCM} (a)), the input feature of the last convolutional layer of the video reconstruction network (Fig.~\ref{fig:TCM} (b)), and the pixel-domain reconstructed frame (Fig.~\ref{fig:TCM} (c)). To explore their performance differences, we design two variant models based on our proposed scheme. The variant models learn temporal contexts from the input feature of the last convolutional layer of the video reconstruction network (last feature) and reconstructed frames, respectively. We list the BD-rate results on the HEVC datasets in Table~\ref{ablation:tcm}. When calculating the BD-rate, we regard our framework (Ours) as the anchor.
The results show that feature-based temporal context mining achieves better compression performance than pixel-based temporal context mining. Among feature-based temporal context mining schemes, learning temporal contexts from the last feature achieves an average 0.74\% BD-rate reduction compared with that from the intermediate feature. However, as listed in Table~\ref{ablation:context_time}, when oriented to downstream vision tasks, learning temporal contexts from the last feature still needs much decoding complexity. Even if we use the last feature to perform downstream tasks, only the complexity of the last convolutional layer can be saved. 
\begin{table}[!t]
\caption{Ablation study on different ways to learn temporal context. Compression efficiency difference is measured by BD-rate (\%).}
  \centering
\begin{tabular}{l|c|c|c|c|c}
\toprule[1.5pt]
\diagbox{Schemes}{Datasets} & B   & C  &  D & E &RGB \\ \hline
intermediate feature (Ours)    & 0.0 & 0.0& 0.0& 0.0& 0.0\\ \hline
last feature  & --0.6 & --2.6& --3.5&3.3& --0.3\\ \hline
reconstructed frame   & 2.2 &0.7&1.3&3.2&1.7\\
\bottomrule[1.5pt]
\end{tabular}
\label{ablation:tcm}
\end{table}
\begin{table}[t]
\caption{Average decoding time (entropy decoding time is included) and decoding GFLOPs for a 1080p frame when oriented to human and machine vision tasks.}
  \centering
\begin{tabular}{l|c|c}
\toprule[1.5pt]
Schemes & Dec T (s) & Dec G\\ \hline
intermediate feature (Ours)    & 0.936 & 4709.17\\ \hline
last feature  & 0.960 & 6745.61\\ \hline
reconstructed frame   & 0.957 &6742.93\\
\bottomrule[1.5pt]
\end{tabular}
\label{ablation:context_time}
\end{table}
\begin{table}[t]
\caption{Ablation study on different ways to estimate motion vectors. Compression efficiency difference is measured by BD-rate (\%).}
  \centering
\begin{tabular}{l|c|c|c|c|c}
\toprule[1.5pt]
\diagbox{Schemes}{Datasets} & B   & C  &  D & E &RGB \\ \hline
cross-domain (Ours)    & 0.0 & 0.0& 0.0& 0.0& 0.0\\ \hline
pixel-domain  & --1.6 & --1.0& --2.1& 2.9& 1.2\\ \hline
feature-domain   & 2.3 &0.7&0.1&4.1&1.8\\
\bottomrule[1.5pt]
\end{tabular}
\label{ablation:cross}
\end{table}
\begin{table}[t]
\caption{Average encoding time (entropy encoding time is included) and encoding GFLOPs for a 1080p frame.}
  \centering
\begin{tabular}{l|c|c}
\toprule[1.5pt]
Schemes & Enc T (s) & Enc G\\ \hline
cross-domain (Ours)    & 1.336 & 4606.08\\ \hline
pixel-domain  & 2.075 & 7909.90\\ \hline
feature-domain   & 1.871 &6730.34\\
\bottomrule[1.5pt]
\end{tabular}
\label{ablation:cross_time}
\end{table}
\subsection{Analysis of Cross-Domain Motion Encoder and Decoder}
In addition to saving the computational complexity of the decoder, we find that using the decoded intermediate feature as a reference feature can also save a lot of computational complexity for the encoder. As illustrated in Fig.~\ref{fig:mv_encoder_decoder}, there are three kinds of ways to estimate and compress motion vectors, i.e. cross-domain-based schemes (Ours), pixel-domain-based schemes, and feature-domain-based schemes. 
We build two variant models to verify the effectiveness of our proposed cross-domain motion encoder and decoder. For the pixel-domain-based variant, following previous work~\cite {sheng2022temporal,li2021deep}, we use a pre-trained Spynet~\cite{ranjan2017optical} to estimate the motion vector from the current input frame $x_t$ and pixel-domain reference frame $\hat{x}_{t-1}$. The estimated motion vectors $v_t$ are compressed and then decoded to $\hat{v}_t$ using the moiton encoder-decoder as shown in Fig.\ref{fig:mv_encoder_decoder} (b). For the feature-domain-based variant model, we use convolutional layers to transform $x_t$ and $\hat{x}_{t-1}$ into the feature-domain first. The number of feature channels is 64. Then we feed the transformed features into the motion encoder-decoder to compress and reconstruct motion vectors, as shown in Fig.\ref{fig:mv_encoder_decoder} (c).  The number of the input channel of the motion encoder is 128 and the number of the output channel of the motion decoder is 2. We list the BD-rate results of three contrasting models on the HEVC dataset in Table~\ref{ablation:cross} and their encoding time/GFLOPs in Table~\ref{ablation:cross_time}.  The experimental results show that the pixel-domain-based variant model gets a better compression performance on a part of videos but it needs a large computational cost for Spynet and pixel-domain video reconstruction in the encoder.
The feature-domain variant model achieves the worst compression performance. Although it also does not need Spynet, it needs pixel-domain video reconstruction in the encoder, which also brings much encoding complexity.
Our proposed cross-domain motion encoder and decoder bring a little compression performance drop on average but it greatly reduces the encoding complexity. \par

\section{Conclusion}\label{sec:conclusion}
In this paper, we propose a versatile neural video coding framework (VNVC) for efficient human and machine vision. The framework enables the use of a single-bitstream, compact coded representation, targeting video reconstruction, video enhancement, and video analysis tasks simultaneously. We propose to partially decode the compact coded representation into an intermediate feature that can be directly fed into the downstream video enhancement and video analysis algorithms, without the need for pixel-domain reconstructed videos. Experimental results show that our framework can achieve high compression efficiency for video reconstruction and satisfactory performances for various downstream video enhancement and video analysis tasks with lower decoding complexities.
In the future, we will focus on further improving the compression performance of our framework and reducing the computational complexity for practical applications.

\bibliographystyle{IEEEtran}
\bibliography{bare_jrnl_compsoc}

\begin{thebibliography}{10}
\providecommand{\url}[1]{#1}
\csname url@samestyle\endcsname
\providecommand{\newblock}{\relax}
\providecommand{\bibinfo}[2]{#2}
\providecommand{\BIBentrySTDinterwordspacing}{\spaceskip=0pt\relax}
\providecommand{\BIBentryALTinterwordstretchfactor}{4}
\providecommand{\BIBentryALTinterwordspacing}{\spaceskip=\fontdimen2\font plus
\BIBentryALTinterwordstretchfactor\fontdimen3\font minus
  \fontdimen4\font\relax}
\providecommand{\BIBforeignlanguage}[2]{{%
\expandafter\ifx\csname l@#1\endcsname\relax
\typeout{** WARNING: IEEEtran.bst: No hyphenation pattern has been}%
\typeout{** loaded for the language `#1'. Using the pattern for}%
\typeout{** the default language instead.}%
\else
\language=\csname l@#1\endcsname
\fi
#2}}
\providecommand{\BIBdecl}{\relax}
\BIBdecl

\bibitem{barnett2018cisco}
T.~Barnett, S.~Jain, U.~Andra, and T.~Khurana, ``Cisco visual networking index
  (vni) complete forecast update, 2017--2022,'' \emph{Americas/EMEAR Cisco
  Knowledge Network (CKN) Presentation}, pp. 1--30, 2018.

\bibitem{gao2021recent}
W.~Gao, S.~Liu, X.~Xu, M.~Rafie, Y.~Zhang, and I.~Curcio, ``Recent standard
  development activities on video coding for machines,'' \emph{arXiv preprint
  arXiv:2105.12653}, 2021.

\bibitem{tassano2019dvd}
M.~Tassano, J.~Delon, and T.~Veit, ``Dvdnet: A fast network for deep video
  denoising,'' in \emph{2019 IEEE International Conference on Image Processing
  (ICIP)}, 2019, pp. 1805--1809.

\bibitem{tassano2020fastdvdnet}
------, ``Fastdvdnet: Towards real-time deep video denoising without flow
  estimation,'' in \emph{Proceedings of the IEEE/CVF Conference on Computer
  Vision and Pattern Recognition (CVPR)}, 2020, pp. 1354--1363.

\bibitem{tian2020tdan}
Y.~Tian, Y.~Zhang, Y.~Fu, and C.~Xu, ``Tdan: Temporally-deformable alignment
  network for video super-resolution,'' in \emph{Proceedings of the IEEE/CVF
  Conference on Computer Vision and Pattern Recognition (CVPR)}, 2020, pp.
  3360--3369.

\bibitem{wang2019edvr}
X.~Wang, K.~C. Chan, K.~Yu, C.~Dong, and C.~Change~Loy, ``Edvr: Video
  restoration with enhanced deformable convolutional networks,'' in
  \emph{Proceedings of the IEEE/CVF Conference on Computer Vision and Pattern
  Recognition Workshops}, 2019, pp. 0--0.

\bibitem{pan2021deep}
J.~Pan, H.~Bai, J.~Dong, J.~Zhang, and J.~Tang, ``Deep blind video
  super-resolution,'' in \emph{Proceedings of the IEEE/CVF International
  Conference on Computer Vision}, 2021, pp. 4811--4820.

\bibitem{pan2023cascaded}
J.~Pan, B.~Xu, H.~Bai, J.~Tang, and M.-H. Yang, ``Cascaded deep video
  deblurring using temporal sharpness prior and non-local spatial-temporal
  similarity,'' \emph{IEEE Transactions on Pattern Analysis and Machine
  Intelligence}, 2023.

\bibitem{zhu2022deep}
C.~Zhu, H.~Dong, J.~Pan, B.~Liang, Y.~Huang, L.~Fu, and F.~Wang, ``Deep
  recurrent neural network with multi-scale bi-directional propagation for
  video deblurring,'' in \emph{Proceedings of the AAAI conference on artificial
  intelligence}, vol.~36, no.~3, 2022, pp. 3598--3607.

\bibitem{lin2020tsm}
J.~Lin, C.~Gan, K.~Wang, and S.~Han, ``Tsm: Temporal shift module for efficient
  and scalable video understanding on edge devices,'' \emph{IEEE Transactions
  on Pattern Analysis and Machine Intelligence}, 2020.

\bibitem{carreira2017quo}
J.~Carreira and A.~Zisserman, ``Quo vadis, action recognition? a new model and
  the kinetics dataset,'' in \emph{proceedings of the IEEE Conference on
  Computer Vision and Pattern Recognition (CVPR)}, 2017, pp. 6299--6308.

\bibitem{feichtenhofer2019slowfast}
C.~Feichtenhofer, H.~Fan, J.~Malik, and K.~He, ``Slowfast networks for video
  recognition,'' in \emph{Proceedings of the IEEE/CVF International Conference
  on Computer Vision (ICCV)}, 2019, pp. 6202--6211.

\bibitem{zhu2017flow}
X.~Zhu, Y.~Wang, J.~Dai, L.~Yuan, and Y.~Wei, ``Flow-guided feature aggregation
  for video object detection,'' in \emph{Proceedings of the IEEE International
  Conference on Computer Vision (ICCV)}, 2017, pp. 408--417.

\bibitem{sun2020semantic}
S.~Sun, T.~He, and Z.~Chen, ``Semantic structured image coding framework for
  multiple intelligent applications,'' \emph{IEEE Transactions on Circuits and
  Systems for Video Technology}, vol.~31, no.~9, pp. 3631--3642, 2020.

\bibitem{pang2021quasi}
J.~Pang, L.~Qiu, X.~Li, H.~Chen, Q.~Li, T.~Darrell, and F.~Yu, ``Quasi-dense
  similarity learning for multiple object tracking,'' in \emph{Proceedings of
  the IEEE/CVF conference on computer vision and pattern recognition}, 2021,
  pp. 164--173.

\bibitem{zhang2022bytetrack}
Y.~Zhang, P.~Sun, Y.~Jiang, D.~Yu, F.~Weng, Z.~Yuan, P.~Luo, W.~Liu, and
  X.~Wang, ``Bytetrack: Multi-object tracking by associating every detection
  box,'' in \emph{European Conference on Computer Vision}.\hskip 1em plus 0.5em
  minus 0.4em\relax Springer, 2022, pp. 1--21.

\bibitem{yang2021aot}
Z.~Yang, Y.~Wei, and Y.~Yang, ``Associating objects with transformers for video
  object segmentation,'' in \emph{Advances in Neural Information Processing
  Systems (NeurIPS)}, 2021.

\bibitem{caelles2017one}
S.~Caelles, K.-K. Maninis, J.~Pont-Tuset, L.~Leal-Taix{\'e}, D.~Cremers, and
  L.~Van~Gool, ``One-shot video object segmentation,'' in \emph{Proceedings of
  the IEEE conference on computer vision and pattern recognition}, 2017, pp.
  221--230.

\bibitem{ascenso2023jpeg}
J.~Ascenso, E.~Alshina, and T.~Ebrahimi, ``The jpeg ai standard: Providing
  efficient human and machine visual data consumption,'' \emph{Ieee
  Multimedia}, vol.~30, no.~1, pp. 100--111, 2023.

\bibitem{ascenso2021jpeg}
J.~Ascenso, ``Jpeg ai use cases and requirements,'' \emph{ISO/IEC JTC1/SC29/WG1
  M90014}, 2021.

\bibitem{duan2015overview}
L.-Y. Duan, V.~Chandrasekhar, J.~Chen, J.~Lin, Z.~Wang, T.~Huang, B.~Girod, and
  W.~Gao, ``Overview of the mpeg-cdvs standard,'' \emph{IEEE Transactions on
  Image Processing}, vol.~25, no.~1, pp. 179--194, 2015.

\bibitem{duan2018compact}
L.-Y. Duan, Y.~Lou, Y.~Bai, T.~Huang, W.~Gao, V.~Chandrasekhar, J.~Lin,
  S.~Wang, and A.~C. Kot, ``Compact descriptors for video analysis: The
  emerging mpeg standard,'' \emph{IEEE MultiMedia}, vol.~26, no.~2, pp. 44--54,
  2018.

\bibitem{duan2020video}
L.~Duan, J.~Liu, W.~Yang, T.~Huang, and W.~Gao, ``Video coding for machines: A
  paradigm of collaborative compression and intelligent analytics,'' \emph{IEEE
  Transactions on Image Processing}, vol.~29, pp. 8680--8695, 2020.

\bibitem{Xia2020AnEC}
S.~Xia, K.~Liang, W.~Yang, L.~yu~Duan, and J.~Liu, ``An emerging coding
  paradigm vcm: A scalable coding approach beyond feature and signal,''
  \emph{2020 IEEE International Conference on Multimedia and Expo (ICME)}, pp.
  1--6, 2020.

\bibitem{jin2022semantically}
X.~Jin, R.~Feng, S.~Sun, R.~Feng, T.~He, and Z.~Chen, ``Semantically video
  coding: Instill static-dynamic clues into structured bitstream for ai
  tasks,'' \emph{arXiv preprint arXiv:2201.10162}, 2022.

\bibitem{choi2022scalable}
H.~Choi and I.~V. Baji{\'c}, ``Scalable video coding for humans and machines,''
  in \emph{2022 IEEE 24th International Workshop on Multimedia Signal
  Processing (MMSP)}.\hskip 1em plus 0.5em minus 0.4em\relax IEEE, 2022, pp.
  1--6.

\bibitem{lu2020end}
G.~Lu, X.~Zhang, W.~Ouyang, L.~Chen, Z.~Gao, and D.~Xu, ``An end-to-end
  learning framework for video compression,'' \emph{IEEE Transactions on
  Pattern Analysis and Machine Intelligence}, 2020.

\bibitem{lin2020m}
J.~Lin, D.~Liu, H.~Li, and F.~Wu, ``{M-LVC}: Multiple frames prediction for
  learned video compression,'' in \emph{Proceedings of the IEEE/CVF Conference
  on Computer Vision and Pattern Recognition (CVPR)}, 2020, pp. 3546--3554.

\bibitem{hu2022fvc}
Z.~Hu, D.~Xu, G.~Lu, W.~Jiang, W.~Wang, and S.~Liu, ``Fvc: An end-to-end
  framework towards deep video compression in feature space,'' \emph{IEEE
  Transactions on Pattern Analysis and Machine Intelligence}, 2022.

\bibitem{li2021deep}
J.~Li, B.~Li, and Y.~Lu, ``Deep contextual video compression,'' \emph{Advances
  in Neural Information Processing Systems}, vol.~34, pp. 18\,114--18\,125,
  2021.

\bibitem{sheng2022temporal}
X.~Sheng, J.~Li, B.~Li, L.~Li, D.~Liu, and Y.~Lu, ``Temporal context mining for
  learned video compression,'' \emph{IEEE Transactions on Multimedia}, 2022.

\bibitem{shi2022alphavc}
Y.~Shi, Y.~Ge, J.~Wang, and J.~Mao, ``{AlphaVC}: High-performance and efficient
  learned video compression,'' in \emph{European Conference on Computer Vision
  (ECCV)}.\hskip 1em plus 0.5em minus 0.4em\relax Springer, 2022, pp. 616--631.

\bibitem{canfvc}
Y.-H. Ho, C.-P. Chang, P.-Y. Chen, A.~Gnutti, and W.-H. Peng, ``{CANF-VC}:
  Conditional augmented normalizing flows for video compression,''
  \emph{European Conference on Computer Vision (ECCV)}, vol. 13676, pp.
  207--223, 2022.

\bibitem{li2022hybrid}
J.~Li, B.~Li, and Y.~Lu, ``Hybrid spatial-temporal entropy modelling for neural
  video compression,'' in \emph{Proceedings of the 30th ACM International
  Conference on Multimedia}, 2022, pp. 1503--1511.

\bibitem{lowe2004distinctive}
D.~G. Lowe, ``Distinctive image features from scale-invariant keypoints,''
  \emph{International journal of computer vision}, vol.~60, pp. 91--110, 2004.

\bibitem{PMID:31562087}
Z.~Chen, K.~Fan, S.~Wang, L.~Duan, W.~Lin, and A.~C. Kot, ``Intermediate deep
  feature compression: Toward intelligent sensing,'' \emph{IEEE transactions on
  image processing : a publication of the IEEE Signal Processing Society},
  September 2019.

\bibitem{zhang2021use}
Y.~Zhang, M.~Rafie, and S.~Liu, ``Use cases and requirements for video coding
  for machines,'' \emph{ISO/IEC JTC}, vol.~1, 2021.

\bibitem{wu2018video}
C.-Y. Wu, N.~Singhal, and P.~Krahenbuhl, ``Video compression through image
  interpolation,'' in \emph{Proceedings of the European Conference on Computer
  Vision (ECCV)}, 2018, pp. 416--431.

\bibitem{liu2020conditional}
J.~Liu, S.~Wang, W.-C. Ma, M.~Shah, R.~Hu, P.~Dhawan, and R.~Urtasun,
  ``Conditional entropy coding for efficient video compression,'' in
  \emph{European Conference on Computer Vision (ECCV)}.\hskip 1em plus 0.5em
  minus 0.4em\relax Springer, 2020, pp. 453--468.

\bibitem{mentzer2022vct}
F.~Mentzer, G.~Toderici, D.~Minnen, S.-J. Hwang, S.~Caelles, M.~Lucic, and
  E.~Agustsson, ``Vct: A video compression transformer,'' \emph{arXiv preprint
  arXiv:2206.07307}, 2022.

\bibitem{Habibian_2019_ICCV}
A.~Habibian, T.~v. Rozendaal, J.~M. Tomczak, and T.~S. Cohen, ``Video
  compression with rate-distortion autoencoders,'' in \emph{Proceedings of the
  IEEE/CVF International Conference on Computer Vision (ICCV)}, October 2019.

\bibitem{sun2020high}
W.~Sun, C.~Tang, W.~Li, Z.~Yuan, H.~Yang, and Y.~Liu, ``High-quality
  single-model deep video compression with frame-conv3d and multi-frame
  differential modulation,'' in \emph{European Conference on Computer Vision
  (ECCV)}.\hskip 1em plus 0.5em minus 0.4em\relax Springer, 2020, pp. 239--254.

\bibitem{sullivan2012overview}
G.~J. Sullivan, J.-R. Ohm, W.-J. Han, and T.~Wiegand, ``Overview of the high
  efficiency video coding ({HEVC}) standard,'' \emph{IEEE Transactions on
  circuits and systems for video technology}, vol.~22, no.~12, pp. 1649--1668,
  2012.

\bibitem{wiegand2003overview}
T.~Wiegand, G.~J. Sullivan, G.~Bjontegaard, and A.~Luthra, ``Overview of the
  {H.264/AVC} video coding standard,'' \emph{IEEE Transactions on Circuits and
  Systems for Video Technology}, vol.~13, no.~7, pp. 560--576, 2003.

\bibitem{bross2021overview}
B.~Bross, Y.-K. Wang, Y.~Ye, S.~Liu, J.~Chen, G.~J. Sullivan, and J.-R. Ohm,
  ``Overview of the versatile video coding ({VVC}) standard and its
  applications,'' \emph{IEEE Transactions on Circuits and Systems for Video
  Technology}, 2021.

\bibitem{balle2018variational}
J.~Ball{\'{e}}, D.~Minnen, S.~Singh, S.~J. Hwang, and N.~Johnston,
  ``Variational image compression with a scale hyperprior,'' in
  \emph{International Conference on Learning Representations (ICLR)}, 2018.

\bibitem{ranjan2017optical}
A.~Ranjan and M.~J. Black, ``Optical flow estimation using a spatial pyramid
  network,'' in \emph{Proceedings of the IEEE/CVF Conference on Computer Vision
  and Pattern Recognition (CVPR)}, 2017, pp. 4161--4170.

\bibitem{ilg2017flownet}
E.~Ilg, N.~Mayer, T.~Saikia, M.~Keuper, A.~Dosovitskiy, and T.~Brox, ``Flownet
  2.0: Evolution of optical flow estimation with deep networks,'' in
  \emph{Proceedings of the IEEE conference on computer vision and pattern
  recognition (CVPR)}, 2017, pp. 2462--2470.

\bibitem{dai2017deformable}
J.~Dai, H.~Qi, Y.~Xiong, Y.~Li, G.~Zhang, H.~Hu, and Y.~Wei, ``Deformable
  convolutional networks,'' in \emph{Proceedings of the IEEE international
  conference on computer vision}, 2017, pp. 764--773.

\bibitem{DBLP:conf/iclr/BalleLS17}
J.~Ball{\'{e}}, V.~Laparra, and E.~P. Simoncelli, ``End-to-end optimized image
  compression,'' in \emph{International Conference on Learning Representations,
  (ICLR) 2017, Toulon, France, April 24-26, 2017, Conference Track
  Proceedings}, 2017.

\bibitem{NEURIPS2018_53edebc5}
D.~Minnen, J.~Ball\'{e}, and G.~D. Toderici, ``Joint autoregressive and
  hierarchical priors for learned image compression,'' in \emph{Advances in
  Neural Information Processing Systems}, vol.~31.\hskip 1em plus 0.5em minus
  0.4em\relax Curran Associates, Inc., 2018.

\bibitem{8701503}
J.~Hu, L.~Shen, S.~Albanie, G.~Sun, and E.~Wu, ``Squeeze-and-excitation
  networks,'' \emph{IEEE Transactions on Pattern Analysis and Machine
  Intelligence}, vol.~42, no.~8, pp. 2011--2023, 2020.

\bibitem{xia2017w}
X.~Xia and B.~Kulis, ``W-net: A deep model for fully unsupervised image
  segmentation,'' \emph{arXiv preprint arXiv:1711.08506}, 2017.

\bibitem{Liang2021SwinIRIR}
J.~Liang, J.~Cao, G.~Sun, K.~Zhang, L.~V. Gool, and R.~Timofte, ``Swinir: Image
  restoration using swin transformer,'' \emph{2021 IEEE/CVF International
  Conference on Computer Vision Workshops (ICCVW)}, pp. 1833--1844, 2021.

\bibitem{liu2021Swin}
Z.~Liu, Y.~Lin, Y.~Cao, H.~Hu, Y.~Wei, Z.~Zhang, S.~Lin, and B.~Guo, ``Swin
  transformer: Hierarchical vision transformer using shifted windows,'' in
  \emph{Proceedings of the IEEE/CVF International Conference on Computer Vision
  (ICCV)}, 2021.

\bibitem{he2016deep}
K.~He, X.~Zhang, S.~Ren, and J.~Sun, ``Deep residual learning for image
  recognition,'' in \emph{Proceedings of the IEEE Conference on Computer Vision
  and Pattern Recognition (CVPR)}, 2016, pp. 770--778.

\bibitem{wu2019sequence}
H.~Wu, Y.~Chen, N.~Wang, and Z.~Zhang, ``Sequence level semantics aggregation
  for video object detection,'' in \emph{Proceedings of the IEEE/CVF
  International Conference on Computer Vision (ICCV)}, 2019, pp. 9217--9225.

\bibitem{7485869}
S.~Ren, K.~He, R.~Girshick, and J.~Sun, ``Faster r-cnn: Towards real-time
  object detection with region proposal networks,'' \emph{IEEE Transactions on
  Pattern Analysis and Machine Intelligence}, vol.~39, no.~6, pp. 1137--1149,
  2017.

\bibitem{qdtrack_conf}
J.~Pang, L.~Qiu, X.~Li, H.~Chen, Q.~Li, T.~Darrell, and F.~Yu, ``Quasi-dense
  similarity learning for multiple object tracking,'' in \emph{IEEE/CVF
  Conference on Computer Vision and Pattern Recognition}, June 2021.

\bibitem{xue2019video}
T.~Xue, B.~Chen, J.~Wu, D.~Wei, and W.~T. Freeman, ``Video enhancement with
  task-oriented flow,'' \emph{International Journal of Computer Vision}, vol.
  127, no.~8, pp. 1106--1125, 2019.

\bibitem{mercat2020uvg}
A.~Mercat, M.~Viitanen, and J.~Vanne, ``Uvg dataset: 50/120fps 4k sequences for
  video codec analysis and development,'' in \emph{Proceedings of the 11th ACM
  Multimedia Systems Conference}, 2020, pp. 297--302.

\bibitem{wang2016mcl}
H.~Wang, W.~Gan, S.~Hu, J.~Y. Lin, L.~Jin, L.~Song, P.~Wang, I.~Katsavounidis,
  A.~Aaron, and C.-C.~J. Kuo, ``{MCL-JCV}: a jnd-based {H.264/AVC} video
  quality assessment dataset,'' in \emph{2016 IEEE international conference on
  image processing (ICIP)}.\hskip 1em plus 0.5em minus 0.4em\relax IEEE, 2016,
  pp. 1509--1513.

\bibitem{david2013common}
F.~David, S.~Karl, and R.~Chris, ``{Doc. JCTVC-N1006}: Common test conditions
  and software reference configurations for {HEVC} range extensions,''
  \emph{Joint Collaborative Team on Video Coding (JCT-VC) of ITU-T SG}, 2013.

\bibitem{perazzi2016benchmark}
F.~Perazzi, J.~Pont-Tuset, B.~McWilliams, L.~Van~Gool, M.~Gross, and
  A.~Sorkine-Hornung, ``A benchmark dataset and evaluation methodology for
  video object segmentation,'' in \emph{Proceedings of the IEEE Conference on
  Computer Vision and Pattern Recognition (CVPR)}, 2016, pp. 724--732.

\bibitem{nah2019ntire}
S.~Nah, S.~Baik, S.~Hong, G.~Moon, S.~Son, R.~Timofte, and K.~Mu~Lee, ``Ntire
  2019 challenge on video deblurring and super-resolution: Dataset and study,''
  in \emph{Proceedings of the IEEE/CVF Conference on Computer Vision and
  Pattern Recognition Workshops}, 2019, pp. 0--0.

\bibitem{liu2013bayesian}
C.~Liu and D.~Sun, ``On bayesian adaptive video super resolution,'' \emph{IEEE
  Transactions on Pattern Analysis and Machine Intelligence}, vol.~36, no.~2,
  pp. 346--360, 2013.

\bibitem{yang2021real}
X.~YANG, W.~Xiang, H.~Zeng, and L.~Zhang, ``Real-world video super-resolution:
  A benchmark dataset and a decomposition based learning scheme,'' 2021.

\bibitem{soomro2012ucf101}
K.~Soomro, A.~R. Zamir, and M.~Shah, ``{UCF101}: A dataset of 101 human actions
  classes from videos in the wild,'' \emph{arXiv preprint arXiv:1212.0402},
  2012.

\bibitem{goyal2017something}
R.~Goyal, S.~E. Kahou, V.~Michalski, J.~Materzyńska, S.~Westphal, H.~Kim,
  V.~Haenel, I.~Fruend, P.~Yianilos, M.~Mueller-Freitag, F.~Hoppe, C.~Thurau,
  I.~Bax, and R.~Memisevic, ``The "something something" video database for
  learning and evaluating visual common sense,'' 2017.

\bibitem{idrees2017thumos}
H.~Idrees, A.~R. Zamir, Y.-G. Jiang, A.~Gorban, I.~Laptev, R.~Sukthankar, and
  M.~Shah, ``The thumos challenge on action recognition for videos “in the
  wild”,'' \emph{Computer Vision and Image Understanding}, vol. 155, pp.
  1--23, 2017.

\bibitem{russakovsky2015imagenet}
O.~Russakovsky, J.~Deng, H.~Su, J.~Krause, S.~Satheesh, S.~Ma, Z.~Huang,
  A.~Karpathy, A.~Khosla, M.~Bernstein \emph{et~al.}, ``Imagenet large scale
  visual recognition challenge,'' \emph{International Journal of Computer
  Vision}, vol. 115, no.~3, pp. 211--252, 2015.

\bibitem{milan2016mot16}
A.~Milan, L.~Leal-Taix{\'e}, I.~Reid, S.~Roth, and K.~Schindler, ``Mot16: A
  benchmark for multi-object tracking,'' \emph{arXiv preprint
  arXiv:1603.00831}, 2016.

\bibitem{Loshchilov2019DecoupledWD}
I.~Loshchilov and F.~Hutter, ``Decoupled weight decay regularization,'' in
  \emph{International Conference on Learning Representations (ICLR)}, 2019.

\bibitem{wang2003multiscale}
Z.~Wang, E.~P. Simoncelli, and A.~C. Bovik, ``Multiscale structural similarity
  for image quality assessment,'' in \emph{The Thrity-Seventh Asilomar
  Conference on Signals, Systems \& Computers, 2003}, vol.~2.\hskip 1em plus
  0.5em minus 0.4em\relax Ieee, 2003, pp. 1398--1402.

\bibitem{2020mmaction2}
{MMAction2 Contributors}, ``Openmmlab's next generation video understanding
  toolbox and benchmark,'' \url{https://github.com/open-mmlab/mmaction2}, 2020.

\bibitem{mmtrack2020}
{MMTracking Contributors}, ``{MMTracking: OpenMMLab} video perception toolbox
  and benchmark,'' \url{https://github.com/open-mmlab/mmtracking}, 2020.

\bibitem{huang2022neural}
C.~Huang, J.~Li, B.~Li, D.~Liu, and Y.~Lu, ``Neural compression-based feature
  learning for video restoration,'' in \emph{Proceedings of the IEEE/CVF
  Conference on Computer Vision and Pattern Recognition (CVPR)}, 2022, pp.
  5872--5881.

\bibitem{mallikarachchi2020decoding}
T.~Mallikarachchi, D.~Talagala, H.~Kodikara~Arachchi, C.~Hewage, and
  A.~Fernando, ``A decoding-complexity and rate-controlled video-coding
  algorithm for {HEVC},'' \emph{Future Internet}, vol.~12, no.~7, p. 120, 2020.

\end{thebibliography}
\end{document}


\title{Supplemental Material of \\ VNVC: A Versatile Neural Video Coding Framework for Efficient Human-Machine Vision}

\author{Xihua~Sheng,
        Li~Li,~\IEEEmembership{Member,~IEEE,} Dong~Liu,~\IEEEmembership{Senior~Member,~IEEE}, and Houqiang Li,~\IEEEmembership{Fellow,~IEEE}
\IEEEcompsocitemizethanks{\IEEEcompsocthanksitem Date of current version Jan 18, 2023. (Corresponding author: Li Li). 
\IEEEcompsocthanksitem The authors are with the CAS Key Laboratory of Technology in Geo-Spatial Information Processing and Application System, University of Science and Technology of China, Hefei 230027, China (e-mail: xhsheng@mail.ustc.edu.cn; lil1@ustc.edu.cn; dongeliu@ustc.edu.cn; lihq@ustc.edu.cn). \protect\\}

}
\markboth{}%
{Sheng \MakeLowercase{\textit{et al.}}: VNVC: A Versatile Neural Video Coding Framework for Efficient Human-Machine Vision}
\maketitle
\IEEEdisplaynontitleabstractindextext
\IEEEpeerreviewmaketitle
\section{Subjective Results for Video Denoising}
\begin{figure*}[htb]
  \centering
   \includegraphics[width=0.75\linewidth]{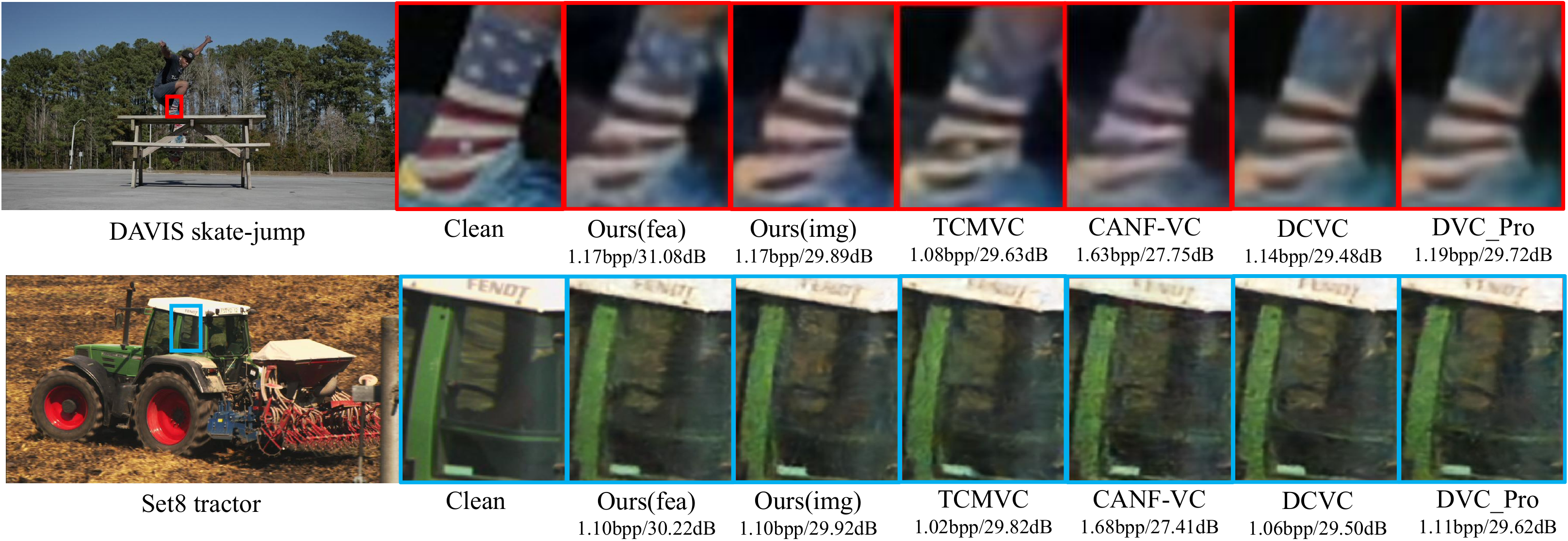}
   \caption{Subjective results for non-blind video denoising on DAVIS \emph{skate-jump} sequence and blind video denoising on Set8 \emph{tractor} sequence when $\delta=40$.}
\label{fig:subjective_denoising_non_blind_blind}
\end{figure*}

We illustrate the subjective results of non-blind/blind video denoising in Fig.~\ref{fig:subjective_denoising_non_blind_blind}. Experimental results show that the videos denoised by Ours(fea) can reduce more high-frequency noise while retaining more texture details.
\section{Subjective Results for Video Super-resolution}
\begin{figure*}[htb]
  \centering
   \includegraphics[width=0.7\linewidth]{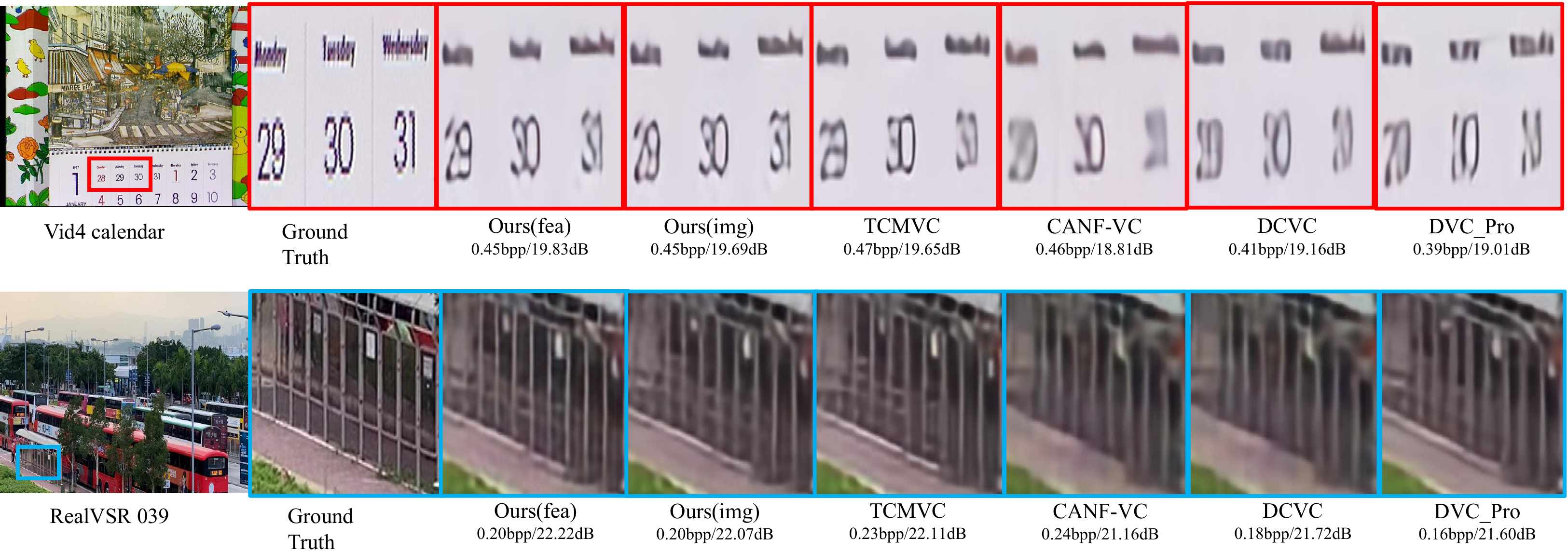}
   \caption{Subjective results for video super-resolution on Vid4 \emph{calendar} sequence and RealVSR \emph{039} sequence.}
   \label{fig:subjective_sr}
\end{figure*}

We illustrate the subjective results of video super-resolution in Fig.~\ref{fig:subjective_sr}. Experimental results show that the high-resolution videos generated by Ours(fea) have clearer texture details.

\section{Architecture for Video Reconstruction Network}
\begin{figure*}[htb]
  \centering
   \includegraphics[width=0.9\linewidth]{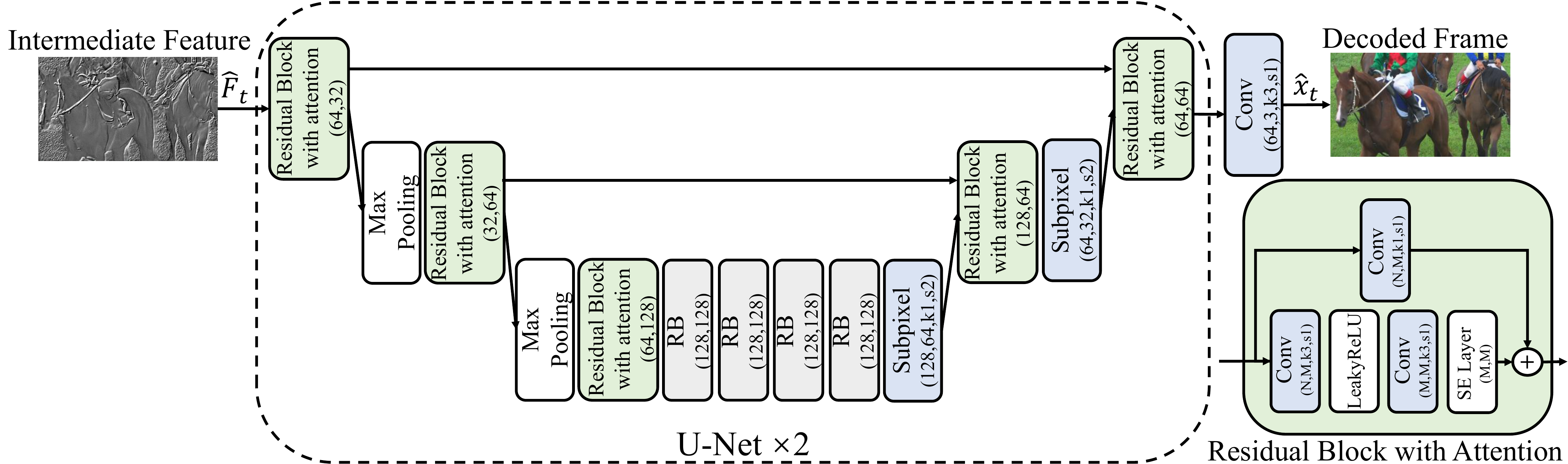}
   \caption{Architecture for the video reconstruction network. }
   \label{fig:frame_generator}
\end{figure*}

We illustrate the detailed architecture for the video reconstruction network in Fig.~\ref{fig:frame_generator}.
\section{Architecture for the Video Denoising and Video Super-Resolution Networks}
\begin{figure}[htb]
  \centering
   \includegraphics[width=0.9\linewidth]{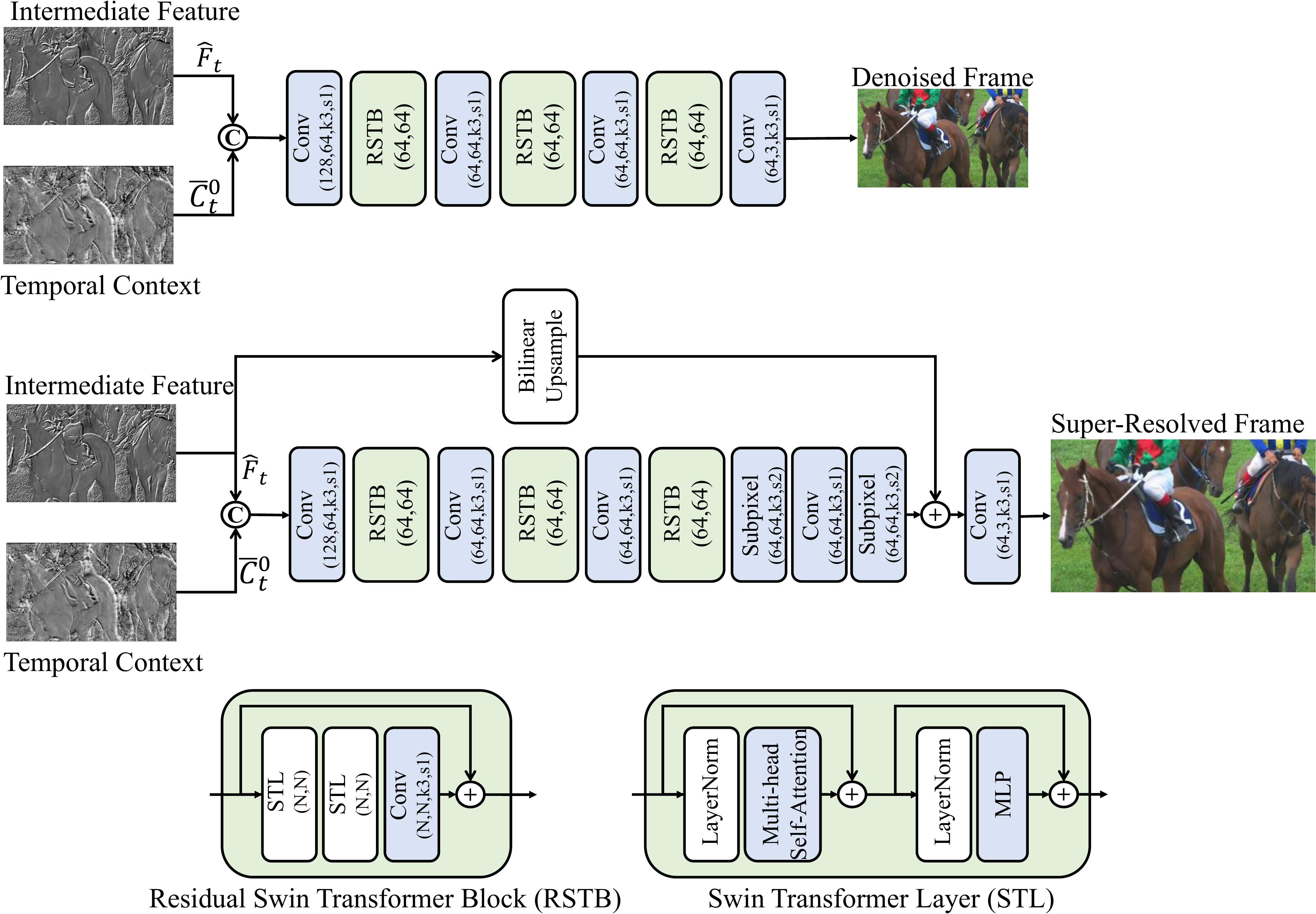}
   \caption{Architecture for video denoising and video super-resolution networks.}
   \label{fig:denoise_sr}
\end{figure}

We illustrate the detailed architectures for the video denoising networks and video super-resolution networks in Fig.~\ref{fig:denoise_sr}. 

\section{Architecture Comparison of the Video Denoising Networks of Ours(fea) and Ours(img)}
\begin{figure*}[htb]
  \centering
   \includegraphics[width=0.9\linewidth]{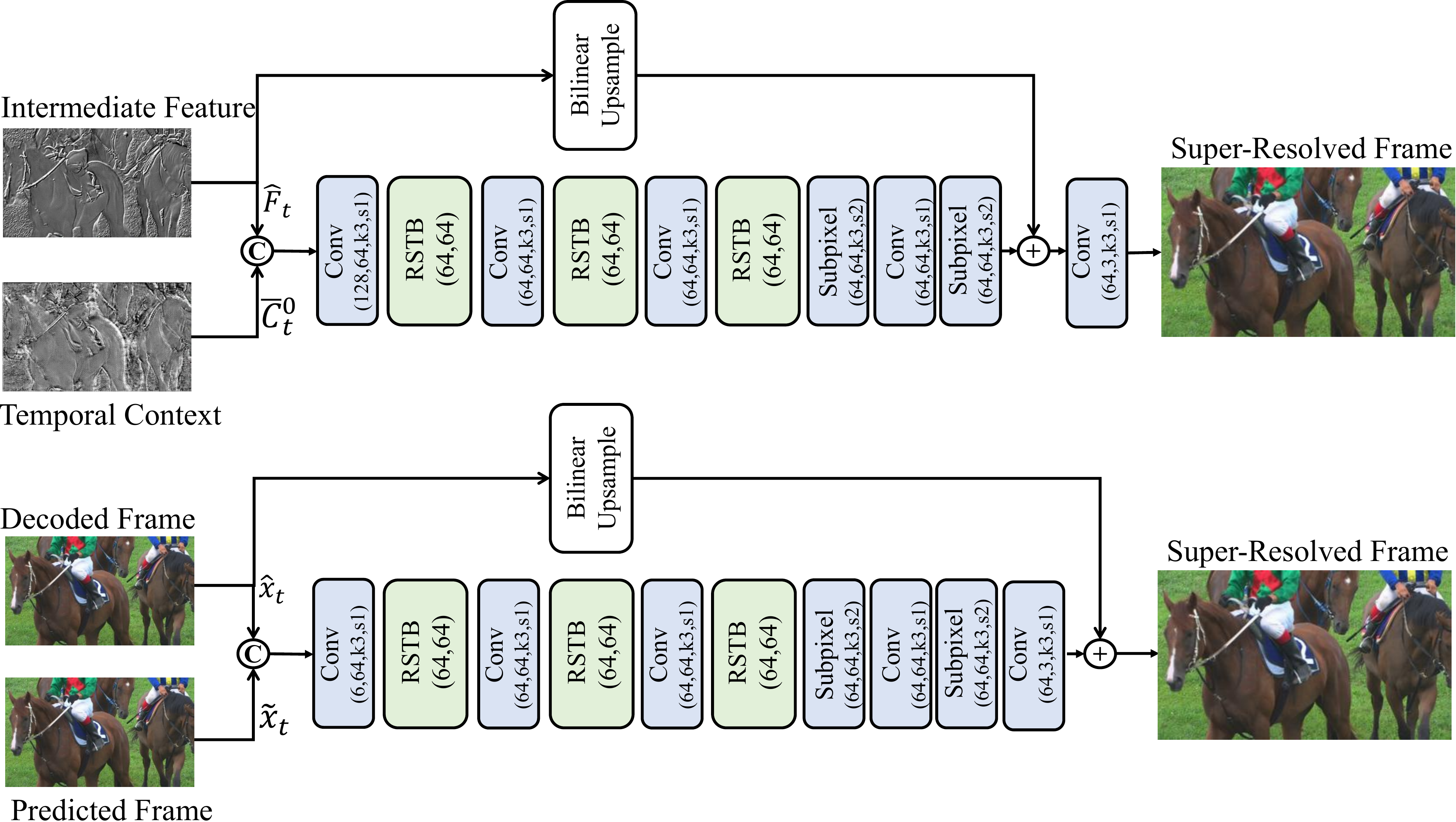}
   \caption{Architecture comparison of the video super-resolution networks of Ours(fea) and Ours(img).}
   \label{fig:sr_difference}
\end{figure*}

We compare the architectures of the video denoising networks of Ours(fea) and Ours(img) in Fig.~\ref{fig:sr_difference}. DVC\_Pro, DCVC, CANF-VC, and TCMVC also use decoded pixel-domain frames to perform video denoising using the networks with the same architecture as Ours(img).
\begin{figure*}[htb]
  \centering
   \includegraphics[width=0.8\linewidth]{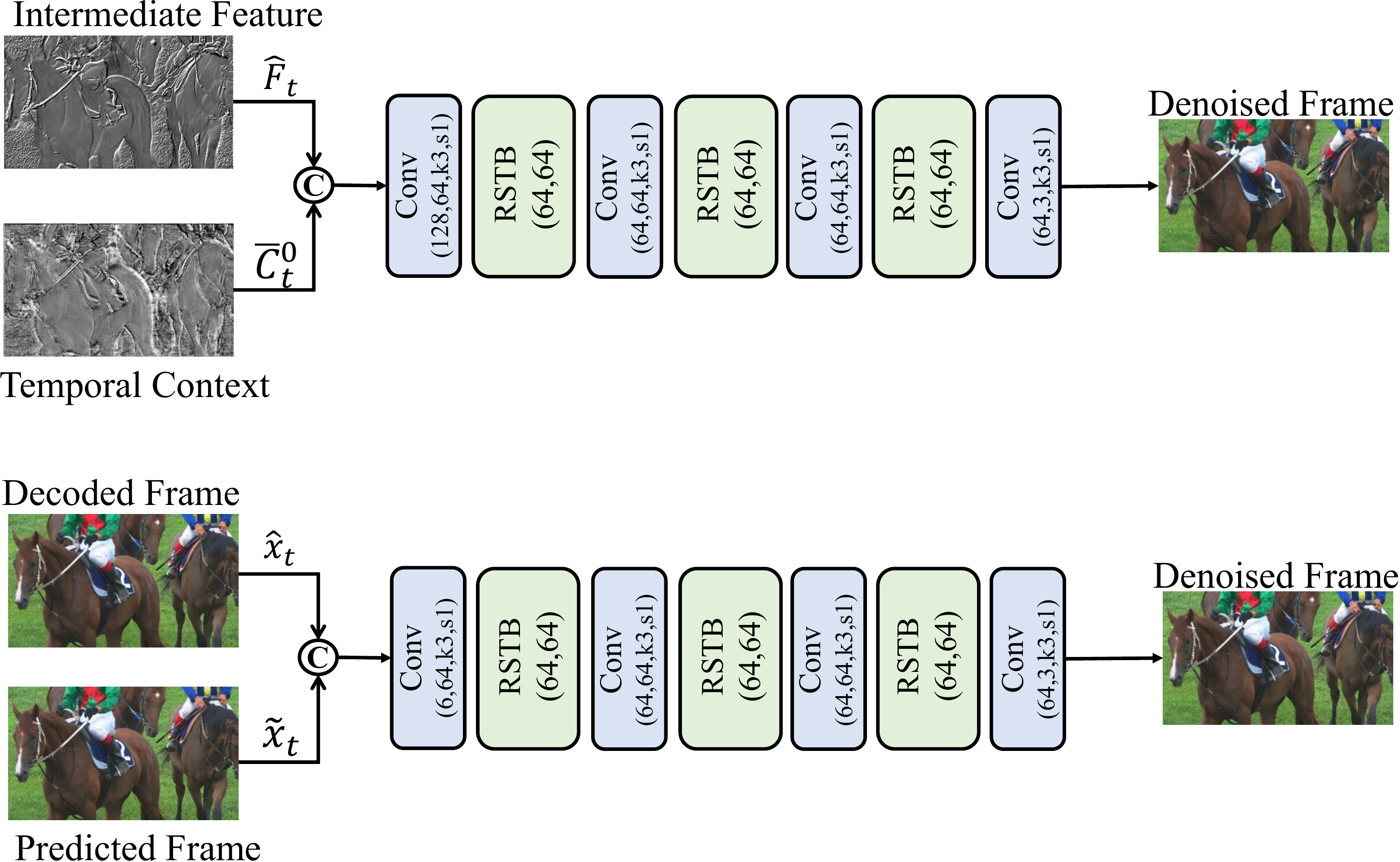}
   \caption{Architecture comparison of the video denoising networks of Ours(fea) and Ours(img).}
   \label{fig:denoise_difference}
\end{figure*}
\section{Architecture Comparison of the Video Super-Resolution Network of Ours(fea) and Ours(img)}
We compare the architectures of the video super-resolution networks of Ours(fea) and Ours(img) in Fig.~\ref{fig:denoise_difference}. DVC\_Pro, DCVC, CANF-VC, and TCMVC also use decoded pixel-domain frames to perform video super-resolution using the networks with the same architecture as Ours(img).
